# Status Report of the DPHEP Study Group: Towards a Global Effort for Sustainable Data Preservation in High Energy Physics

www.dphep.org

## Abstract

Data from high-energy physics (HEP) experiments are collected with significant financial and human effort and are mostly unique. An inter-experimental study group on HEP data preservation and long-term analysis was convened as a panel of the International Committee for Future Accelerators (ICFA). The group was formed by large collider-based experiments and investigated the technical and organisational aspects of HEP data preservation. An intermediate report was released in November 2009 addressing the general issues of data preservation in HEP. This paper includes and extends the intermediate report. It provides an analysis of the research case for data preservation and a detailed description of the various projects at experiment, laboratory and international levels. In addition, the paper provides a concrete proposal for an international organisation in charge of the data management and policies in high-energy physics.

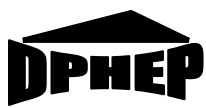

Study Group for Data Preservation and Long Term Analysis in High Energy Physics

# The DPHEP Study Group

Zaven Akopov, Deutsches Elektronen Synchrotron, DESY, Hamburg
Silvia Amerio, INFN Padova, Italy
David Asner, Pacific Northwest National Laboratory, USA
Eduard Avetisyan, Deutsches Elektronen Synchrotron, DESY, Hamburg
Olof Barring, CERN, Geneva, Switzerland
James Beacham, New York University, USA
Matthew Bellis, Stanford University, USA
Gregorio Bernardi, LPNHE, University Paris VI
Siegfried Bethke, MPI for Physics, Munich, Germany
Amber Boehnlein, Stanford Linear Accelerator Laboratory, SLAC, USA
Travis Brooks, Stanford Linear Accelerator Laboratory, SLAC, USA
Thomas Browder, Hawaii University, USA
Rene Brun, CERN, Geneva, Switzerland
Concetta Cartaro, Stanford Linear Accelerator Laboratory, SLAC, USA
Marco Cattaneo, CERN, Geneva, Switzerland
Gang Chen, IHEP, Beijing, China
David Corney, Rutherford Appleton Laboratory, STFC, UK
Kyle Cranmer, New York University, USA
Ray Culbertson, Fermi National Laboratory, FNAL, USA
Sunje Dallmeier-Tiessen , CERN, Geneva, Switzerland
Dmitri Denisov, Fermi National Laboratory, FNAL, USA
Cristinel Diaconu, CPPM, Aix-Marseille University, France and DESY, Germany
Vitaliy Dodonov, Deutsches Elektronen Synchrotron, DESY, Hamburg, Germany
Tony Doyle, University of Glasgow, SUPA, UK and CERN, Geneva, Switzerland
Gregory Dubois-Felsmann, Stanford Linear Accelerator Laboratory, SLAC, USA
Michael Ernst, Brookhaven National Laboratory, BNL, USA
Martin Gasthuber, Deutsches Elektronen Synchrotron, DESY, Hamburg, Germany
Achim Geiser, Deutsches Elektronen Synchrotron, DESY, Hamburg, Germany
Fabiola Gianotti, CERN, Geneva, Switzerland
Paolo Giubellino, CERN, Geneva, Switzerland
Andrey Golutvin, CERN, Geneva, Switzerland
John Gordon, Rutherford Appleton Laboratory, STFC, UK
Volker Guelzow, Deutsches Elektronen Synchrotron, DESY, Hamburg, Germany
Takanori Hara, KEK, Tsukuba, Japan
Hisaki Hayashii, Nara Women's University, Japan
Andreas Heiss, Karlsruhe Institute for Technology, Germany
Frederic Hemmer, CERN, Geneva, Switzerland
Fabio Hernandez, Computing Centre IN2P3, France
Graham Heyes, Jefferson National Laboratory, JLAB, USA
Andre Holzner, CERN, Geneva, Switzerland
Peter Igo-Kemenes, CERN, Geneva, Switzerland
Toru Iijima, Nagoya University, Japan
Joe Incandela, University of California Santa Barbara, USA and CERN, Geneva, Switzerland
Roger Jones, CERN, Geneva, Switzerland
Yves Kemp, Deutsches Elektronen Synchrotron, DESY, Hamburg, Germany
Kerstin Kleese van Dam, Pacific Northwest National Laboratory, PNNL, USA
Juergen Knobloch, CERN, Geneva, Switzerland
David Kreincik, Cornell University, USA
Kati Lassila-Perini, Helsinki Institute of Physics, Finland
Francois Le Diberder, Laboratoire de l'Accelerateur Lineaire, LAL/IN2P3, Orsay,, France
Sergey Levonian, Deutsches Elektronen Synchrotron, DESY, Hamburg, Germany
Aharon Levy, Tel Aviv University, Israel

Qizhong Li, Fermi National Laboratory, FNAL, USA
Bogdan Lobodzinski, Deutsches Elektronen Synchrotron, DESY, Hamburg
Marcello Maggi, INFN Bari, Italy
Janusz Malka, Deutsches Elektronen Synchrotron, DESY, Hamburg, Germany
Salvatore Mele, CERN, Geneva, Switzerland
Richard Mount, Stanford Linear Accelerator Laboratory, SLAC, USA
Homer Neal, Stanford Linear Accelerator Laboratory, SLAC, USA
Jan Olsson, Deutsches Elektronen Synchrotron, DESY, Hamburg, Germany
Dmitri Ozerov, Deutsches Elektronen Synchrotron, DESY, Hamburg, Germany
Leo Piilonen, Virginia Polytechnic Institute, VPI, USA
Giovanni Punzi, Fermi National Laboratory, FNAL, USA
Kevin Regimbal, Pacific Northwest National Laboratory, PNNL, USA
Daniel Riley, Cornell University, USA
Michael Roney, University of Victoria, Canada
Robert Roser, Fermi National Laboratory, FNAL, USA
Thomas Ruf, CERN, Geneva, Switzerland
Yoshihide Sakai, KEK, Tsukuba, Japan
Takashi Sasaki, KEK, Tsukuba, Japan
Gunar Schnell, University of the Basque Country UPV/EHU, Bilbao, Spain
Matthias Schroeder, CERN, Geneva, Switzerland
Yves Schutz, Subatech IN2P3, Nantes, France
Jamie Shiers, CERN, Geneva, Switzerland
Tim Smith, CERN, Geneva, Switzerland
Rick Snider, Fermi National Laboratory, FNAL, USA
David M. South, Deutsches Elektronen Synchrotron, DESY, Hamburg, Germany
Rick St. Denis, University of Glasgow, UK
Michael Steder, Deutsches Elektronen Synchrotron, DESY, Hamburg, Germany
Jos Van Wezel, Karlsruhe Institute for Technology, Germany
Erich Varnes, University of Arizona, USA
Margaret Votava, Fermi National Laboratory, FNAL, USA
Yifang Wang, IHEP, Beijing, China
Dennis Weygand, Jefferson National Laboratory, JLAB, USA
Vicky White, Fermi National Laboratory, FNAL, USA
Katarzyna Wichmann, Deutsches Elektronen Synchrotron, DESY, Hamburg, Germany
Stephen Wolbers, Fermi National Laboratory, FNAL, USA
Masanori Yamauchi, KEK, Tsukuba, Japan
Itay Yavin, McMaster University and Perimeter Institute, Canada
Hans von der Schmitt, MPI Munich, Germany

# Executive Summary

The activity of the ICFA study group on Data Preservation and Long Term Analysis in High Energy Physics (DPHEP) is focused on a global study on long-term data analysis as a way to maximise the scientific return for investment in large-scale accelerator facilities. The first recommendations were included in an intermediate report[1] and address general issues related to data analysis and management beyond the lifetime of collaborations. The report was produced following two workshops held in 2009 at DESY in January and SLAC in May and contains the following recommendations:

- Urgent action is needed for data preservation in HEP
- The preservation of the full capacity to do analysis is recommended such that new scientific output is made possible using the archived data
- The stewardship of the preserved data should be clearly defined and taken in charge by data archivists, a new position to be defined in host laboratories
- A synergic action of all stakeholders appears as necessary
- The activity is best steered by a lightweight organisation at international level

ICFA welcomed this intermediate report at its meeting on August 19, 2009, and suggested DPHEP further its understanding. The third DPHEP workshop took place at CERN in the following December, and was preceded by a half day open Symposium, where general arguments for data preservation were presented. The strategy was agreed and further developed during the fourth and fifth workshops held at KEK in July 2010 and Fermilab in May 2011, respectively.

In the last year an encouraging tendency to initiate concrete projects within the participating experiments and laboratories has been observed. This tendency, largely triggered by awareness of the DPHEP study group, supports the multi-laboratory approach, an ideal approach to combine scarce resources and to improve the flow of information. Multi-experiment common projects now appear to be necessary and the formation of such projects proceeds rapidly, often based on concrete R&D projects already planned within individual experiments.

This paper constitutes the first concrete collection of projects related to HEP data preservation. The projects span all levels of collaborations: within experiments, at the laboratory level and at the international level. This coherence of methods is expected to continue in the future. The document is therefore built as a blueprint for HEP data preservation and addresses the following areas:

1) The physics case for preservation
2) Experiment-level and laboratory-level strategy
3) Global projects for HEP-wide sustainable data preservation
4) International co-ordination of data preservation activities

---

[1] DPHEP study group, "Data preservation in high energy physics", arXiv:0912.0255.

1) The physics case for long-term data preservation is investigated in detail and specific cases for long-term analysis are evaluated. Concrete examples exist for most experiments examined. Besides the now well-known example of the re-use of JADE data, more recent re-analyses of old data have been presented or are planned. The participating experiments provide examples of long-term analyses that will be made possible by their data preservation initiatives beyond the planned end of the collaborations. Examples of possible analyses with LEP data are described, as well as an inventory of analyses that are not covered at present for experiments approaching the end of the analysis phase (those experiments at HERA and BaBar). Some analyses of the Tevatron and Belle experiments that may take longer after the end of the data taking or involve combination with the next generation of experiments are presented. Some other, more isolated cases are also described, where data analysis was or will be made possible by preserving data from previous experiments. Based on the given examples, a generic classification of potential research benefits is proposed as well. While this is the first in-depth attempt to investigate and classify the physics case for data preservation, it cannot be claimed to be exhaustive at this stage.

2) Specific proposals of preservation models of individual experiments are presented and evaluated. The current analysis models and their evolution towards long-term infrastructures are described. Babar has developed the most advanced initiatives and a systematic multi-experiment approach is being pursued by the HERA experiments. The resources and costing models for these endeavours are estimated. The first, preliminary feedback from the experiments indicate of order 2-3 dedicated (FTE) high level scientists and/or engineers are required to participate in the first phase R&D data preservation programmes, mainly devoted to the consolidation of previous systems and migration to frameworks dedicated to long-term preservation. It should be stressed that this additional person-power is part of and should act coherently with the existing computing teams in each experiment. Besides the preparatory R&D phase, custodianship is needed for the lifetime of the data sets and should be ensured by a dedicated data archivist position, estimated to be equivalent to 0.5-1 FTE in each associated laboratory.

3) International projects involving the cooperation of multiple experiments are being defined and cover several areas. Technologies for data preservation are investigated such as virtualisation and virtual repositories, data and analysis migration procedures, data validation suites and archival infrastructures. The management of information and its storage is also examined, including the extension of documentation in the public domain and the enhancement of information by storing figures, data, notes and internal legacy material in collaboration with the INSPIRE service. The integration of preservation and outreach is also explored, concerning standard formats, tools, recasting methods, and communication techniques.

4) A concrete proposal for an international forum on data preservation in HEP is being developed. Following the strong recommendation by all reference forums of the study group (Advisory Committee, ICFA, HEPAP and others), a proposal for an international organisation to oversee and guide data preservation in HEP will be defined with the aim to obtain the necessary funding and to officially install the organisation as soon as possible. The management and the associated operational model are defined. The connections to the supporting bodies (experiments, laboratories and funding agencies) are proposed, as well as the connections with similar initiatives in other fields. Funding models of the organisation will be investigated and specific programmes (both inside and outside HEP specific budget lines) will be explored, with the aim of competing for funds from various national and international sources. The experience of the last three years of cooperation between experiments and computing centres indicates a need for a dedicated person empowered to initiate, develop and sustain the international organisation.

The official start-up of the DPHEP organisation should include employing a Project Manager, hosted in a large laboratory. The Project Manager will steer the activity of the study group and ensure permanent relations with the Data Archivists to be installed in major laboratories, with the experiments and with the governance bodies of DPHEP: Chair, Steering Committee, Advisory Committee.In summary, the DPHEP study group identified the following priorities, in order of urgency:

- **Priority 1: Experiment Level Projects in Data Preservation.** Large laboratories should define and establish data preservation projects in order to avoid catastrophic loss of data once major collaborations come to an end. The recent expertise gained during the last three years indicate that an extension of the computing effort within experiments with a person-power of the order of 2-3 FTEs leads to a significant improvement in the ability to move to a long-term data preservation phase. Such initiatives exist already or are being defined in the participating laboratories and are followed attentively by the study group.

- **Priority 2: International Organisation DPHEP.** The efforts are best exploited by a common organisation at the international level. The installation of this body, to be based on the existing ICFA study group, requires a Project Manager (1 FTE) to be employed as soon as possible. The effort is a joint request of the study group and could be assumed by rotation among the participating laboratories.

- **Priority 3: Common R&D projects.** Common requirements on data preservation are likely to evolve into inter-experimental R&D projects (three concrete examples are given above, each involving 1-2 dedicated FTE, across several laboratories). The projects will optimise the development effort and have the potential to improve the degree of standardisation in HEP computing in the longer term. Concrete requests will be formulated in common by the experiments to the funding agencies and the activity of these projects will be steered by the DPHEP organisation.

These priorities could be enacted with a funding model implying synergies from the three regions (Europe, America, Asia) and strong connections with laboratories hosting the data samples.

This document is seen by its authors as a conclusion to the initial reflection period and as a first step in an new period where data preservation in HEP will develop at international level, with strong synergies with other scientific fields and with the ambitious goal of enhancing the potential of the HEP data by explicitly using a global, long-perspective and flexible access approach.

# 1. Introduction

## 1.1 The Context for Data Preservation in HEP

Since the 1950s, physicists have constructed particle accelerators to study the building blocks of matter, where technological advances, as well as experimental discoveries, have resulted in the construction of bigger and more powerful accelerators. In most cases the next generation collider operates at a higher energy frontier or intensity than the previous one. This feature is displayed in figure 1, which shows the last 50 years in particle physics, where the clear trend to higher energies is visible in both hadron–hadron and $e^+e^-$ colliders[2]. At the end of the first decade of the 21st century, the focus is firmly on the Large Hadron Collider (LHC) at CERN, which operates mainly as a pp collider, and currently at a centre–of–mass energy of 8 TeV. At the same time, a generation of other high-energy physics (HEP) experiments are concluding their data taking and winding up their physics programmes. These include H1 and ZEUS at the world's only $e^{\pm}p$ collider HERA (data taking ended July 2007), BaBar at the PEP-II $e^+e^-$ collider at SLAC (ended April 2008) and the Tevatron experiments DØ and CDF (ended September 2011). The Belle experiment also recently concluded data taking at the KEK $e^+e^-$ collider, where upgrades are now on going until 2014.

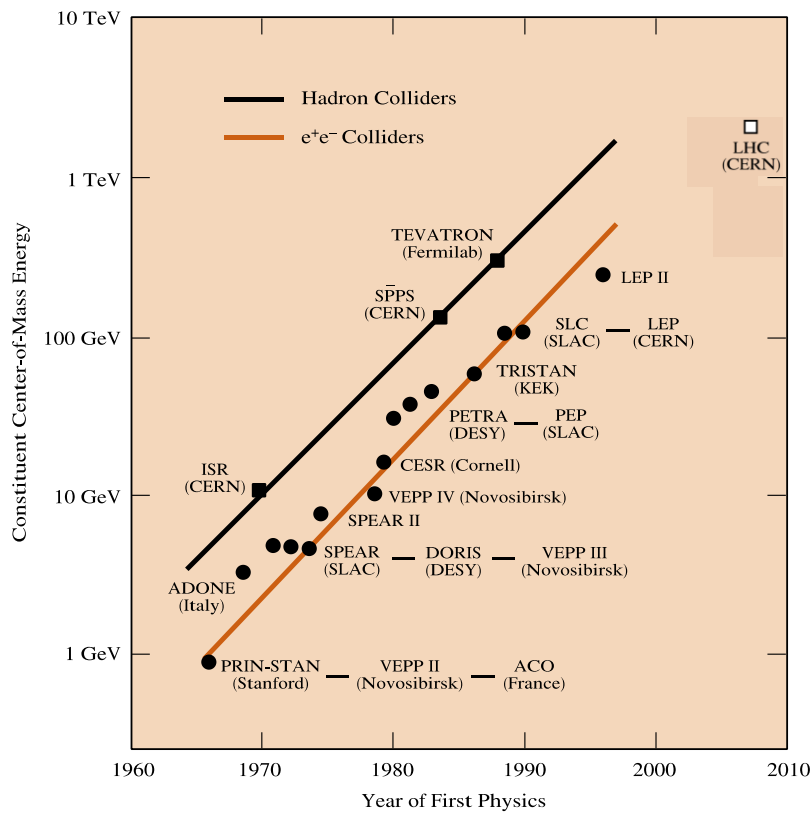

*Figure 1: A history of the constituent centre-of-mass energy of the electron-positron and hadron colliders as a function of the year of first physics.*

---

[2] W. K. H. Panofsky, "The evolution of particle accelerators and colliders", Beam Line, Vol. **27**, No. 1 (1997) eds. M. Riordan et al. p 36 (LHC start date modified accordingly).

The experimental data from these experiments still has much to tell us from the on going analyses that remain to be completed, but it may also contain things we do not yet know about. The scientific value of long-term analysis was examined in a recent survey by the PARSE-Insight project[3], where around 70% of over a thousand HEP physicists regarded data preservation as very important or even crucial, as shown in figure 2. Moreover, the data from in particular the HERA and Tevatron experiments are unique in terms of the initial state particles and are unlikely to be superseded anytime soon, even considering such future projects as the LHeC[4].

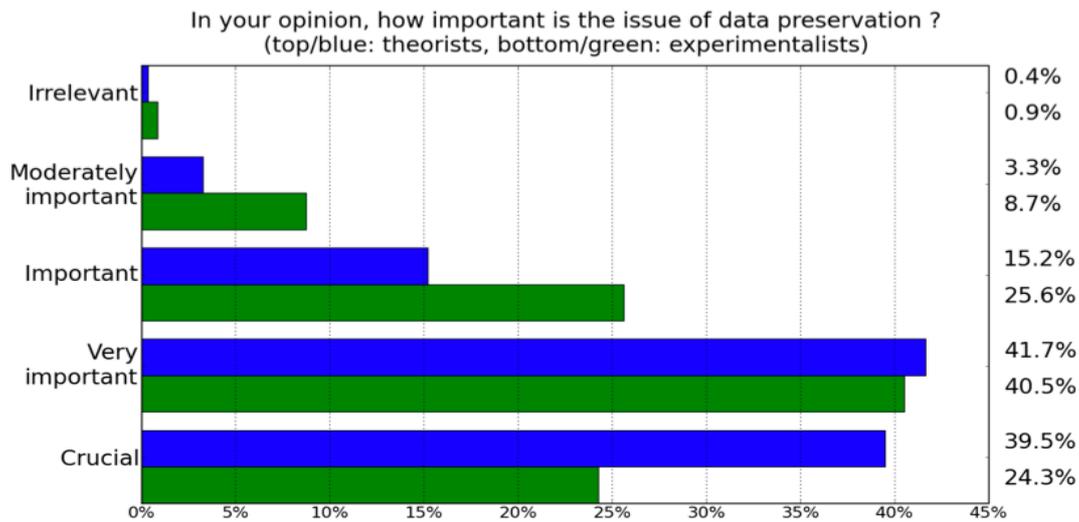

*Figure 2: One of the results of the PARSE-Insight survey of particle physicists on the subject of data preservation. The opinions of theorists and experimentalists are displayed separately.*

It would therefore be prudent for such experiments to envisage some form of conservation of their respective data sets. However, HEP has little or no tradition or clear current model of long-term preservation of data in a meaningful and useful way. It is likely that the majority of older HEP experiments are unable to analyse the original datasets due to a combination of a lack of planning to preserve the capacity to do so, a lack of person-power to carry on the analysis, a lack of compelling physics topics at a given moment in time, a lack of money to carry out analysis and no plan to maintain access to the datasets.

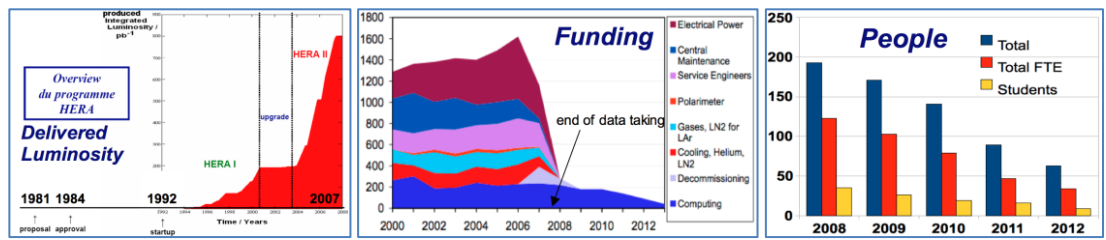

*Figure 3: Illustrative luminosity profile (left), funding (centre) and person-power (right) resources available to a high-energy physics experiment.*

The preservation of and supported long term access to the data is generally not part of the planning, software design or budget of a HEP experiment. This results in a lack of available key resources just as they are needed, as illustrated in figure 3. Accelerators typically deliver the best data towards the end of data taking, which can be seen in the left figure for the HERA accelerator. However, as the centre and right figures show, this contrasts with the reduction in overall funding and available person-power. Any attempts to allocate already limited resources to data preservation at this point have most often proven to be unsuccessful.

---
[3] PARSE-Insight FP7 Project: http://www.parse-insight.eu
[4] For more information on the LHeC, see http://www.ep.ph.bham.ac.uk/exp/LHeC

For the few known preserved HEP data examples, in general the exercise has not been a planned initiative by the collaboration but a push by knowledgeable people, usually at a later date. The distribution of the data complicates the task, with potential headaches arising from ageing hardware where the data themselves are stored, as well as from unmaintained and out-dated software, which tends to be under the control of the (defunct) experiments rather than the associated HEP computing centres. Indeed past attempts of data preservation by the LEP experiments[5], of SLD data[6] at SLAC and of JADE data[7] from the PETRA collider at DESY have had mixed results, where technical and practical difficulties have not always been insurmountable.

## 1.2 Data preservation in other disciplines

Experiences in other disciplines are widespread and highlight various approaches and advances. Although different needs within the research communities exist due to specific research questions, methodologies, analytical and processing steps, similarities can be observed in areas concerning data preservation. Many of the disciplines have already obtained dedicated data preservation and sharing platforms, with various sizes and complexities, for example via dedicated data repositories, data curators or the usage of specific services and tools. A thorough and inspiring analysis of the economic models that can be used for data preservation is available in the report of the Blue Ribbon Task Force[8]. The ODE project has recently highlighted a variety of opportunities for data preservation and re-use[9].

In order to offer a context for the on going discussion within HEP, four other research areas are chosen to highlight discipline specific aspects in regard to data preservation and sharing: astrophysics, molecular biology, geosciences and humanities and social sciences. In the following, initiatives and solutions are used to demonstrate the needs and solutions existing in the field of data preservation.

The astrophysics community face not only an exponential increase in data volume but also in data complexity. Given the nature of the experiments, the data have been made available to communities beyond the experimentalists in charge of the detectors. A general format called FITS was installed in the 1970s at NASA, which lead to further developments and extensions, in particular concerning the analysis tools. It is worth noting that the standardisation of data formats is supported in common by the most important funding agencies[10]. A generic model called OAIS (Open Archival Information System) has emerged as a framework for access to a preserved data set, defining the relationships among the users, producers and custodians[11]. A pioneering example for scientific data sharing is SDSS[12], now in its third version and proposing the SkyServer project[13], with an outstanding education and outreach programme (Galaxy Zoo and the Zoo-Universe). Furthermore, as the observational landscape multiplies by large factors every year, an added value clearly emerges from the joint analysis of tens or hundreds of different data sets. In the 1990s this situation gave rise to the so-called virtual observatories[14], which are able to store and make available astronomical data. An international initiative, the International Virtual Observatory for Astrophysics, emerged as a natural framework to develop the national facilities[15]. The IVOA mission is "to facilitate the international coordination and collaboration necessary for the development and deployment of the tools, systems and organisational structures necessary to enable the international utilisation of astronomical archives as an integrated and interoperating virtual observatory"[16]. The virtual observatories have demonstrated that common data samples made available for larger communities are, by themselves, new experiments and contain new science, beyond the individual scientific programmes of the participating experiments.

---

[5] A. Holzner et al., "Data preservation at LEP", arXiv:0912.1803.
[6] T. Johnson, "SLD data preservation", 2nd DPHEP Workshop, SLAC, May 2009.
[7] S. Bethke et al., "Experience from re-analysis of PETRA (and LEP) data", 1st DPHEP Workshop, DESY, January 2009; J. Olsson, "The Preservation of JADE data (and software)", JADE Meeting, DESY, August 2009.
[8] The Blue Ribbon Task Force on sustainable digital preservation and access, http://brtf.sdsc.edu/about.html. More details can be found in the 2010 report of the group: "Sustainable economics for a digital planet: Ensuring long-term access to digital information".
[9] The Opportunities for Data Exchange Project, http://www.alliancepermanentaccess.org/index.php/current-projects/ode, part of the Alliance for Permanent Access. See the recent report "Ten tales of drivers & barriers in data sharing".

Molecular biology can be considered a traditional data sharing community. As researchers got interested in the datasets behind the publications, a first data library and database was developed very early on, first based on extracting data from journals, but soon with an independent data submission. From the beginning data handling was done on an international scale to serve a global research community and ever since data repositories have been synchronised globally. Better and faster methodologies have increased the amount and complexity of the data output in this field tremendously in the past decades. Thus, the discipline can be considered very data intensive, with an emerging field of projects solely focused on data production or data reuse. With the Bermuda Principles[17] the community decided to provide standardised, public and rapid access to the results of the Human Genome Project. Standard procedures include rapid open access through a global and interoperable database network, community identifier system and reference system. The agreement was supported by publishers, journals and funding bodies, and was followed by many community actions. Today's users are able to deal with the complex data deluge in this field due to the drive to share data and new methodologies at hand. One initiative worth mentioning is the ELIXIR[18] plan for a common pan-European infrastructure for biological information to "manage and safeguard the massive amounts of data being generated every day", which is partly led by EMBL-EBI[19].

---

[10] The Consultative Committee for Space Data Systems: http://public.ccsds.org/default.aspx
[11] Description of the OAIS model, "Reference model for an Open Archival Information System".
[12] Sloan Digital Sky Survey III: http://www.sdss3.org/index.php
[13] SkyServer Project: http://skyserver.sdss.org/public/en
[14] The National Virtual Observatory NVO, http://www.us-vo.org; The European Virtual Observatory EURO-VO: http://www.euro-vo.org
[15] The International Virtual Observatory Alliance, http://www.ivoa.net/
[16] F. Pasian, "Management of astronomical data archives and their interoperability through the Virtual Observatory standards", 1st DPHEP Workshop, DESY, January 2009; R. Hanisch, "Standards in astrophysics", 3rd DPHEP Workshop, CERN, December 2009.
[17] The Human Genome Project, "Policies on release of human genomic sequence data".
[18] Elixir, Data for Life: http://www.elixir-europe.org

The field of earth science or geosciences is also fairly advanced in regard to research data preservation. Research projects in this domain may be very specific to a methodology or research question, but may also be of a cross-disciplinary nature when studying for example ecosystems or climate change. This highlights the need for data preservation; one can only detect variations over time when preserving the records over time. With the growing awareness in this regard many initiatives that foster cross-discipline data access (and by nature data preservation) have been established in the past decades. The data repository Pangaea has done pioneering work using DOIs (digital object identifiers) as a persistent identification and citation facilitator for research data[20]. Moreover, the integration of datasets and publication information has meant that both may now be cited[21] and count in research assessments. This data repository (which is part of the ICSU World Data System[22]) also showcases how a data repository and a publisher can collaborate in terms of data preservation and sharing successfully. The success of data repositories such as Pangaea indicates the emerging awareness within the community for the need of data preservation that is met by corresponding preservation platforms and services.

In the domain of the humanities and social sciences, research data is not only relevant for present and future scientific endeavours, but also highly relevant to policy decision makers. To be accessible and understandable for such re-use, the data must be prepared accordingly and long-term support for data producers is needed with regard to data preservation. This is mainly due to certain methodologies and materials that deal with personal data, in that it needs to be held anonymously, or require consent to be prepared, signed and preserved. Even though easy access to the data might be foreseen, it might not be feasible to provide it and access restrictions on re-use may apply. Many of these services are based on dedicated data repositories for data preservation and all require personal effort and a commitment to funding in both the mid- and long-term[23].

This tour across selected disciplines highlights different approaches to dealing with data preservation and data sharing as a whole. However while the actual implementations vary, many of the underlying concepts and processes are the same across the communities. These have been documented by the UK Digital Curation Centre's Curation Lifecycle Model[24]. In this framework a number of key concepts have emerged to support communities in regard to the Lifecycle model:

- Collect as much information as possible about your data at the time of creation and processing, when rich information is available and might be automatically captured
- Appraise your data and select what is really worthwhile preserving

---

- Ingest, secure and maintain both the physical data as well as its content, syntax and semantics
- Data and tools will need to evolve to keep pace with both IT technological developments, but also scientific demands including data transformation, new analytics, changed descriptions and so on
- Plans for data preservation need to be regularly reviewed and updated

Altogether, it is evident that large amounts of complex data, comparable to HEP in many aspects, also exist in other scientific fields. While the challenges might be multi-faceted, data preservation and data sharing are major undertakings in many disciplines and the resulting projects are often considered as a laboratory for more science. Such projects need and often benefit from a coordinated effort to support dedicated and adequate frameworks and technological projects. While some of the steps accomplished recently in other disciplines are commonplace within HEP experiments, others aspects have been discussed as part of a specific examination of data preservation in HEP and will be described in this document.

## 1.3 A study group on data preservation in HEP

A study group on Data Preservation and Long Term Analysis in High Energy Physics, DPHEP, was formed at the end of 2008 to address the issue in a systematic way. The aims of the study group include to confront the data models, clarify the concepts, set a common language, investigate the technical aspects, and to compare with other fields such as astrophysics and those handling large data sets. The experiments BaBar, Belle, BES-III, CLAS, CLEO, CDF, DØ, H1, HERMES and ZEUS are represented in DPHEP, and were joined recently by the LHC experiments ALICE, ATLAS, CMS and LHCb. The associated computing centres at CERN (Switzerland/France), DESY (Germany), Fermilab (USA), IHEP (China), JLAB (USA), KEK (Japan) and SLAC (USA) are all also represented in DPHEP.

A series of workshops have taken place over the last three years, beginning at DESY in January 2009 and most recently at Fermilab in May 2011. The study group is officially endorsed with a mandate by the International Committee for Future Accelerators, ICFA and the first DPHEP recommendations were published in 2009, summarising the initial findings and setting out future working directions. The aims of the study group have also been presented to a wider physics audience via seminars, conferences and publications in periodicals. The role of the DPHEP study group is to provide international coordination of data preservation efforts in high-energy physics and to provide a set of recommendations for past, present and future HEP experiments. The study group has presented its intermediate conclusions to several oversight bodies (the DESY Physics Review Committee, HEPAP (DOE/NSF) Panel, CERN Scientific Policy Committee, IN2P3 Scientific Council) and has unanimously received very positive feedback and support.

# 2. The Scientific Potential of Data Preservation in High Energy Physics

The physics case for data preservation can be deduced from the unexploited potential of the existing data sets as presented below. It should be noted that the last generation of experiments, generically named as "multi-purpose", developed complex physics programmes over more that one decade. The intermediate developments conjugated with the dynamics of the resources, often lead to a class of subjects not sufficiently addressed in the scientific output.

To illustrate the scientific potential of existing data sets, we consider first a few examples based on past experiments at the PETRA and LEP colliders. In spite of the fact that data preservation had not been planned, the usefulness of long-term access to the data and to the analysis frameworks has been demonstrated. A second set of examples comes from the experiments that presently pursue the final analysis phase after the end of collisions: the B-factories, HERA and the Tevatron. These experiments have in principle the opportunity to perform all imaginable analyses. However, due to the decrease of the human resources, some of the known subjects will not be covered or will be not treated with the full potential precision. Beyond this two-fold approach, issued from direct observations of the physics planning within the experiments, a generic classification (taken from the 2009 document) of the potential of the preserved data is also proposed at the end of this section.

## 2.1 Physics Potential of Former Experiments

### 2.1.1 The JADE experiment at PETRA

The recent resurrection and re-analysis of data from JADE, an experiment that operated at the PETRA $e^+e^-$ collider between 1979 and 1986 is now well known. The advances in theoretical knowledge and analysis methods with respect to those available in the 1980s, in particular for the modelling of hadronic final states, have lead to an improved measurement in a unique energy domain, no longer available or reproducible, despite the higher energy and luminosity offered by LEP. Enhanced and more profound theoretical knowledge, more sophisticated Monte Carlo (MC) and hadronisation models, improved and optimised experimental observables and methods, and a much deeper understanding and precise knowledge of the Standard Model (SM) of electroweak and strong interactions make it mandatory and beneficial to re-analyse old data and to significantly improve their scientific impact.

The impact of this well-known re-analysis is best illustrated in figure 4. A measurement with rather poor precision but otherwise unique in its energy range[25] could be converted to several precise measurements proving the running of the strong coupling constant[26]. The efforts of a small team of people over several years could resurrect and improve the outcome of a data set collected 15 years before with significant effort and which is non-reproducible amongst other experiments.

---

[25] G. Altarelli, "Experimental tests of perturbative QCD", Ann. Rev. Nucl. Part. Sci. **39** (1989) 357.
[26] S. Bethke et al., "Determination of the strong coupling $\alpha_S$ from hadronic event shapes and NNLO QCD predictions using JADE data", Eur. Phys. J. C **64** (2009) 351 [arXiv:0810.1389].

However, its important contribution to the precise determination of $\alpha_S$ should not hide that this success entailed huge individual commitment and some elements of good luck. Furthermore, there were more actual topics extracted from PETRA data than the running coupling $\alpha_S$. For example, analyses of the energy independence of the hadronisation process were performed (not feasible during PETRA times), as well as measurements of the longitudinal and transverse cross-sections, determinations of event-shape observables introduced later and, more generally, studies of the moments of event-shape observables. The JADE data in its renewed form is preserved at MPI Munich, following the effort described above, although a long-term solution is still lacking.

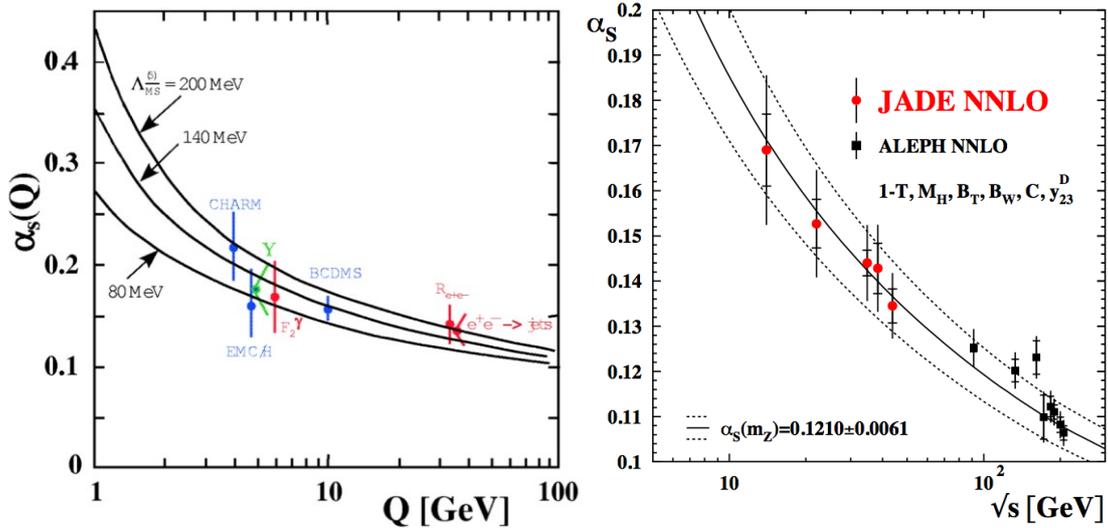

*Figure 4: Re-analysis of the JADE data has lead to a significant improvement in precision and to a differential measurement that can prove the running of the strong coupling in $e^+e^-$ experiments. The plot on the left shows the status of various $\alpha_S$ measurements in 1989, where all $e^+e^-$ inclusive measurements appear in the point marked $R_{e+e-}$. Using modern day analysis techniques and predictions, as shown in the right plot from a 2009 publication, the JADE data alone are able to demonstrate the scaling of $\alpha_S$, as well as provide a measurement in a unique energy range.*

## 2.1.2 The LEP experiments

The $e^+e^-$ collider LEP at CERN demonstrates a clear example of the long tail of physics output typical of HEP experiments, which extends well beyond the end of data taking, as illustrated in figure 5. LEP was running from 1989 to 1995 at centre-of-mass energies around the nominal mass of the $Z^0$ resonance (91 GeV), after which the energy was increased in steps to a maximal energy of 209 GeV, reached shortly before the final shutdown in November 2000. Since then, the four LEP experiments, ALEPH, DELPHI, L3 and OPAL, have continued to publish scientific papers.

The analyses of LEP data, performed during the decade after the shutdown, were made possible by ensuring some level of access to data, analysis tools and documentation. They mostly belong to the core part of the LEP physics programme, are based on the full statistics of the experiments, use the most sophisticated analysis procedures, and take advantage of the best understanding of the systematic uncertainties. Some of these publications must therefore be regarded as part of the LEP legacy.

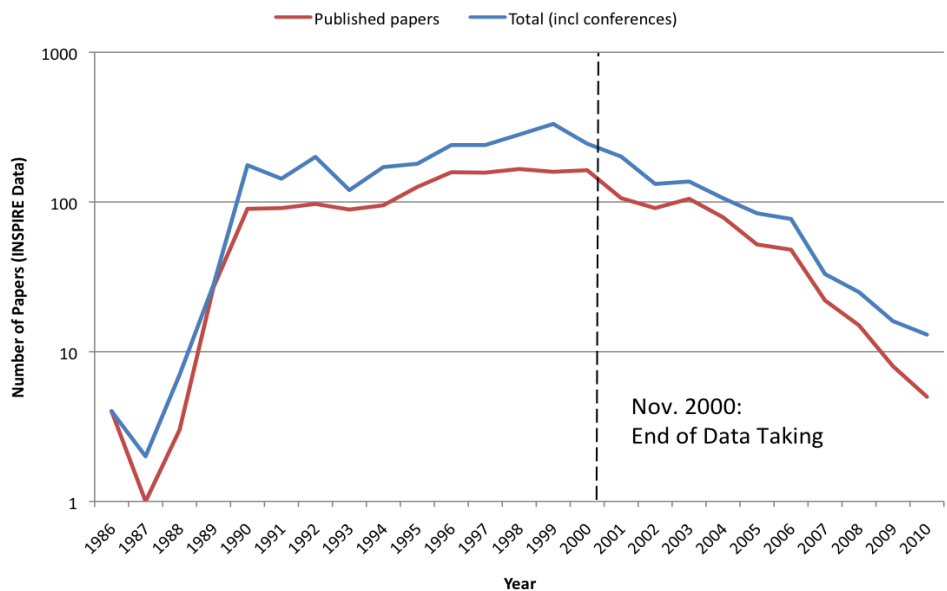

*Figure 5: Publication profile of papers of the LEP collaborations.*

Considering that the strength of the collaborations strongly diminished within 3 years after the end of data taking, 2004 and beyond are commonly taken as the appropriate period to assess the scientific production of the archival period of LEP. The papers published in the period 2004-2009 total 349, as detailed in table 1, and represent 13.5% of all LEP papers. All important physics subjects are covered among them, as detailed in table 2. The most outstanding are Electroweak and QCD measurements, followed by searches for Higgs bosons, SUSY particles and other exotica, physics of specific flavours (b, c, τ) and exclusive production channels. After the LEP shutdown, three of the four experiments were converted for the observation of cosmic rays and produced in total 12 publications on this subject. About 30% of the analyses were on searches. One can argue that some of these searches were exploring phase space corners and could later be superseded at higher energy machines. However, approximately 100 publications belong to the core programme, although they were concluded only in the long tail of physics analyses. These late publications include some of the most important LEP legacy papers, which combine the full statistics of the four experiments, such as precision electroweak measurements on the $Z^0$ resonance[27], searches for neutral MSSM Higgs bosons[28] and about 10 extended reviews on the most important subjects, among others electroweak couplings, QCD, Higgs and SUSY searches, as well as physics of the W boson.

While some analyses are presently on going (e.g. NNLO analyses aiming at the precise determination of $α_S$, as in OPAL) some new analyses could be triggered in light of discoveries at the LHC, like performing specific searches (for example SUSY), and some others would deserve attention irrespective of new findings. To give a few examples of the latter, one may consider a return to the LEP data to try to solve long standing anomalies such as the $A^{FB}_b$ asymmetry, the subject of recent review[29], to determine basic quantum numbers such as the spin of the gluon, which did not receive enough attention, or to perform an analysis of the angular ordering of proto-jets in the hadronisation cascades to probe QCD coherence effects. A few examples of the most recent LEP publications are described in the following.

---

[27] S. Schael et al. [ALEPH, DELPHI, L3, OPAL and SLD Collaborations and the LEP Electroweak Working Group and SLD Electroweak Group and SLD Heavy Flavour Group], "Precision electroweak measurements on the Z resonance", Phys. Rept. **427** (2006) 257 [hep-ex/0509008].
[28] S. Schael et al. [ALEPH, DELPHI, L3, OPAL and the LEP Working Group for Higgs Boson Searches], "Search for neutral MSSM Higgs bosons at LEP", Eur. Phys. J. C **47** (2006) 547 [hep-ex/0602042].

|  | 2001 | 2002 | 2003 | 2004 | 2005 | 2006 | 2007 | 2008 | 2009 | Total | 2004-2009 |
|---|---|---|---|---|---|---|---|---|---|---|---|
| **ALEPH** | 46 | 42 | 24 | 34 | 12 | 9 | 4 | 4 | 2 | **607** | 65 |
| **DELPHI** | 64 | 30 | 31 | 58 | 21 | 19 | 7 | 7 | 2 | **678** | 114 |
| **L3** | 51 | 40 | 23 | 52 | 16 | 11 | 5 | 2 | 0 | **578** | 86 |
| **OPAL** | 61 | 38 | 32 | 55 | 9 | 11 | 4 | 3 | 2 | **675** | 84 |
| **All** | 222 | 150 | 110 | 199 | 58 | 50 | 20 | 16 | 6 | **2538** | 349 |

*Table 1: Statistics of peer-reviewed publications of the LEP collaborations.*

| Papers 2004-2009 | ALEPH | DELPHI | L3 | OPAL | All |
|---|---|---|---|---|---|
| **Electroweak** | 17 | 26 | 22 | 24 | 89 |
| **QCD** | 19 | 25 | 19 | 22 | 85 |
| **Higgs Searches** | 6 | 14 | 8 | 9 | 37 |
| **SUSY Searches** | 4 | 7 | 5 | 9 | 25 |
| **Exotica Searches** | 5 | 12 | 10 | 7 | 34 |
| **Flavour Physics** | 6 | 15 | 4 | 5 | 30 |
| **Exclusive Channels** | 3 | 8 | 8 | 2 | 21 |
| **Cosmo-LEP** | 3 | 3 | 6 | 0 | 12 |
| **Other** | 2 | 4 | 4 | 6 | 16 |
| **Total** | **65** | **114** | **86** | **84** | **349** |

*Table 2: Distribution of physics topics in LEP publications in the years 2004-2009.*

## Recent search for a Higgs boson with resurrected ALEPH data

Despite a large and comprehensive set of searches done during the experiments lifetime, new models or a better understanding of the theoretical framework may reveal islands of sensitivity that were not explored before. This is the subject of a recent re-analysis[30] of ALEPH data, a search for a low mass Higgs super-symmetric partner that may be produced in pairs and would be able to decay in four tau leptons. This configuration and the corresponding decay channel were not explored during the collaboration lifetime and are now shown to cover a new domain in the parameter space, as illustrated in figure 6. Indeed, a real discovery chance was explored ten years after the data-taking period. The re-analysis involved re-use of the analysis software and a dedicated effort to generate and reconstruct samples of MC events, illustrating the need for preservation of the capabilities to perform complete analyses. A validation step was also necessary, to ensure the correctness of the results. Examining the same production mechanism but with decay channels to hadrons (to gluons or charm quarks), a follow-up ALEPH analysis sensitive to these final states has been performed and is expected to be submitted for publication soon.

---

[29] A. Djouadi et al., "Forward-backward asymmetries of the bottom and top quarks in warped extra-dimensional models: LHC predictions from the LEP and Tevatron anomalies", Phys. Lett. B **701** (2011) 458 [arXiv:1105.3158].

[30] S. Schael et al. [ALEPH Collaboration], "Search for neutral Higgs bosons decaying into four taus at LEP2", JHEP **1005** (2010) 049 [arXiv:1003.0705].

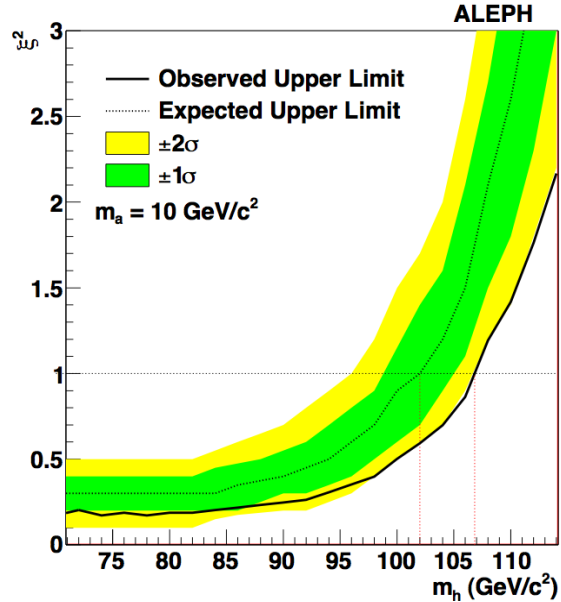

*Figure 6: Observed and expected limits from ALEPH on the combined production cross-section times branching ratio in the search for the process h → 2a → 4τ, as a function of Higgs boson mass $m_h$.*

### Single top production at LEP within a generalised contact interaction model

Searches for new physics are often linked to a specific model where some aspects of a complete or most general theory are ignored or simplified in order to extract a comprehensive message. Further theoretical and experimental studies may encourage and lead to different approaches to the same or similar data analyses and explore new paradigms. For example, the search for single top quark production at LEP2 was performed in the past by all collaborations within a specific model of anomalous couplings. This has recently been extended by the DELPHI collaboration to a more general model based on the contact interactions approach[31], which has been used for the first time in this context.

### Precise results with refined experimental methods

Employing today's commonly used experimental methods may also lead to improved results using old data. A recent DELPHI study[32] of the b-quark fragmentation function has been able to improve the model independence of the results by using two unfolding methods. The refined measurement allowed a better insight into the perturbative and non-perturbative effects and was combined with previous results obtained at LEP and SLD as well.

### Precise strong coupling determination using a recent theoretical calculation

Improved calculations at next-to-next-to-leading-order (NNLO) perturbative expansion matched with the re-summed terms in the next-to-leading-logarithmic-approximation (NLLA) have been used to reanalyse OPAL data and to obtain a new and improved determination of the strong coupling at various energies[33]. Figure 7 presents the comparisons of the new measurements (NNLO+NLLA) from 2011 with the initial determinations from 2005 (NLO) and illustrates the gain in precision obtained from the new theoretical work.

---

[31] J. Abdallah et al. [DELPHI Collaboration], "Search for single top quark production via contact interactions at LEP2", Eur. Phys. J. C **71** (2011) 1555 [arXiv:1102.4455].

[32] J. Abdallah et al. [DELPHI Collaboration], "A study of the b-quark fragmentation function with the DELPHI detector at LEP I and an averaged distribution obtained at the Z Pole", Eur. Phys. J. C **71** (2011) 1557 [arXiv:1102.4748].

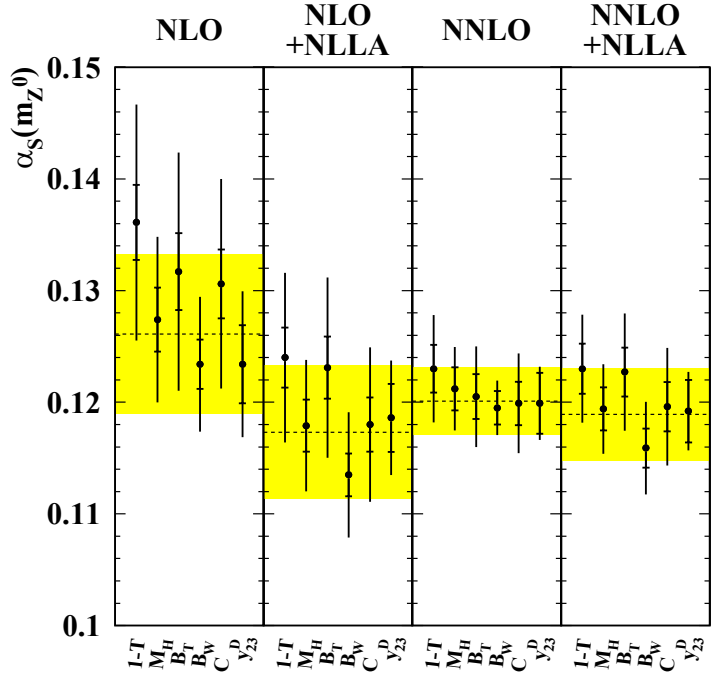

*Figure 7: Determination of the strong coupling using various event shape variables from OPAL data. The determinations done in 2011 and corresponding to an improved theory calculation are compared to previous determinations from 2005 (NLO).*

**Investigations of particle production**

Models of Bose-Einstein correlations among particles produced in the final state are tested with improved accuracy in a recent L3 analysis[34]. Particle production has been analysed in light of a new model that is found to give a better description of the data. Given the universality claimed for the measured effect in the hadronic final states, similar comprehensive analyses in other experimental environments, for instance at hadron or heavy ion colliders, are encouraged and may provide a reason to reanalyse previous data sets.

**Searches exploiting jet substructure**

A set of techniques to take advantage of jet substructure has been developed in recent years and has gathered a considerable amount of attention due to the power they add to searches for new physics[35]. The predictions for jet substructure are challenging, and this complicates the calibration and optimisation of these techniques at the LHC. A recent and surprising idea is that the measurement of thrust and related distributions at LEP offers an excellent handle for jet substructure at the LHC.

---

[33] G. Abbiendi et al. [OPAL Collaboration], "Determination of alpha(s) using OPAL hadronic event shapes at sqrt(s) = 91-209 GeV and re-summed NNLO calculations" Eur. Phys. J. C **71** (2011) 1733 [arXiv:1101.14700].

[34] P. Achard et al. [L3 Collaboration], "Test of the tau-model of Bose-Einstein correlations and reconstruction of the source function in hadronic Z-boson decay at LEP", Eur. Phys. J. C **71** (2011) 1648 [arXiv:1105.4788].

[35] A. Altheimer et al., "Jet Substructure at the Tevatron and LHC: New results, new tools, new benchmarks", arXiv:1201.0008.

### 2.1.3 Other examples of long-term data re-use

Besides the well know and documented cases, examples, sometimes anecdotic, reveal aspects of the usefulness of a systematic approach of data preservation in high-energy physics. The examples listed below do not originate from a complete analysis or from a dedicated large-scale investigation and are therefore by no means a definitive list of opportunities.

**Reanalysis of E852 data from BNL at JLab**

Around 1989 the E852 collaboration began a large experimental programme at the MPS facility at the BNL AGS to observe, via partial wave analysis techniques (PWA), 'exotic' mesons, i.e. mesons beyond the Naive Quark Parton Model that are predicted by models inspired by QCD. Data taking took place from 1994 to 1998. In 1997 the data from 1994 on the $\pi_1(1600)$ and the $\pi_1(1400)$, both with exotic quantum numbers, were published. The further inclusion of the 1998 data added the $\pi_1(1600)$ and $\pi_1(2000)$. However, data taken from 1997 to 1998 were never published. The MPS has been dismantled, and the AGS programme concluded. Now, some 12 years later Jefferson Lab (JLab) is embarking on a new meson spectroscopy programme with the CEBAF upgrade (from 6 GeV to 12 GeV), and Hall D/GlueX. New PWA techniques are being developed for GlueX. It would be very useful to re-examine the data from E852 (both published and never studied) with better PWA analyses. Although the E852 collaboration no longer exists, the software used both for reconstruction and for PWA are still intact, as well as the appropriate calibration databases.

However, this will not remain true much longer. The data also exist, albeit in boxes in the basement of the physics building at BNL, but risk disposal in the near future. There is probably enough hardware around to read the data tapes. A plan was presented to the Physics Department at JLab to move the data tapes from BNL to JLab, and copy them to the JLab tape robot. By today's standards, it is not much data and the cost of new media is trivial. The software library is already at JLab, as well as calibration databases. Finally, and perhaps crucially, there are still enough of the E852 collaborators also working at JLab to make use of the data in a new archival period.

**Dark photons: the relevance of old data for new models**

Another example covers the recent rise in interest in models involving the so-called dark photons[36]. These bosons would result from a special theory extending quantum electrodynamics and leading to a heavy photon interacting boson coupling to the photon. This hypothesis has prompted dedicated small experiments[37], as well as the re-analysis of existing data. Indeed, such a new boson would lead to a change in the branching ratio of neutral pions to photons. These branching ratios are best measured in so-called beam-dump experiments, performed essentially at previous fixed target facilities. The re-analysis of some of these data has led to improved restrictions on such models[38], as illustrated in figure 8. Most of the recent exclusion analyses performed around dark photons models use experimental data that is older than two decades.

---

[36] N. Arkani-Hamed et al., "A theory of dark matter", Phys. Rev. D **79** (2009) 015014. [arXiv:0810.0713].

[37] H. Merkel et al. [A1 Collaboration], "Search for light gauge bosons of the dark sector at the MAMI", Phys. Rev. Lett. **106** (2011) 251802 [arXiv:1101.4091];
S. Andreas and A. Ringwald, "Status of sub-GeV hidden particle searches, [arXiv:1008.4519];
S. Abrahamyan et al., "Search for a new gauge boson in electron-nucleus scattering by the APEX experiment", Phys. Rev. Lett. **107** (2011) 191804 [arXiv:1108.2750].

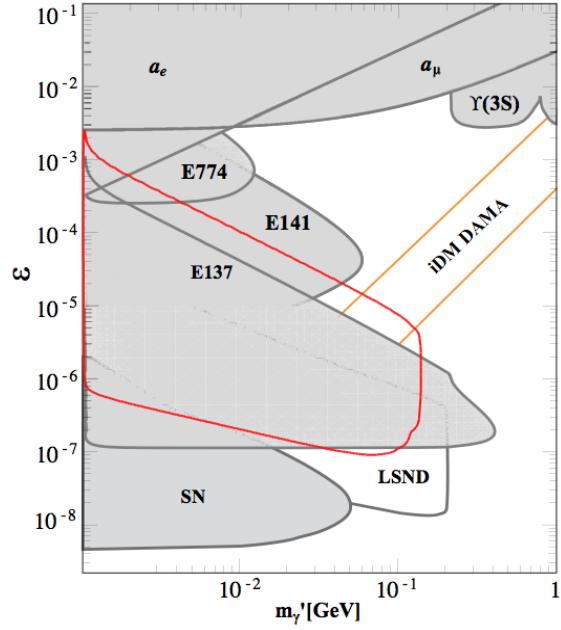

*Figure 8: Limits of exclusions for the dark photon mass $m_{\gamma'}$ as a function of the mixing parameter ε (mixing with the normal photons). The theory proposals and interpretations are mainly recent, while the used data is older than ten years.*

Some of these scenarios can also be addressed with rare decays of the $Z^0$ boson that are accessible at LEP, where it has been estimated that the 17 million on-shell $Z^0$ bosons produced would yield about 150 events of dark fermion pairs[39]. The re-analysis of LEP data is being made possible by the simulation of these new processes, as illustrated in figure 9.

**CPLEAR**

The CPLEAR experiment at CERN collected a large sample of flavour-tagged neutral kaons between 1992 and 1996, produced in antiproton-proton annihilations at rest. This sample was then used to make high precision measurements of T and CPT violations in the neutral kaon system[40]. By studying the time dependence of neutral kaon decays to flavour-tagged and untagged final states, it was possible to put limits on a non-conventional theoretical model involving virtual black holes which alter the time evolution of states compared to the predictions of normal quantum mechanics. It is not excluded that another non-conventional theoretical model comes up in the future, which could have been studied with the data collected by CPLEAR.

---

[38] J. Blümlein and J. Brunner, "Exclusion limits for dark gauge forces from beam-dump data", Phys. Lett. B **701** (2011) 155 [arXiv:1104.2747].
[39] M. Baumgart et al., "Non-abelian dark sectors and their collider signatures", JHEP **0904** (2009) 014 [arXiv:0901.0283].
[40] A. Angelopoulos et al. [CPLEAR Collaboration], "Physics at CPLEAR", Phys. Rep. **374** (2003) 165.

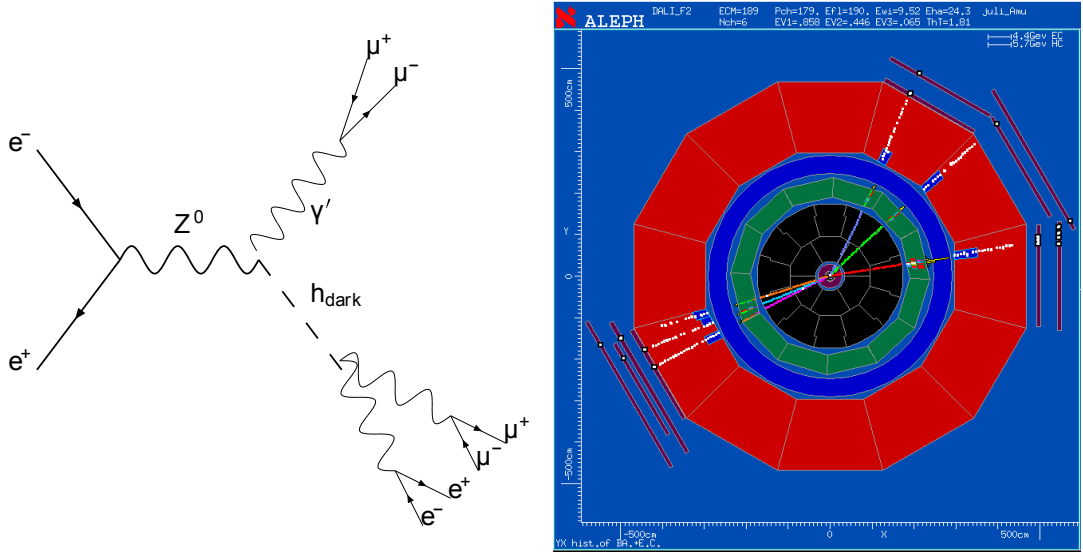

*Figure 9: Feynman diagram (left) and event display (right) of a recently simulated six lepton event in ALEPH, inspired by the advent of dark photon analyses.*

Unfortunately, all of the raw data, about 10000 3480-cartridges corresponding to about 10 TB, was destroyed in 2001 due to space limits at CERN. What remains are histograms, which entered into the publications, theses and nano-DST files of the two-pion decay mode. A search for data of other decay channels, for example semi-leptonic decays, is still on going. All software versions, written in FORTRAN77 and using the "PATCHY car format", are still available. Logbooks from data taking and minutes of analysis meetings survive at the moment in a cupboard at CERN. There is no person-power available to scan the documents. Using the CERNLIB version as shipped with recent Ubuntu distributions, (Ubuntu 11.04), it is still possible to read all of the histogram files using PAW/HBOOK, as well as running the analysis program over the nano-DST files and producing histograms. Since the experimental setup is rather straight forward, one might consider how to recycle the two-pion data for outreach purposes, as an example on how to study CP violation with time dependent flavour-tagged decays.

## 2.2 Long-term Physics Potential of Experiments in the Final Analysis Phase

### 2.2.1 The B-factories

To be specific, we take here the example of the BaBar collaboration, but a similar physics case could be presented for the Belle and Cleo collaborations. The BaBar collaboration will move into a new organisation model towards the end of 2012, in what is referred to as the Archival Period. At that point it is expected that most of the scientific programme will have been covered, but not all.

The likely candidates for analyses to be performed in the Archival Period belong to various categories, and it is difficult to determine precisely what their topics will be. Ideas for unforeseen analyses, like searches for signals for new physics that may have been missed, also cannot be ruled out. Other examples of this type of unforeseen analyses are crosschecks of results performed on upcoming data from the Super B-Factory projects, in the event of them finding something unexpected.

However, to give an idea of how many analyses could in principle be performed, one may consider the number of identified analyses that are presently covered in the collaboration. As of February 2012 there are 85 analyses on track for publication many of which are so-called core analyses, pertaining to the core programme of the collaboration, and will likely be covered before the Archival Period. About another 30 exist in their preliminary phase or are less active and will take longer to publish. More potential for new analyses have been identified and will enter the long-term programme.

These figures indicate that the physics potential offered by the Archival Period stretches beyond what can actually be achieved considering the current person-power, even if one only takes into account analysis subjects already identified. Based on the evolution of analyst person-power, it is presently projected that the Archival Period will allow the publication of more than 50 papers.

The expected BaBar publication time-profile is shown in figure 10. Archival Period analyses may belong to topics where a limited expert person-power is already saturated by on going analyses, but where the teams have a long-term programme in mind. In BaBar, these analyses are spectroscopy of various sorts and ISR/QED/QCD physics. In addition, there will likely also be some more conventional B/tau/charm physics activities as well.

Analysis in the Archival Period can also be of the type implying lengthy analysis effort, implying tough systematics or extremely sophisticated fits. As a concrete example, Initial State Radiation (ISR) implies delicate analyses aiming at the determination of the cross-sections of the $e^+e^-$ annihilation into multi pions and/or kaons final states, using fully the BaBar data set, with the goal of providing a complete set of cross-sections for $e^+e^-$ annihilation into hadrons, at low energies, where the contribution to g-2 is the most critical. The most important part, $e^+e^- \rightarrow \pi^+\pi^-$, has been dealt with, but not with the full data set: in particular, the higher quality data have not yet been used. Other examples are provided by BaBar-Belle joint analyses involving multi-parameter fits, such as Dalitz-plot analyses (for example, for D-mixing or for B decays). The technical solution for allowing such combined analyses is available and has been exercised already, thanks in particular to a very fruitful collaboration with CERN. It is worth noting that slow pace analyses may also result from an approach based on multiple undergraduate students, working in turn for their completion.

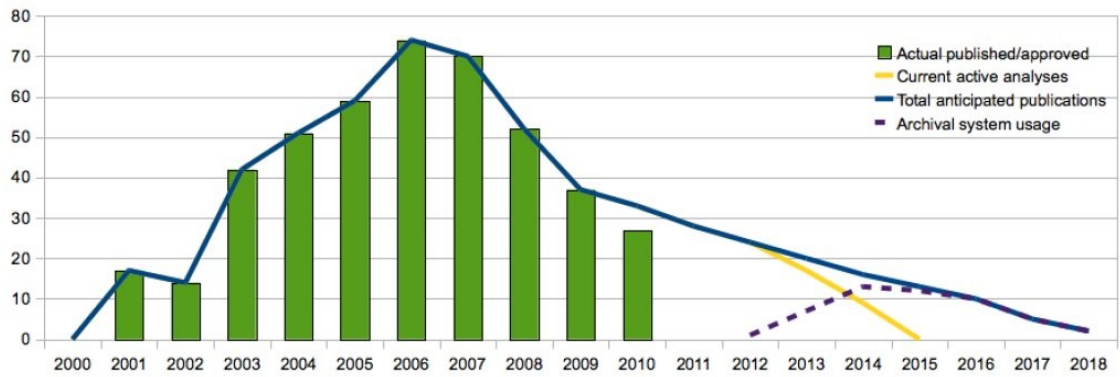

*Figure 10: Publications of the BaBar Collaboration as a function of year.*

## 2.2.2 The HERA experiments

The HERA data themselves represent a unique achievement in HEP, and are unlikely to be superseded in the near future. The core of the HERA physics programme, the so-called flagship measurements, is given constant attention and the full potential of the collected data will be exploited in such analyses. It is also certain that the majority of the obtained results will not be repeated.

On top of the analyses that are currently foreseen, the potential of the data can be further exploited if new developments in theory or new experimental methods are able to enhance the precision of existing measurements. Despite careful planning, the end of the analysis period is affected by the decrease of the person-power in the experiments and the migration to other projects.

The scientific production at HERA is largely driven by the profile of the data taking, with a significant increase in the accumulated luminosity in the last two years of the collisions programme. The exploitation of these data required dedicated studies to achieve the required precision and a final reprocessing. At the same time, the number of effectively contributing collaborators has observed a linear decrease, where most of the collaboration members in 2012 are also engaged in other experimental programmes more than 50% of their time. This configuration induces a longer analysis time for the on going papers. The collaborations have succeeded nevertheless to publish papers with an approximately constant rate, as illustrated in figure 11.

The intensive analysis phase will continue for at least 3 more years until 2015 and will include a few very important results that are now in the convergence phase. The publication plans of the HERA experiments include another 20-40 papers and the collaborations count in addition of order 10-20 results that are considered important but are not covered at present due to the lack of person-power. These results are the main candidates for the long-term analysis phase. It is anticipated that some analyses addressing so far uncovered topics will continue in the case of ZEUS up to 2030 or even later. This timescale also mirrors that at which new experiments at the ep energy frontier may start taking data at the LHeC.

In the following, a number of examples of analysis topics are given that will either be worked on for an extended period, or that, due to external events, might attract renewed interest long after the official HERA data analysis ends. Some of these will occur with certainty, but for some others it is a matter of speculation whether interest will actually arise. The list of topics described below is by no means exhaustive.

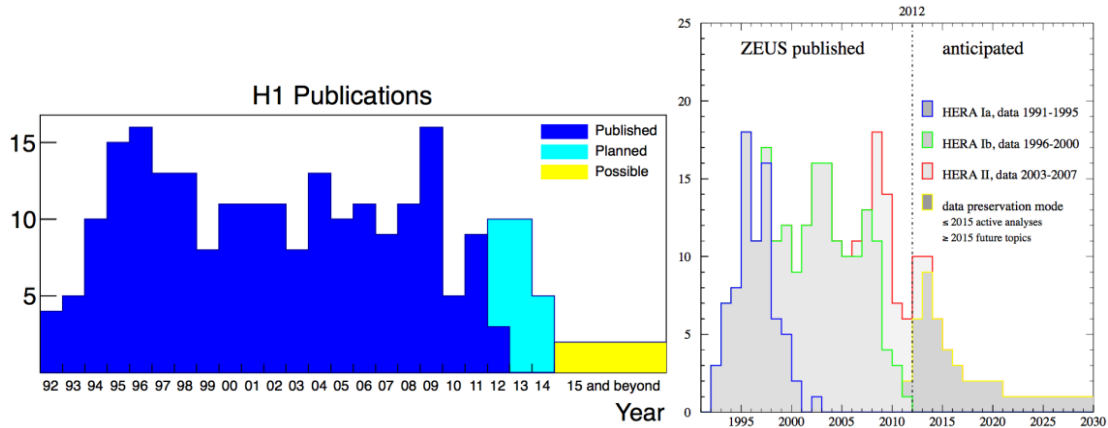

*Figure 11: Publications of the H1 (left) and ZEUS (right) Collaborations as a function of year.*

**Combinations and QCD analysis of different observables**

The HERA experiments have embarked on an ambitious programme to combine their data and extract the maximum phenomenological information. The first products of this effort are the first three combination papers published by H1 and ZEUS in 2009[41]. One of these, the combination of the neutral and charged current cross section measurements using data from the HERA I phase, has demonstrated most prominently the improvements that can be obtained both experimentally and phenomenologically from this effort. However, this is only a first step in a long-term programme that will extend well beyond the currently planned end of HERA data analysis. Eventually it should comprise all information on inclusive cross sections, on jet production, on heavy flavour production and on diffractive processes. In order to achieve this goal, further theoretical developments will be needed and significant experimental and computational difficulties have to be surmounted.

HERA measurements are considered as crucial input to the field of high-energy physics. The measurements of the collaborations are close to final, but the combinations and the interpretation in terms of QCD fits will extend for a few more years. A common H1 and ZEUS initiative on this issue has started and it is expected that the combination of data together with the associated fits will provide valuable input to the corresponding LHC measurements. It is worth mentioning that within this effort the HERA collaborations have recently released their code for QCD analyses into an open access framework[42]. The transition from internal usage to open access has received encouraging feedback from the theorists and from the LHC experimentalists interested in precision measurements related to the proton structure and may be a prototype for the long term knowledge transfer to the next generation of experimental programmes.

There is also significant interest in Generalised Parton Densities (GPDs), where the theoretical framework is comparatively recent and still being developed. Currently the available information on GPDs is rudimentary but much more could be gained from the HERMES data, in addition to the H1 and ZEUS data, which are the only available sources of information at small values of $x_{Bjorken}$. It is almost certain that a combination of H1 and ZEUS data together with data from HERMES for the purpose of GPD analysis, for instance on deeply virtual Compton scattering or on diffractive vector meson production, will happen during the archival period of the data. In particular, the various pure targets used at HERMES and the availability of the complete set of combinations of target and beam polarisation and beam charge make this data set unique in the world. Furthermore, a large fraction of data had been taken during the last two years of HERA with a recoil-proton detector in place, vital for clean exclusive event samples.

---
[41] Combined HERA results can be found here: http://www.desy.de/h1zeus
[42] More information on HERA-fitter can be found on the homepage: http://herafitter.hepforge.org

Concerning un-integrated (or transverse-momentum-dependent) parton distributions, the analysis frameworks are being developed only now. New MC generators need to be written and together with existing ones have to be refined and tuned to data. The improved theoretical understanding in turn will help to reduce the often still-large systematic uncertainties attached to present results.

**A high precision measurement of the strong coupling constant**

Current extractions of the strong coupling constant at HERA suffer from the fact that state-of-the-art theoretical calculations for jet production in DIS are limited to NLO. There is significant activity on going to extend these calculations in the DIS regime. As these efforts mature this will renew the interest in precision jet and other final state measurements. This will allow precision tests in kinematic regions, in particular at small values of $Q^2$, where up to now the uncertainties were large. This might be of particular interest should new physics be found at the LHC which would modify the behaviour of the running coupling. The HERA extractions allow the study of the scale dependence over a significant range of the scale including small values where the lever arm is an important benefit to understand the very high-energy behaviour. Such a study can be done directly without recourse to other experiments, which would introduce complicated correlations of the uncertainties.

**Further long-term studies**

In addition to the specific examples given above, there will be a number of studies that will continue on a low level for an extended period of time. These might be carried out systematically and over a long period improve the uncertainties. These studies might also be driven by individual interest or the motivation for these may be educational. Examples for future analysis are the study of hadronic final states with the goal to improve the modelling in MC generators, or more exotic topics such as the search for the presence of certain non-perturbative effects as the QCD instanton, or investigations of the (sometimes contradictory) pentaquark observations. If new discoveries are made at new machines, the production mechanism in ep collisions may supply new and unique information.

The complexity and the precision of the HERA data suggest that new observables may be used in the future. Examples of such observables are: new jet algorithms; usage of different global event shapes; the study of angular correlations; atypical signatures of new particles. At the experimental level, many analyses have efficiencies or acceptances varying strongly within measurement bins, but where finer binning is not possible due to the available statistics, like in D* or J/psi analyses, or due to the experimental resolution, leading to large model errors from MC modelling. Some present measurements are already dominated by uncertainties related to the theory (hadronisation, signal extraction, acceptance and so on). If the model error is dominant, such an analysis could be repeated with better MC modelling, when basic processes are constrained by other experiments or more sophisticated predictions lead to a significant improvement in the simulation.

### 2.2.3 The Tevatron

The CDF and DØ experiments at the Tevatron collected data with an integrated luminosity of around 10 fb$^{-1}$ during Run II. These data are still yielding valuable information on the nature of physics phenomena and will continue to do so in the future, especially in light of new discoveries by LHC or other experiments and advances in theoretical models. Additionally, as no plans are foreseen for a proton-antiproton collider in the future, Tevatron data will be unique for a very long time in terms of initial state particles; measurements of effects enhanced by $q\bar{q}$ production will remain competitive with respect to the LHC, for example in inclusive jet production at large $x_{Bjorken}$.

As an example within the realm of top physics, $t\bar{t}$ production asymmetry measurements have shown a discrepancy with the SM, which could hint at new physics[43]. At the LHC such an effect is more difficult to observe, as symmetric gluon-gluon-initiated events dominate top pair production. Moreover, differing production mechanisms in the two environments test distinct aspects of the SM and require different analysis strategies, as for example in $t\bar{t}$ spin correlation measurements[44]. Tevatron data will also be of importance for single-top searches[45], particularly in the s-channel: the s-channel search is more challenging at LHC than at Tevatron, since the relative cross-section is much smaller. The mass of the top quark is now known with a relative precision of 0.54%, limited by the systematic uncertainties, which are dominated by the jet energy scale uncertainty[46]. This is the result of the combination of several measurements made by CDF and DØ in different $t\bar{t}$ decay channels on data samples with an integrated luminosity of up to 5.8 fb$^{-1}$. The uncertainty on the jet energy scale is expected to improve as analysis are performed on the full data samples, since analysis techniques constrain the jet energy scale using kinematical information from W→ qq' decays. For the first time the total uncertainty of the combination is below 1 GeV. Such a level of precision urges a more detailed theoretical study of the exact renormalisation scheme definition corresponding to the current top mass measurements. The era of precision measurements in the top sector has arrived, and the Tevatron will provide a substantial contribution to the top mass world average for many years to come.

---

[43] CDF Collaboration, "Study of the top quark production asymmetry and its mass and rapidity dependence in the full Run II Tevatron dataset", CDF Note 10807;
V. Abazov et al. [DØ Collaboration], "Forward-backward asymmetry in top quark-antiquark production", Phys. Rev. D **84** (2011) 112005 [arXiv:1107.4995].
[44] CDF Collaboration, "Measurement of $t\bar{t}$ spin correlations coefficient in 5.1 fb$^{-1}$ dilepton candidates", CDF Note 10719;
V. M. Abazov et al. [DØ Collaboration], "Evidence for spin correlation in ttbar production", Phys. Rev. Lett. **108** (2012) 032004 [arXiv:1110.4194].
[45] CDF Collaboration, "Measurement of single top quark production in 7.5 fb$^{-1}$ of CDF data Using neural networks", CDF Note 10793;
V. M. Abazov et al. [DØ Collaboration], "Measurements of single top quark production cross sections and |Vtb| in ppbar collisions at sqrt{s}=1.96 TeV", Phys. Rev. D **84** (2011) 12001 [arXiv:1108.3091].
[46] The Tevatron Electroweak Working Group and the CDF and DØ Collaborations, "Combination of CDF and DØ results in the mass of the top quark using up to 5.8 fb$^{-1}$ of data", arXiv:1107.5255.

In the electroweak sector, one of the most important measurements made at the Tevatron is the precise determination of the W mass. In conjunction with top mass, the W boson mass constrains the mass of the Higgs boson, as well as the mass of possible new particles beyond the SM. The measurement is very challenging due to presence of an undetected neutrino from the W decay, which makes it impossibly to fully reconstruct the final state unambiguously.

The CDF and DØ experiments have recently measured the most precise values of the W mass to date[47], achieving a total uncertainty of 19 and 23 MeV respectively, dominated by the uncertainty on parton distribution functions (PDFs). It will be difficult for the LHC to supersede this precision for at least several years. Current measurements are performed on 2.2 and 4.3 fb$^{-1}$ of data by CDF and DØ respectively. With the full data sample, these measurements could constrain systematic uncertainties and, in principle, reach a precision of 10 MeV. Attaining this precision will require considerable effort, especially in reducing the uncertainty on the PDFs.

Heavy flavour physics has several potential analyses that may be carried out in the coming years. Some of the ideas that have emerged include measuring $A_{SL}$ in $B^0$ and $B^0_S$ decays, studying the forward-backward asymmetry in charm and bottom production, measurements of the interference between scalar and vector resonances in B decays, and measuring production cross sections and polarisations (where possible) for as many heavy flavour states as possible. There are numerous decay modes that can be extracted from the data, some which will likely be overlooked by other experiments. More generally, the Tevatron data might be useful to cross check a result from another experiment. This may be particularly important in the light of new discoveries at the LHC, which may require CDF data to be revisited, possibly with new, more advanced analysis techniques. This was recently demonstrated with the evidence for a CP asymmetry difference between $D^0 \rightarrow K^+K^-$ and $D^0 \rightarrow \pi^+\pi^-$ decays from the LHCb experiment[48], which was soon after confirmed by CDF[49].

It should also be stressed that Tevatron measurements are made in a unique energy domain, which will be no longer available; therefore QCD measurements performed on Tevatron data will continue to be as valuable in understanding QCD as those made using LHC data. Examples of such measurements include the di-photon production cross section, Z/W + jets production, underlying and minimum bias events characteristics, and diffractive W and Z production. For the same reason, the Tevatron experiments collected data at two different energy points, 900 and 300 GeV, before the final shutdown of the accelerator. These data samples, collected by minimum-bias and selective triggers, will provide some valuable legacy measurements in non-perturbative QCD, soft and strong interactions.

These are only some highlights of the enormous potential of Tevatron data over timescales ranging from the near future to many years in the future. Importantly, the Tevatron will serve as a fundamental point of comparison for the LHC. Just as the

---

[47] R. C. Lopes de Sa (on behalf of the CDF and DØ Collaborations), "Precise measurements of the W mass at the Tevatron and indirect constraints on the Higgs mass", arXiv:1204.3260.

[48] R. Aaij et al. [LHCb Collaboration], "Evidence for CP violation in time-integrated $D^0 \rightarrow$ h-h+ decay rates", Phys. Rev. Lett. **108** (2012) 111602 [arXiv:1112.0938].

[49] CDF Collaboration, "Improved Measurement of the Difference between time–integrated CP asymmetries in $D^0 \rightarrow K^+K^-$ and $D^0 \rightarrow \pi^+\pi^-$ decays at CDF", CDF Note 10784.

Tevatron has produced a wealth of important results to date, we expect that, if coupled to a strong data preservation effort, the Tevatron will continue to produce high quality scientific results in the coming years

## 2.3 Generic Classification of the Physics Potential of Preserved Data

Long term preservation of HEP data is crucial to preserve the ability of addressing a wide range of scientific challenges and questions at times long after the end of the experiment that collected the data. In many cases, these data are and will continue to be unique in their energy range, process dynamics and experimental techniques. New, improved and refined scientific questions may require a re-analysis of such data sets. Some scientific opportunities for data preservation are summarised in the following points.

**Long-term completion and extension of scientific programmes**

This involves the natural continuation of the physics programme of the individual experiments, although at a slower pace, to ensure a full exploitation of the physics potential of the data, at a time when the strength of the collaboration (analyst person-power as well as internal organisation) has diminished. It is estimated that the scientific output gained by the possibility to maintain long-term analysis capabilities represents roughly 5 to 10% of the total scientific production of the collaborations. More important than the number of publications is the nature of these additional analyses. Typically, such analyses are the most sophisticated and benefit from the entire statistical power of the data as well as the most precise data reprocessing and control of systematic effects.

**Cross-collaboration analyses**

The comprehensive analysis of data from several experiments at once opens appealing scientific opportunities to either reduce statistical and/or systematic uncertainties of single experiments, or to permit entirely new analyses that would otherwise be impossible. Indeed, ground breaking combinations of experimental results have been performed at LEP, HERA and the Tevatron, during the collaborations' lifetime, providing new insight in precision measurements of fundamental quantities, and extending the ranges for search of new physics. Preserved data sets may further enhance the physics potential of experimental programmes, by offering the possibility of combinations which would not be otherwise possible. A particular case are the experiments which are in general superseded by new programmes but present nevertheless unique features: unique geometric acceptance regions, special trigger conditions, more favourable background. Data from facilities where no active collaboration is operating would be available for combination with new data. At the same time, well-documented preserved data would also enhance opportunities for combinations among current experiments, which may be otherwise prevented by the lack of standards leading to insurmountable technical or scientific problems. The HEP community comprises sub-communities of experts in various fields such as flavour physics, neutrino physics, and so on. These expert communities would greatly benefit from having simultaneous access to data sets from relevant experiments. For example, B-physics experts could devise analyses simultaneously using data from BaBar, Belle, and Cleo. Such an effort to combine analyses is already on going, for example between the H1 and ZEUS collaborations, and an evaluation of such an approach is underway between the Belle and BaBar collaborations. An effort in standardising and/or documenting data sets for long-term preservation would have an immediate return in facilitating these combinations.

**Data re-use**

Several scientific opportunities could be seized by re-using data from past experiments. For instance, new theoretical developments could allow new analyses leading to a significant increase in precision for the determination of physical observables. Theoretical progress can also lead to new predictions (for example of new physics effects) that were not probed when an experiment was running and are not accessible at present-day facilities. Similarly, new experimental insights (for example breakthroughs in MC simulation of detector response, new calibration techniques) or new analysis techniques (for example better statistical methods, multivariate analysis tools, greater computing capabilities) could allow improved analyses of preserved data, with a potential well beyond the one of the published analyses. Results at future experimental facilities may require a re-analysis of preserved data (for example because of inconsistent determinations of physical observables, or observation of new phenomena which may/should have been observed before). Results from the LHC experiments may very well induce re-analysis of LEP, Tevatron or HERA data.

**Education, training and outreach**

Preserving data opens new opportunities in training, education, and outreach. It permits data analysis by undergraduate or graduate students, without restriction to institutes that collaborated to the experiments, opening new opportunities for institutes in developing countries to initiate and develop HEP research. The benefit to the field is the ability to attract and train the best inquisitive minds. It also gives unprecedented opportunities to teach hands-on classes in particle physics, experimental techniques, statistics, and to explore physics topics that would not have been otherwise covered. High schools students could be exposed to simplified and highly visual analyses (similar to the successful IPPOG[50] master classes which use special sub-sets of the LEP and now LHC data), in order to further enhance the general public interest in the field and to attract new students to physics.

---

[50] The International Particle Physics Outreach Group: http://ippog.web.cern.ch

# 3. Models of HEP Data Preservation

## 3.1 Data Preservation Levels

Different preservation models can be organised in levels of increased complexity. Each level is associated with one or more use cases. The preservation model of an experiment should reflect the level of the use cases to be enabled in the future, and the whole aim of the preservation exercise. A survey of a few computing models revealed that the amount of data for preservation (including simulated data) of current experiments is between 0.5 PB and 10 PB, which is a significant but manageable size[51]. The costs related to maintaining and migrating the software and the data analysis infrastructures, to effectively preserve them, are model dependent and are difficult to estimate. Nevertheless, it is expected that the cost of various preservation models is primarily driven by person-power requirements rather than the cost of data storage.

| Preservation Model | Use case |
| --- | --- |
| 1. Provide additional documentation | Publication-related information search |
| 2. Preserve the data in a simplified format | Outreach, simple training analyses |
| 3. Preserve the analysis level software and data format | Full scientific analysis based on existing reconstruction |
| 4. Preserve the reconstruction and simulation software and basic level data | Full potential of the experimental data |

*Table 3: Various preservation models, listed in order of increasing complexity.*

Different preservation models are summarised in table 3 and presented in the following, with remarks on the associated cost estimates and benefits. The implementation of these models at the beginning of the lifetime of an experiment will greatly increase the likelihood of success, minimise the effort and ease the use of the data in the final years of the collaborations.

**Level 1: Provide additional documentation**

A model of preservation, without actually preserving the data, would be to provide additional documentation. Such a practice is also a recommendation to any preservation effort, and as such the guidelines in this section apply to all models. Additional documentation may include: more information associated with, or embedded in, publications (extra data tables, high-level analysis code, etc.); internal collaboration notes; meta-data related to the running conditions; technical drawings; general experimental studies (for example on systematic correlations); an expert information database (for instance minutes, slides, news); documents available on paper only that could be digitised and stored in electronic format. Care should be taken to tag the redundant information that often appears during the analysis (for instance intermediate, non-validated hypotheses or non-pursued technical solutions that may appear in the daily exchanges but be irrelevant for the final analysis configuration).

In the process of preparing documentation for preservation, global information infrastructures in the community as well as those within experimental collaborations may be beneficial for a robust preservation project. An organised internal documentation migration to a HEP community information system like INSPIRE[52] would be one way to achieve this goal. Auto-documentation tools like those included in ROOT[53] should be used to their maximum ability. Day-to-day documentation within the collaboration may be stored in a wiki that also has the advantage of a simple, text-based preservation option. A common format for popular tools such as electronic log books would also be useful, enabling such metadata to be preserved in a similar way. It would be beneficial to experiments to consult with a professional archivist who is aware of the standards within the HEP community and elsewhere. In particular, the proper storage and possible digitisation of paper documents should be pursued in collaboration with libraries. For a new experiment, the costs of a more coherent and preservation-oriented documentation strategy would be minimal, if applied from the beginning, whereas the benefits for future use are clearly significant.

---

[51] A Petabyte (PB) = $10^{15}$ bytes; for comparison, the four LHC experiments will produce about 15 PB of data per year and Google computing centres process 20 PB per day.

To preserve nothing beyond the publications and the associated, improved documentation may be an option only if the belief is that the data are no longer of any scientific use, such that they have been superseded by a new experiment or the full potential of the physics programme has been extracted. Past experience demonstrates that this is rarely the case, and concrete examples of scientific benefits of data preservation are given in the previous section. However, an effort to preserve lower level information needed for a full analysis certainly provides an added value to the scientific reach of an experiment.

**Level 2: Preserve the data in a simplified format**

An economic means of preserving the real and simulated data without the need for any experiment-specific software would be to just preserve the basic, event-level, four-vectors describing the detected particles. This should be done in as simple a structure as possible in order to facilitate future interpretation and re-use of this data. A simple four-vector format can be very useful in terms of a model for outreach and education purposes. However, this format will in general not be sufficient to perform a full physics analysis, except for a few particular cases. It is likely that a dedicated effort would be needed within each collaboration to decide on the physics content of the data format. Further options, like preserving the capability to perform simplified error propagation, may also be implemented. In terms of the required person-power, this option would require a dedicated effort of the experts to define the relevant information, with a relatively simple technical implementation and modest requirements for long-term maintenance.

**Level 3: Preserve the analysis level software and data format**

This option includes the preservation of analysis level experiment-specific software as well as the analysis level data format, and is sufficient to perform complete analyses when the existing detector reconstruction and simulated data sets are adequate for the pursued goal. With respect to level 2 this introduces a supplementary dependence on the longevity of the experiment-specific software and may require a thorough study of the computing environment to identify external dependencies such as ROOT or CERNLIB. More effort than level 2 would most likely be required for the preparation and maintenance of this dataset, especially if backwards compatibility issues arise. However, the benefits of this level of preservation are the ease of analysis and access to extra features and improvements from the software, while often greatly reducing the reliance on external dependencies compared to a level 4-type preservation scheme.

---

[52] INSPIRE is the new information platform for HEP, which replaces and enhances the popular SPIRES system and is realised by CERN, DESY, FERMILAB and SLAC: http://inspirehep.net. More information can be found at http://www.projecthepinspire.net.
[53] ROOT is the analysis software framework based on the C++ programming language (widely used in high physics analyses, in particular at the LHC experiments: http://root.cern.ch.

**Level 4: Preserve the reconstruction and simulation software as well as the basic level data**

Certain analyses may require the production of new simulated signals or even require a re-reconstruction of the real and/or simulated data. For these, the ability to recompile the full reconstruction and simulation software needs to be preserved. This may or may not require basic (raw) level data, depending on what is stored at the more abstract level (usually called DST), which is experiment specific. Generally for greater flexibility all data should be preserved. By preserving the full analysis chain, one retains the ability to newly derive associated corrections, studies of efficiencies and acceptances, and to perform a full systematic error analysis. Special care should be given to the protection of the sensitive components (official calibrations, simulation tuning etc.) that should not be redone unless high-level experts are involved. It is clear that level 4 preservation will necessarily include the full range of both experiment-specific and external software dependencies, although attempts to minimise the latter should be carried out in the initial step. As such, significant resources will be needed for this preservation model during the preparation (R&D) and maintenance (archived data) phases. However the clear benefit of such a model is that the full physics analysis chain is available and full flexibility is retained for future use.

## 3.2 Preparing a Data Preservation Project in HEP

The preservation preparation and planning should be taken into account as early as possible in the computing strategy, such that the transition to the archival phase is done with a reduced effort. Guiding principles have been made available in the framework of a few generic initiatives[54], described in section 1.2. Based on the chosen data preservation scenario, the computing team of the experiment can then plan and implement the preservation effort. To maximise the efficiency of the preservation project, a collaboration should employ as much centralised software as possible. This also benefits the collaboration by fostering the adoption of common code and results in a more efficient use of person-power. It is likely that the experiments close to the end of the data analysis phase need a dedicated effort to achieve a reliable implementation of their chosen preservation model. A necessary component of any preservation project is the implementation of robust validation procedures, which should be able to indicate the status of the data analysis capabilities without physics expert intervention. The validation software should be seen as an essential component for the preparation of the technological steps such as storage upgrades or operating system migrations, which are the critical moments of a data preservation project.

The R&D effort of the data preservation project should have significant overlap with the collaboration lifetime and should benefit, in addition to the dedicated human resources, from the general expertise in data analysis within the collaboration. In the longer term, the preservation project should be seen as a permanent activity, implemented in the associated host laboratories or computing centres, aimed at maintaining and optimising the exploitation of the preserved data analysis facility. A schematic view of the transition from a full analysis environment to an analysis based on the preservation model is presented in figure 12.

---

[54]See for example: http://www.dcc.ac.uk/resources/tools-and-applications/data-asset-framework

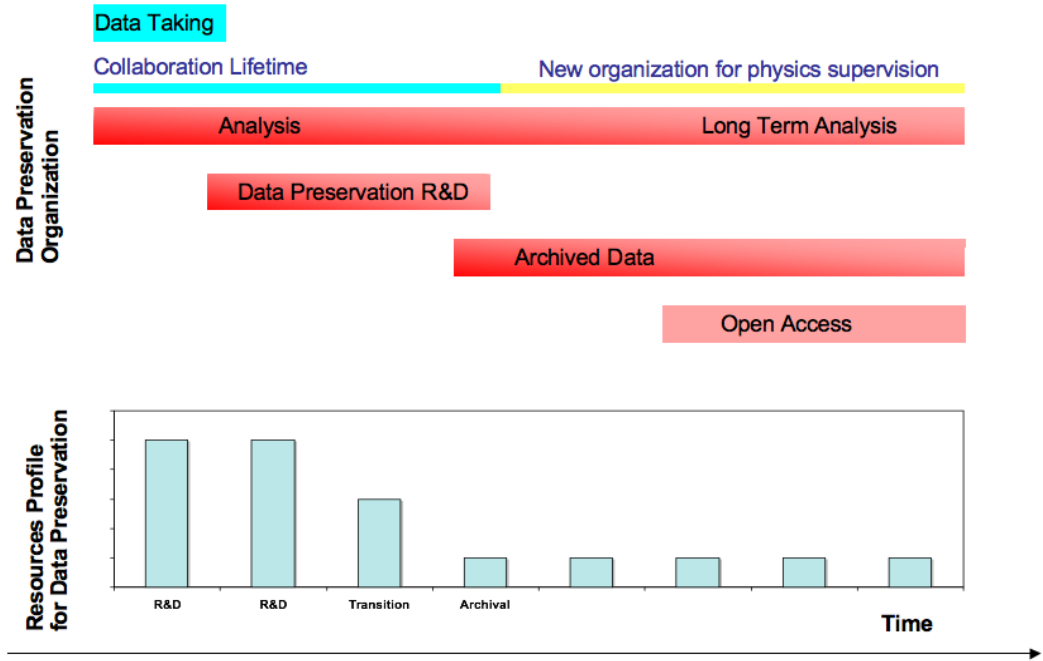

*Figure 12: A model for data preservation organisation and resources presented as the milestones of the organisation and the resources evolution as a function of time. The data preservation project should start as early as possible, ideally during the data-taking period, and should be coherent with the computing project and the analysis activities.*

**Long term organisation**

A data preservation project is void in absence of a form of supervision or connection to the original collaboration. Indeed, the publication of physics results during the lifetime of a collaboration follows rigorous procedures, exercised over many years. In order to ensure a proper usage of the preserved data, a process of scrutiny of the physics output obtained from this data should be defined. Certification mechanisms ensuring the correctness of the produced results should be implemented, reflecting the quality requirements specific to the level of detail used in the analysis.

The collaboration should therefore foresee a lightweight form of organisation for the long-term period, which should allow a rapid contact to the experts in various areas in order to prevent catastrophic loss of the ability to access and use data. In addition, the revival capability should be ensured, meaning simple procedures to signal new analysis to the largest possible community and mechanisms to ensure peer-review during a phase when such activities can only be considered as sporadic should be taken into account.

### Authorship

The adopted scheme should also include consideration of the authorship of potential new papers. The author lists of the HEP publications are defined according to internal mechanisms and usually include all members of the collaboration. Beyond the collaboration lifetime, the authorship rules for the scientific papers obtained using the preserved data sets should be clearly defined such that data analysis is encouraged and the proper credits are allocated to the collaboration that collected the data. The authorship rules should be linked to the physics supervision process.

### Supervision of the data preservation process

Data preservation is likely to include complex technical aspects which can be affected by their time profile: intense activity corresponding to major technological operation (for example changes in storage media or operating systems), or periods with reduced activity, with a risk of gradual dispersion of the know-how. Data archivists are likely to constitute a minority in the participating computing centres and in some cases they may be partially allocated to other tasks. The organisation of the preservation process should therefore include supervision mechanisms aimed at enforcing the contracts between the collaborations and the computing centres or facilities and to make sure that the necessary level of expertise is maintained. The Open Archival Information System (OAIS) reference model provides guidance for the type of processes and players required in the long-term data preservation process. The technical actions defined in the preservation model should be constantly reviewed. The ability of the data archivist to direct potential analysers to the proper documentation of calibration and analysis procedures, as well as giving advice on whom to contact for the resolution of non-documented analysis issues, would be of great value.

### Access to preserved data

The key motivation for the long-term preservation of HEP data is the unique scientific opportunities opened up by their re-use. In some scientific cases this re-use is by members of the collaborations who originally took the data. In other scientific cases, new opportunities are generated by the re-use of data by scientists not originally involved in the collaboration. Eventually, Open Access could generate further opportunities to use preserved HEP data. The PARSE.Insight study of over a thousand HEP scientists found that 54% of the theorists and 44% of experimentalists think that access to data from past experiments could have improved their scientific results. While Open Access to preserved HEP data, immediate or at a later stage, generates new opportunities, it also raises new issues in our community, such as the scientific responsibility for results obtained from preserved data sets. The survey also found that 45% of the respondents are "very concerned" or "gravely concerned" that re-use of data may in general lead to an inflation of incorrect results. At the same time, as many as 53% of the respondents are concerned about incorrect results due to a misinterpretation of the preserved data.

Open Access to data, albeit of often vastly larger simplicity, is sometimes the norm in other disciplines. HEP colleagues who started programmes in ground or satellite-based astro-particle physics have met and adapted to these different realities. The opportunities held by Open Access to preserved data have to be evaluated against the concerns that are raised, so that informed policy decisions can be made at the highest level. These considerations should be decoupled from pursuing the unavoidable and necessary steps in data preservation, and addressed through a parallel, wider debate.

## 3.3 Hardware Requirements and Computing Centre Issues

Plans for computer centre resources must be made to ensure the short and long-term availability of proper data archiving and data serving and analysis. These plans should include a proper accounting for the costs, effort and maintenance of the systems that are required. For a successful programme of long-term data archiving the necessary resources will have to be provided for the duration of the final analysis, archive, and open access phases of the project. It is assumed that the bulk of the resources will be at a single or possibly multiple data centres. However, models that utilise more distributed resources are possible and are potentially quite interesting, as long as the basic requirements are met.

**Data size and data integrity**

The size of the data to be stored, the structure of that data, the likely realistic access patterns, and the capability of the computer centre to store and provide computing resources and provide access to that data are all important considerations when making plans and estimating costs for long-term archiving and access to the data. For example, infrequent bulk transfers of large datasets have different requirements for the architecture of the storage system than frequent access to small files. Other factors such as local versus wide-area access can make the difference between the system architecture requiring a large disk buffer, or tape only. Requirements on worldwide distribution and access of data are also a critical factor.

A further important consideration is data integrity, in particular if data is preserved over long periods of time the computing centre will need to guard against a number of key challenges: bit rot, transfer and transformation errors and media decay. Regular checks of the data can help to identify errors introduced to the data through bit rot, and strategies need to be in place to deal with such incidents, either by replacing the damaged data with a backup copy (if data volumes allow) or by marking the data as damaged, so that future users are warned. Similarly, errors can be introduced into the data during transfer (from one media to another) or transformation (to a new format required to keep the data usable by up to date software), and well-documented procedures and stringent checks can minimise the risk of introducing such errors. Finally, and in particular for long-term preservation efforts, one has to consider the lifespan of different storage media and devise a migration strategy that balances increased risk of data damage or loss through media at the end of their predicted lifespan, with financial constraints at the host organisation. Multiple copies at one or many locations can guard against data loss. If used, recovery mechanisms should be put into place to automatically recover or reduplicate information that are lost in the normal operation of a storage system over time.

**Costs of the data storage in the computing centres**

The costs of data storage include the cost of the media and disks, the tape robot, the cost to migrate the data to new media and disks when necessary, tape drives, compute servers, networks, software and maintenance for the software to interface to the data, people to maintain and run the systems, user support, and space, power and cooling. Many of these components are shared with other facilities in the computer centres, meaning that the costs are shared and also depend on the details of the configuration of the computer centres and the storage systems. Nevertheless, it is important to consider all of the costs when planning for the archiving of data sets. It is not possible to produce a universal precise cost estimate given the variations in pricing. The cost of maintaining an archived data set is not simply the cost of the media and disk systems but includes substantial costs of person-power and all of the other components listed above. As described above, the critical parameters are volume, access patterns and period of archive.

The exact form of the technologies for preserving the data is not determined at this time. It is strongly suggested that all technology investigations be made in collaborations that include the experiments and the computer centres. The computing centre will support the final technology choices and they will need flexibility and control over many aspects of the service. The experiments will need to provide detailed requirements and the computer centres the technical solutions that can be installed and supported over the long timescales required.

Technologies that are being investigated by many people for possible use for long-term data archiving include virtualisation and cloud computing. These are new technologies and are rapidly changing. The large centres and the experiments should work closely together on R&D, exchanging information and knowledge, before any final technology choices are made.

One option mentioned by many experiments is the option of maintaining all or some of the archived data on disk only, with multiple copies on separate servers and/or locations to achieve the necessary redundancy. This is an interesting option and should be investigated carefully both at the initial phases of investigation and also by the computer centres as the technology of mass storage and access change. It may also be useful to study how other data archives are managed and served to learn about good and bad experiences. The SDSS data releases and other similar data releases may give useful information about costs, access, redundancy and sharing of responsibility for data release and user support.

# 4. Experimental Level and Laboratory Level Strategies

## 4.1 The Data Preservation Programme at BaBar

Data collection ended for the BaBar experiment at PEPII on April $7^{th}$, 2008. The experiment was conceived 16 years ago (LOI 1994) and collected data at a centre of mass energy corresponding to the mass of the Upsilon(4S) and eventually the mass of the Upsilon(3S) and Upsilon(2S) over almost a decade through the efforts of 10 countries and about 600 collaborators.

Based on the projected analysis load, the future of BaBar was divided into three epochs. These phases are the Intense Analysis Period: (up to the $3^{rd}$ quarter of 2010), Steady Analysis Period: ($3^{rd}$ quarter of 2010 until 2012), Archival Analysis Period: (2013 and beyond). This was the result of the BaBar Beyond 2010 Task Force surveys and projections of the projected number of active analyses and person-power resources. A following Task Force (TFIII) refined the projections for the productivity of the project and found a significant need for access to the data extending out to 2018. Over the years, these projections have proven to be consistent with the actual publication output, analysis activity and computing load. While pressure has been mounting on reducing the computing load on the central computing resources at SLAC, the actual usage has remained almost constant as of late 2011.

The data preservation of BaBar has involved intense development of the archival system during the Intense Analysis Period, thorough testing of the prototype archival system and implementation of the ultimate archival system during the Steady Analysis Period and the full switch to complete dependence on the archival system during the Archival Analysis Period. As of now, the full archival system is functioning and has entered the production phase with more than 20 users and there is pressure to accommodate users at a rate faster than initially foreseen.

**Description of the BaBar data model**

BaBar code releases are contained in about 900 packages mostly written in C++ and include some Python scripts. Some Java appears for event displays and GUIs. Some Fortran is present from the event generators. The releases contain about 3 million lines of offline code and 0.5 million lines of online code. A major release involved a full rebuild due to major code changes; major either in the number of changes or very fundamental structures being altered. Unmodified packages in minor releases were just links back to the proceeding major (or base) release. At the time of writing, 48 major releases exist and 101 minor releases exist.

Data flows of 5-10 TB/day between SLAC and the major international BaBar sites were common to support distributed production and analysis. There were ~20 BaBar and ~20 laboratory computing FTEs at the peak. BaBar computing was divided among a set of "Tier-A" sites: SLAC, CC-IN2P3 (France), INFN Padova & CNAF (Italy), GridKa (Germany), RAL (UK). Responsibility for core computing (CPU & disk) provision divided as ~2/3 SLAC, ~1/3 non-US. Tier-A countries delivered their shares largely on an in-kind basis at their Tier-A sites. Simulation was also provided by 10 to 20 other sites, mostly universities. Analysis computing was assigned to Tier-A sites by topical group. Skimmed data and MC data files were distributed to sites as needed for this. Specific production tasks were assigned to some sites as well. Roughly half of BaBar computing was off-SLAC in 2004-2007.

The total amount of raw data (XTC files) is around 0.8 PB. Together with all the reconstruction, simulation and skimmed data output, BaBar has stored about 6 PB. However, only the last two reprocessings of the data and corresponding simulation production output will be kept which amounts to 2 PB. This "legacy" data was recently migrated to new silos using T10000 tapes. It is also being copied to CCIN2P3 for backup. So far, the raw data has been transferred and the reconstructed data transfer is in progress.

**Reprocessing strategy and production plans**

BaBar completed its last major reprocessing of the Upsilon(4S) data by the end of 2008 with clean-up production continuing through the first quarter of 2009. The Upsilon(2S) and Upsilon(3S) data as well as the scans above the Upsilon(4S) taken during the last run cycle (December 2007 until 7 April 2008) were last reprocessed in 2010/11. No further reprocessings are foreseen beyond this but significant production of new signals and production of new cycles (skim cycles) of physics analysis streams continues. Whereas before the CPU cycles were equal balanced between production and analysis activities, it is now being dominated by analysis work.

**Analysis and production resources**

In 2008 the computing resources consisted of 160 embedded 300 MHz PPC CPUs (DAQ) and 200 modern x86 cores for software triggering and data quality monitoring. 4000-5000 x86 Linux cores at SLAC, and a similar number elsewhere, for data production and analysis. To accomplish the last major reprocessing of the Upsilon(4S) data the resources at the SLAC computing centre were supplemented with the acquisition of a SUN Black Box[55], which was filled with 252 quad core systems. At the same time an effort was initiated to improve the efficiency of the production code which in itself resulted in an improvement roughly equivalent to adding the capacity of a SUN Black Box.

In 2010, the resources were all for offline production and analysis and consisted of:

- 5700 cores at SLAC accessible to BaBar for general (typically analysis) work
- 1100 dedicated cores principally for BaBar only production
- 450 TB XRootD cluster servers
- XFER data distribution machines
- NFS and AFS servers for release repository, production output, CVS and AWG space

Most of this was available in quad core and dual quad core machines. Approximately 800 TB of disk space (bulk data, production scratch, user space) was available at SLAC with similar amounts among the other sites.

The agreement with SLAC is that the resource level will correspond to the projected needs of BaBar. Cores are gradually being transferred to new projects by rebalancing of the fare-shares in the general queues. The XFER data distribution machines have been recently replaced and the NFS and AFS servers are being replaced. In addition, the analysis working groups have access to significant resources at the following Tier-A sites: SLAC, CCIN2P3, CNAF, GRIDKA, UVIC. Roughly half of the analysis CPU capacity comes from the non-SLAC Tier-A sites.

---
[55] "SLAC prepares for first black-box to expand computing power",
http://today.slac.stanford.edu/a/2007/06-20.htm

**The BaBar Long Term Data Access archival system**

The BaBar Long Term Data Access (LTDA) project aims to preserve both the data and the ability to do analysis until at least 2018 and will provide support for greater than 50 publications foreseen beyond 2012. After considerable research and development, a system involving a combined storage and processing structure was designed. It is isolated from the host institution by a firewall and uses virtualisation to allow the preservation of a stable validated platform. At the time of writing, the complete system, which is pictured in figure 13, has been acquired and setup to the point that tens of users have started using it for their analyses. For easy migration of the users, the system appears to the user very similar to the standard framework provided by the laboratory but the system is a separate site much as the remote analysis and production sites associated with BaBar. The network plan was key in allowing one to operate a system running frozen platforms at a site with a high level of computing security. The network design is illustrated in figure 14.

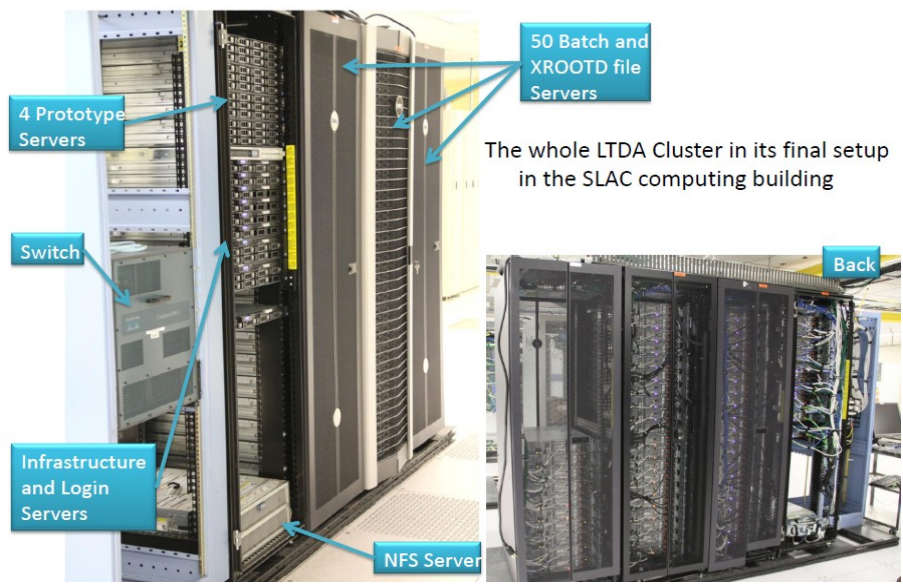

*Figure 13: Photograph of the BaBar Long Term Data Access archival system ready for production in March 2012.*

The hardware employed includes 9 infrastructure servers: 3 login machines, 1 cron-server, 1 test server, 2 authentication servers, and 2 database servers (mirrored). The various services (dhcp, nfs, dns, and so on) running on the servers is shown in figure 14. The virtual machines on the batch servers are isolated in the "back-end" by a firewall while the servers themselves are unprotected because they use only managed platforms that pose no security threat beyond those of regular laboratory systems. The back-end uses 54 DELL R510 systems with each with: 12x3.1 GHz cores (24 with hyper-threading, with approximately 60% gain for full load), 24 TB of storage on each server with 2 TB used for scratch and the rest for XRootD storage. Virtual machines jobs as well as regular jobs are controlled through the batch system using Torque with the MAUI system used for load balancing. Images for SL6 and SL5 virtual machines are used but other platforms are not a-priori excluded. The BaBar infrastructure is all embodied in the images but the common libraries, code and storage is all accessed from NFS areas not in the image. For protection, various access rules such as not being able to write from the virtual machines to the home areas are enforced.

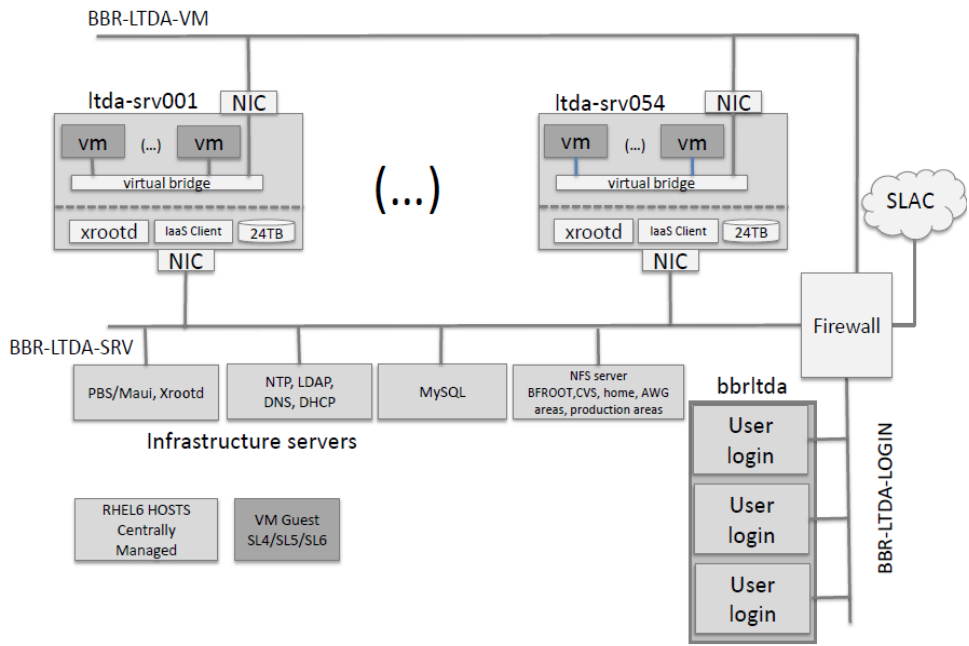

*Figure 14: The networking schematic used for the BaBar LTDA archival system with the primary features being the firewall between the virtual machine backend and the server and login networks. Note the connection between the backend machines and the infrastructure servers for shared storage access.*

The production phase for the BaBar archival system will officially start March 2012 but there are already more than 20 users doing analysis work on it and it is also being used for simulation production.

For such systems it was discovered that a carefully designed network was essential to obtaining a secure design acceptable for the host laboratory. Since the virtual machines are using operating systems that are no longer secure the virtual machines are considered to be a-priori compromised. Such rules such as not allowing virtual machines to write to the home areas and not allowing connections from the virtual machines beyond the systems firewall had to be implemented. Another lesson was to be careful of the licensing costs which if the virtual system design is not carefully done can result in high significant extra costs.

**Other preservation efforts taken by BaBar**

BABAR's raw and most recently reconstructed data and physics streams (the legacy data) were migrated to new media. Old tapes were stored away safely but old drives would have to be purchased to read them. All of the legacy data is being copied to the CCIN2P3 centre in Lyon, France, and is expected to take about 1 year to transfer. To continue to be able to access the raw data copy at Padova, Italy, the tape library was restored. The analysis storage areas that were previously left to the analysts to perform backups themselves are now automatically backed up. The history of both the BaBar and Belle analyses are being preserved in the joint BaBar/Belle physics of B-factories book (PBF) project. BaBar is also looking into using INSPIRE for long-term archiving of analysis, appropriate support documents and data taking logs.

**Lifetime**

The BaBar archival system will be needed until the data samples of one of the future SuperB/BelleII experiments has accumulated as much data as BaBar. The dates for this range from 2016 to 2018, at the earliest. However, the need will likely extend far beyond that as crosschecks between future experiments and BaBar will likely be needed.

**Timescales**

As of March 21st, 2012 the BaBar archival system is complete and now in its production phase with over 20 users. A procedure to accelerate the migration of analysts to the system is now under study.

**Documentation**

The archival system documentation including documentation of the analyses performed on the archival system are on a Wiki setup explicitly for this purpose. Much of the old html BaBar documentation has been replace by up-to-date documentation on the Wiki. The scope of the effort was such that a documentation working group (DWG) was formed which not only had experts on it but also new students who are able to more easily catch the holes in the old documentation. Other incentives have also been employed to encourage experts to work on documenting their area of expertise. An example of an incentive, service credit could be assigned to this task so that contributors will have a higher chance of getting offered conference talks.

**Person-power**

The projection for the number of FTEs needed for supporting BaBar Computing not including centralised services (managing LSF, HPSS, basic hardware maintenance of batch, WWW and database servers and so on) is 10 FTE (2009), 7 FTE (2010), 4 FTE (2011), 2 FTE (2012) and 0.5 FTE (2013 and beyond). This is a mixture of two categories of FTEs called physicists (no special skills needed) and computing professionals that can include not only purely computing scientists but also physicists with special computing skills. For the data preservation effort contributions were received from individuals with the following skills:

- Code, virtualisation and cloud expertise
- Systems, conditions database, XRootD expertise
- Production, data quality and database expertise
- Joint fitting, data monitoring, production expertise
- Reconstruction and general code expertise
- Wiki, outreach, production, python expertise
- Documentation and job manager expertise
- Computing and offline coordination

Each of the individuals associated with the above expertise spends roughly 0.25 FTE on the effort. In 2011 funding from the DOE for a research associate was granted, which turned out to be of tremendous help. In addition to the core group presented above, the participation of the following bodies was very important:

- An LTDA Advisory committee formed mostly of members external to the collaboration
- Central SLAC computing personnel
- Support from SLAC, the International Finance Committee and the DOE in the form of access to resources, funding and awarding of a post-doc position to contribute to the effort

**Future Organisation**

A task force addressing the future organisation of Babar is now installed. Among the major issues includes the means of continuing the strict analysis review procedures to maintain the high quality and confidence in results published on BaBar data. At the current time, the collaboration activity level has not diminished significantly and the current governance is suitable for the near future.

## 4.2 The Data Preservation Programme at H1

The physics programme of the H1 experiment, which is well defined and regularly updated by the collaboration, is scheduled to come to an end by 2013/14, after more than 20 years of study of the ep data from the HERA collider. The computing model for the end of the analysis period at H1 was adopted in 2006 and the allocated resources were approved by the funding agencies in 2008 for the period up until 2013. A regular survey within the H1 Collaboration shows a linear decrease of the person-power from about 250 members in 2008 to about 50 expected in 2013. The computing and data preservation projects are evaluated in this context in order to identify the optimal evolution of the analysis model towards a stable system which should have the best chance for a long-term perspective, after the end of the collaboration in its present form.

**H1 data and MC samples**

The preservation of the data themselves is in fact only a small part of the project, and is relatively easy and inexpensive. The H1 raw data (made up of good and medium quality runs) comprises around 75 TB and is the basic format to be preserved. A full set of Compressed Data Storage Tape (CDST) data for the 1996-2007 period is about 15-20 TB, and the analysis level files (H1OO) are around 4 TB. Although only 1996-2007 has been regularly reprocessed (the final iteration was DST 7), the early collision data from 1994-1995 will also be secured in the raw data format. Other data, such as random trigger streams, noise files, cosmic-data, luminosity-monitor and other calibration data amounts to only a few TB. Standard MC sets for preservation will also be defined, where the total data volume is likely to be similar to real data. This includes generator files as well as the larger simulated and reconstructed files. The total preservation volume, including MC and non-collision data, is conservatively estimated to about 0.5 PB, and certainly no more than 1 PB, which is an order of magnitude lower than the predicted yearly data output of the LHC experiments. The availability and associated security of the different types of preserved H1 data will be defined in cooperation with DESY-IT, who will continue to host the data in the long term.

### Reconstruction and simulation software

The H1 reconstruction and simulation software, which creates DSTs from the raw data is written mainly in Fortran, but also contains some C and C++. The MC simulation takes the generator files as input and passes them through GEANT 3, a Fortran-based simulation of the H1 detector, taking the relevant run conditions from a database, to produce MC events in the same format as the data, with some additional information. The same reconstruction software as used on the data is then applied to the simulated MC events. The desired future physics output defines which capabilities should be preserved, but the full potential for improvements is retained only if the full simulation and reconstruction chain is available for analysis, corresponding to DPHEP level 4 preservation, as defined in section 3. As new theory or new experimental methods are likely to be the prime reasons for re-analysing the H1 data, scenarios may arise where only a full preservation model will provide the necessary ingredients, for example if a cut in the current reconstruction turns out to have been too harsh, or a new simulation model, written in an alternative computing language, requires a new interface to the existing code.

### Analysis level software

The majority of physics analysis performed by the H1 Collaboration is done using the same C++ analysis framework, H1OO. This has had huge benefits in terms of shared analysis code, expert knowledge and calibrations, working environments and, perhaps most importantly, handling the actual data, where the whole collaboration uses the same file format, and more often than not the same physical files. The common H1OO files comprise in reality of two persistent file formats: the *H1 Analysis Tag* (HAT), containing simple variables for use in a fast selection and the larger *micro Object Data Store* (mODS), which contains information on identified particles. A third H1OO file format, the *Object Data Store* (ODS) is accessed transiently during analysis and is equivalent in content to the DST. The H1OO framework is based on ROOT, and uses its functionality for I/O, data handling, producing histograms, visualisation and so on. ROOT also provides attractive solutions for code documentation, which are fully utilised by H1. Given the level of use in the HEP community, especially at the LHC, it is expected that ROOT will continue to be supported in the long term. Major development of the analysis software is essentially completed with the recent 4.0 release series, which was developed for and in parallel to DST 7.

### Databases and other external software

As well as ROOT several other external software dependencies exist within the H1 software. ORACLE is used in several key areas: in the NDB database of run conditions, the detector slow control database, registration of generated and simulated MC events and within the H1 webpages for storing information about the collaboration. At some point in the future, it may be possible to freeze the database contents and employ so called snapshots of the last versions. Further external dependencies include: CERNLIB, which is used in analysis level executables; the FASTJET program, which is used in the H1OO jet finder; NEUROBAYES is a neural network utility, which is used in the H1OO cluster separation algorithm; GKS is a graphics interface used in the (old) event display. Non-supported dependencies or packages without available source code will be replaced if possible, although it has already been established that GKS is not available in SLD5, and therefore a small number of SLD4 resources are retained by H1 at some level for the short term.

**Operating system**

The main OS used within H1 is Scientific Linux DESY 5 (SLD5), following a successful migration campaign in 2011. This OS is fully supported by DESY IT and support is expected from the distributor until 2017. The migration to SLD5, which is the base level for the preservation project, was used to streamline the software and identify potential future problems. The default for H1 is SLD5/32-bit, although the recently established validation project (see section 5.1) has allowed SLD5/64-bit compatibility to be evaluated. This system will be rigorously tested in the coming months against SLD6, which will only be available in 64-bit, demonstrating the value of the current evaluation.

**Mass storage**

The main resource for mass storage used by H1 is the HERA dCache, which uses the DESY-IT tape-robot duplicate system. Disk pools totalling around 280 TB are used for the most commonly accessed files. The complimentary H1 dCache system is around 90 TB and has the benefit of being a disk only system, allowing faster access. The increase in storage capacity of working group servers, where the latest models contain 24 TB of usable disk space, has resulted in an increase in the use of such systems for analysis level file storage.

**Batch farm system**

The H1 batch farm is an integral part of analysis at H1, and most users run parallel analysis jobs on this resource. Some MC production is also performed on the batch farm (simulation/reconstruction as well as H1OO file production), especially when the number of events is small, where typically this means requests with less than 1 million events. The batch software is fully supported by H1, including specific modifications related to access to AFS and Kerberos credentials. The current capacity allows for 1000 jobs to run in parallel and the H1 batch farm is expected to continue until at least 2013/14, albeit with a reduced capacity.

**Data and MC production frameworks and the GRID**

Reprocessing of the H1 data has traditionally been performed on the H1 batch farm resource, although future reprocessing, which is currently not planned, could be done on the GRID. The majority of the large scale MC production is now done on the GRID. Automated production of the analysis level files on the GRID in the correct H1OO software version follows the registration of new DST MC files. The H1 MC production is a success story, with record production rates of more than 500 million events per month, and around 2.5 billion events annually since 2008. It is anticipated that the level of MC production will begin to decrease in 2012, but will continue in the current operational mode until at least 2013, assuming the structure and interface to the GRID does not significantly change. The H1 MC production for the period 1999-2011 is illustrated in figure 15.

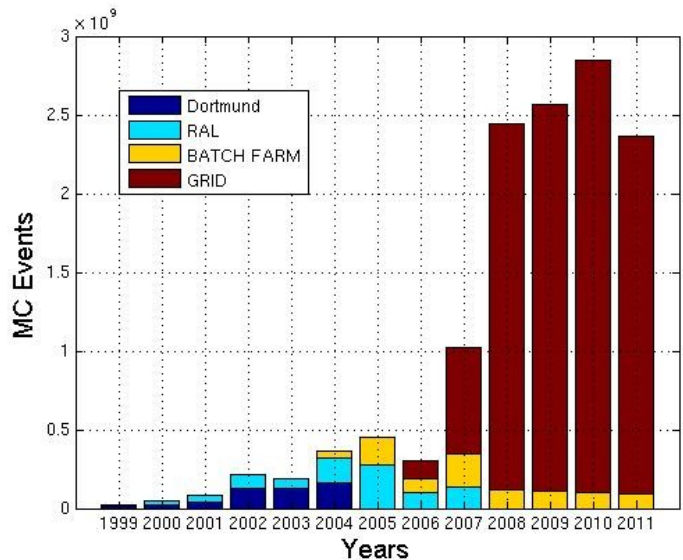

*Figure 15: H1 Monte Carlo production in the years 1999-2011.*

**Current and future person-power requirements**

About 6 people were involved at H1 in the large scale data reprocessing, simulation and reconstruction, and continue to be available for general enquiries and bug fixes. Another 4 key people are involved in the MC production and GRID matters. In terms of the analysis level software, around 3 people are required at the top level, with additional contributions from H1 Collaboration members. A dedicated H1 web-master is also installed, additionally contributing to ORACLE maintenance. The H1 person-power requirements for the long-term phase have been evaluated, where the majority originate from these areas, but others for specialised cases dedicated resources have also been identified.

**Validation tools**

The H1OO software already employs validation tools, so that new or different versions of the analysis software can be easily compared. If future OS or software version transitions are to be considered, investing in the development of more validation tools is necessary, to detect changes and avoid surprises as early as possible. In the case of H1, a complete test suite should be set up for the full software chain, where the cycle will be performed on regular basis to validate all steps against changes in the software environment, beginning with compilation of the basic software and ending a scientific comparison of standard analysis results. In addition, the integrity of the preserved data in a suitable archival storage should be verified. The frequency of such tests can be defined once the validation suite and archival storage systems are in place. The development of both these systems is described in section 5.

**Digital documentation**

A great deal of digital H1 documentation exists, mainly but not exclusively on the official webpages. This includes published papers and preliminary results, review articles and expert notes. In addition, talks from meetings, conferences, lectures, and university courses are also available. There are also many unpublished articles, such as H1-notes and the internal wiki pages that are extensively used by H1. Data quality information (physics and technical) and other electronic documentation like H1 software manuals and notes also contribute. In-house DDL documentation of Fortran software (h1banks) should be updated and/or completed. As mentioned above, the H1OO analysis level software benefits from the automatically generated ROOT documentation in HTML, but only if the code is correctly written, and any missing information should be addressed. Old online shift tools contain much metadata and are particularly vulnerable to loss. Such information was mostly not updated since July 2007 and electronic logbooks (shift, trigger and other detector components) and detailed run information contained in the system supervisor should be secured. Calibration files may still exist on old hardware: in excess of 20 online machines were employed during data taking. Concerning data from the HERA machine group, the status of some of the information about beam conditions, luminosity and polarisation remains unclear.

**Non-digital documents**

A general survey of the state of the non-digital H1 documentation has been performed. There is a great deal of paper documentation: H1 physics and technical talks from pre-web days; detector schematics and blueprints; artefacts from the experimental hall like older logbooks. A future location large enough to store all the documentation for preservation has been secured in the DESY-Library. However, the cataloguing and organisation of large quantities of documentation is also a significant task that can only be done by someone with expert knowledge of the H1 Collaboration. The INSPIRE project has offered, via the DESY-Library, to aid the documentation effort and several pilot projects are underway with the HERA collaborations: including the ingestion of internal notes, digitisation of theses and electronically cataloguing the publication histories (preliminary results, T0 and referee reports, versions of the paper draft). More details on these projects can be found in section 5. The larger scale digitisation of older H1 documentation is also underway, although given the volume of material, this again has required prioritisation, where preference has been given to plenary meetings.

**Future governance of the H1 Collaboration**

After detailed discussions within the existing collaboration infrastructures, a new model for long-term governance of the H1 Collaboration was adopted in 2011, where the transition from the current model will take place in July 2012. In the new model, the current H1 Collaboration Board (H1CB), which includes representatives from all participating institutes and the H1 Executive Committee, which is a smaller structure elected by the H1CB to meet more regularly, will be replaced by the H1 Physics Board (H1PB). The mandate of this board, which comprises a broad selection of H1 members from all physics and technical working groups is: to be the general contact point for H1 physics and data beyond the collaboration lifetime; to communicate with the host lab (DESY) and other experiments; to supervise the H1 data: to maintain contact with the global DPHEP initiative; and to overview further publications using H1 data. The initial composition of the H1PB, comprising 32 H1 members with an initial mandate of 3 years, was approved in February 2012. Future membership of the H1PB, along with all major decisions concerning H1 in the new collaboration model, will be subject to election by all H1 members. The new organisational model is illustrated in figure 16.

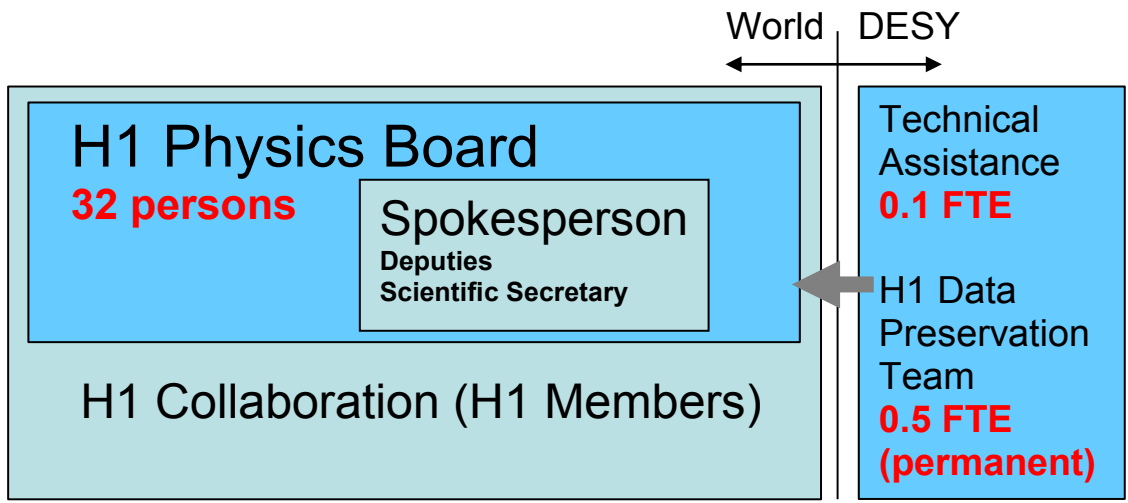

*Figure 16: The new organisational model of the H1 Collaboration, which will be adopted in July 2012.*

### 4.3 The Data Preservation Programme at ZEUS

The first plans regarding the ZEUS computing strategy for the years beyond the end of data taking were described in a document prepared by the ZEUS Computing Board in December 19th, 2006. It defined the collaboration strategy in terms of data reconstruction activities, analysis model, MC production, data preservation, storage and access, cost estimates and funding. The follow-up document, one year later, extended funding till the end of 2013. Further refinements were made until the end of 2010, when the plans were finalised. In the following, the main points of these documents and the implementation of their content are recalled.

**Analysis models of the ZEUS experiment**

The support for the traditional ZEUS analysis model based on MDST (Mini Data Summary Tape) data files will finish by the end of 2012, together with the computing farm operation and access to RAW, MDST data and MDST MC files in the tape library. The current analysis model based on MDST is gradually turning into a so-called Common Ntuple-based model. Instead of each user building his own (different) analysis ntuple from the MDST, the Common Ntuple is designed to be a superset of all these potential ntuples for (almost) any possible physics analysis. Reflecting its relative simplicity and the usage of HEP community tools only (ROOT), the Common Ntuple-based model was chosen as the best candidate for long-term preservation and analysis of ZEUS data (and MC). The possible future incompatibility and lack of support for the software and data formats used in the current analysis model was a major motivation for this choice. The present access and storage methods (dCache and tape library) will be maintained for Common Ntuples, along with some workgroup-server infrastructure for subsample storage and analysis. At the end of 2009 the concept was extended by the decision that the ability to simulate small new MC samples and convert them to Common Ntuple format should be retained in some simplified form.

**The traditional computing model of the ZEUS experiment**

The RAW data collected by ZEUS experiment are kept in ADAMO (Entity-relationship model) structures based on ZEBRA files. The same applies to the reconstructed level (MDST) and MC simulation. Calibration, conditions, geometry and alignment are kept in database-like system called General ADAMO Files (GAFs). Apart from data files, so called Event Collections and Event Tag Database are created during the reconstruction process, allowing fast selection based on trigger and physics quantities. All RAW, MDST and MC data files are registered in an ORACLE database. Data files are stored on tapes with access via the dCache system. Other files are stored either on AFS or on a dedicated NFS server. Average sizes are 125kB/event (RAW) and 75kB/event (MDST). Data samples of HERA-I (1995-2000) and HERA-II (2003-2007) consist of approximately 590 million events, which break down to 68TB (RAW) and 41TB (MDST). MC simulation utilises a detector simulation program based entirely on Geant3. The MC production system is a distributed system among working nodes across many collaborating institutes with central servers located at DESY. Attached to it is a GRID-based system, which allows for a unified submission and retrieval of MC file and covers over 80% of the total production. The current collection of MC samples amounts to about 1 PB, which covers all MC versions.

The reprocessing of HERA II data was finalised in spring 2009, followed by several tests and validation procedures, and subsequently validated through physics analysis. It is considered to be the final reprocessing. The analysis framework is based on a general interface layer allowing access to data files, on top of which a physics analysis layer is built. This layer includes all necessary reconstruction and analysis libraries and has hooks for a user code. Steering is possible via external files. The output consists of either PAW or ROOT ntuples. The reconstruction and analysis software is mainly FORTRAN based with some fraction of C and C++ code, and relies on external libraries (ADAMO, CERNLIB, ROOT, CLHEP). The computing infrastructure consists of an integrated reconstruction and analysis farm, and workgroup servers for ntuple storage and local analysis. The hardware and software resources are maintained by the Offline Group, currently consisting of about 16 people (~8 FTE), which includes the data preservation R&D and implementation efforts. It is divided into subgroups, like Common-Ntuple production, computing farm, MC production, software maintenance, analysis software development, as well as long-term data preservation, validation, and archival.

**Current status and plans for data preservation and long-term analysis**

The archival system deployed by ZEUS is based on the Common Ntuple project developed since 2006. The common ntuples are produced for data and MC within the current analysis framework and with their wide content allow full physics analysis possibilities. This concept fits into preservation model level 3, as described in section 3. The full set of HERA II data and a wide range of MC Common ntuples are now in use by most physics analyses, and first ZEUS physics papers based on this analysis model have been published. Many improvements and additions to the physics analysis tools made their way into the current ($6^{th}$) iteration of Common Ntuples, profiting from the feedback of the large number of users.

In addition, the long-term ability to produce additional new MC samples is being prepared. It is based on a simplified standalone version of the current GRID production concept and includes the whole chain from generation to simulation to common ntuple production enclosed in a virtual environment. The ability to use new generators is also foreseen via an interface to the currently used format. A working prototype of such a virtual environment, developed in collaboration with the DESY IT division and the other HERA experiments, already exists, with all relevant ZEUS software (calibration, conditions, alignment, geometry, executables and steering cards) and running environments detached from external dependencies (AFS, ORACLE, storage). The goal is to keep the ZEUS MC generation, simulation and reconstruction software executables, compiled with the SLD5 operating system, compatible and running with newer operating systems provided by IT, and to provide a validation system for this purpose. For the already existing data and MC Common Ntuples, only compatibility with new ROOT versions and corresponding storage access needs to be validated.

Finally, the current version of the event display, as an essential tool for visualising real and MC events in the ZEUS detector, was an entirely new project developed for HERA II, based on ROOT. It can display the content of the original MDST and MC file, but also has a functioning interface for visualisation of the Common Ntuple content. This interface is under further development with the aim to give users similar functionality as before.

**Digital and non-digital documentation**

Concepts for long-term preservation of the ZEUS digital documentation are being worked out in collaboration with the DESY IT division and library, e.g. based on the INSPIRE system and/or centrally maintained web servers. ZEUS also maintains an extensive archive of non-digital information. This archive includes all ZEUS notes written before 1995, transparencies presented at meetings before 2000, technical drawings and many other things. Most of this archive has already been moved to its final destination, hosted by the DESY library. It is being consolidated, partially digitised, and catalogued taking advantage of the know-how available in the DESY library. At the end of 2012 custody of this archive will be fully handed over to the DESY library.

**Governance**

Along with the change of the analysis and funding model for the post-2013 era, the structure of the ZEUS collaboration is being redefined. This process is to be finalised at the time of writing.

**Long-term Prospects**

Many of the key ZEUS physics results and H1/ZEUS data combinations with full HERA II statistics have already been published and most of the remaining ones enter their final analysis phase now and are expected to be published within the next two years. Nevertheless, the HERA data are and will remain unique for a long time, and not all relevant ZEUS physics topics will be finally covered in this period, also due to shrinking person-power. Furthermore, new physics issues might arise. The Common Ntuple concept implies a significant simplification of the analysis procedure (publishable results are in part already now being produced by Masters students), allowing future analysis with relatively modest resources, provided that some minimal amount of support for data preservation continues to be allocated by the host lab. The concept is designed, both from the technical and the person-power point of view, to provide the possibility for such data (re-)analysis for decades, and it is anticipated that it will be continuously used in this way, as indicated in figure 11 (right).

## 4.4 The Data Preservation Programme at HERMES

HERMES joined the DPHEP initiative in 2009, and intends to aim for a DPHEP level 4 preservation scheme. Several key topics were identified by that time and developed in the subsequent years, such as the maintenance of the digital and non-digital documentation, analysis and production software and data storage, in a close collaboration with the other DPHEP participants. Several alternatives in storage and computing have been tested; in particular the GRID has been actively deployed for MC productions. In the beginning 2012 all final data productions have been done, and the plan for the on going and future analysis as well as MC productions developed.

**Analysis models**

The main analysis format for HERMES is the Distributed ADAMO Database (DAD), a layer allowing ADAMO files to be accessed transparently as local files, Unix sockets or network streams. The same format is used for simulated and reconstructed MC data, as well as reduced micro-DST files that are the main source of the analysis. As such, not only the production software, but also the user analysis codes need to have a link to the DAD libraries to access the micro-DST relational database to access track, cluster and PID information, as well as some low-level data included for data quality and debugging. From the data preservation perspective this imposes specific requirements and external dependencies, such as (unsupported) CERNLIB 2005. Additional analysis, fitting and plotting is performed using CERNLIB and ROOT packages, in the latter no significant version dependence has been observed.

**Software preservation and validation**

To assure flawless compilation and running of future analysis tasks on modern computing platforms, HERMES uses the validation framework being developed by DESY IT, described in section 5.1, where the consistency of software compilation and real physics analysis tasks can be validated. While the full HERMES software tree has been ported to an up-to-date SLD5/32-bit system, various dependency problems have been revealed during the deployment of the validation system. These include the dependency on a particular CERNLIB version, related to a particular compiler version, as well as potential problems in compiling in 64-bit mode. These limitations are however not crucial since there's a high likelihood of support for SLD5/32-bit systems until 2017. Meanwhile, a few tests of HERMES analysis are in development in order to validate results obtained using Fortran, C or C++ based codes.

**Traditional computing**

During the active period HERMES has already undergone a radical change in the computing infrastructure, where the interactive 28-core SGI system has been replaced by a batch Linux cluster, running SLD3, consisting of batch cluster of 80 cores, 2 workgroup servers and 1 central master application server, which also hosted HERMES webpages and mailing list server. Also, 21 NFS fileservers with total disk space of ~100TB served reconstructed and simulated data to the users, as such reducing the usage of the tapes mainly for backup purposes.

**Current status and future plans**

As general analysis activity and person-power got reduced, future long-term alternatives to the existing computing and storage resources have been explored. For the anticipated usage and storage intensity, HERMES has opted for the following set-up:

- IT supported web and wiki servers
- Machine group logbook server (experimental logbook preservation and access)
- IT supported mailing lists
- dCache-based file-service for MC and real data (micro-DST level)
- AFS-based file-service for user and group storage
- IT supported BIRD batch cluster as a computing resource for analyser
- GRID for bulk MC productions

The long-term archiving facility for RAW and not-often used low-level data has not yet been identified in the IT. The space requirements of HERMES will range in the order of 0.5PB for those.

**Digital and non-digital documentation**

Similar to other HERA experiments, HERMES actively deploys the Library and INSPIRE systems for archival and storage of their non-digital and digital documentation, respectively. The dedicated space in the DESY Library is used for catalogued experimental and detector logbooks, early conference and collaboration meeting talks and design notes. All the electronic versions of the internal notes have been made available on INSPIRE in a password-protected area, where HERMES considers to store and cross-link also the earlier paper drafts and relevant information from mailing list archives.

## 4.5 The Data Preservation Programme at Belle

The Belle experiment completed taking data on June 30th, 2010 accumulating an integrated luminosity of more than 1 $ab^{-1}$ after eleven years of operation. The entire data sample was reprocessed in 2010 with an improved tracking algorithm and updated detector constants/parameters. The Belle collaboration has entered the Intense Analysis Phase of this data sample. During this period the collaboration is working to obtain final analysis results based on the full data sets. It is important during this phase to keep the Belle data intact for the Belle collaborators.

In the meantime, the Belle II experiment at the SuperKEKB accelerator has been fully approved in Japanese Fiscal Year (JFY) 2011 and it will be commissioned in 2014. The Belle data will be used to benchmark the early results from Belle II.

The Belle group discussed a policy on data preservation. It was decided that the Belle data will not be released to the public domain until the time the statistics of Belle II supersedes the Belle data and the time all Belle members (and Belle II members) lose interest in Belle data. This situation will likely occur around 2016-17, a couple of years after the SuperKEKB beam commissioning. The plan for Belle data preservation after this period has not yet been determined.

**Data sample**

Belle uses a home-grown bank system as data persistency for all data. Raw data are stored on a tape system that is dedicated to the Belle experiment at KEK and comprises more than 1 PB. Here, the average event size of the raw data is 30-40kB (the size of a typical hadronic event is ~70kB). As a first step, calibration constants are determined from the pre-scaled raw data, the so-called DST data set. After the calibration and alignment constants are determined, the raw data is processed and the derived higher-level information is stored in smaller files (mDST) for the specific physics analyses, for instance, $B$ physics, $\tau$ decays, two-photon processes, and so on. Typical event size of mDST is approximately 30kB. In addition to the real data, generic MC samples corresponding to 6(10) times the statistics of the real Y(4S, 5S) on-resonance (off-resonance) data were created. The output of the MC simulation is stored in mDST format and the size of a MC mDST event is similar to that of real data mDST.

**Software**

Most of the reconstruction software including user analysis is C++. However, because the Belle software was written in C and FORTRAN in the early years, it contains these legacy languages even now. In particular, the full detector simulator is based on GEANT3. Belle developed a dedicated analysis framework "BASF", which has the capabilities of the dynamic linking and the event-by-event parallel processing. The Belle software runs on CentOS 5/64bit and Scientific Linux 5/64bit.

**Computing model**

The original concept of the Belle computing system was to perform all data processing, MC production and physics analysis in a computing facility at KEK. Furthermore, the raw data and all mDST data were intended to be stored there. Eventually, Belle adopted a centralised computing system. As time progressed, thanks to the excellent operation of the KEKB accelerator, the amount of data as a function of the integrated luminosity increased. Additionally, the corresponding Generic MC sample production needed to be produced in a timely manner. In order to cope with these demands, several institutes in the Belle collaboration joined the MC production efforts with local PC farms. Furthermore, since 2007 the Belle Virtual Organization was initiated to utilize grid resources for the MC production.

The computing resources at KEK have been updated several times over the lifetime of the Belle experiment. In February 2012, the latest Belle computing system at KEK, which had a 46 kHepSPEC CPU power, 1.5 PB disk space, and 3.5 PB tape storage, finished the operation and replaced with the new KEK central computing system described in the next paragraph.

**Belle Data Preservation Plan toward the Belle II experiment**

Belle II is required to handle an amount of data corresponding to 50 times the Belle volume in JFY 2020, accordingly huge computing resources relative to Belle are required even in the early stage of the Belle II experiment. The resource requirement of a new computing system up to JFY 2015 was done based on the expected luminosity prospect of the SuperKEKB accelerator, including the resources to continue analysis of the Belle data. At a minimum, all raw data and mDST files from Belle are retained. The new computing system is now being prepared for the official operation from April 2012. The new system has a similar CPU power with the latest Belle computing system, 7PB disk space, and 16PB tape storage. However, it is not a system dedicated only to the Belle/Belle II experiments anymore. The system is shared by the other experiments, such as J-PARC, ILC, and material sciences and so on.

On the new system, the use of CentOS 5/64bit equivalent operation system would make it simple to install the Belle software. Other considerations include the handling of the database for detector constants and how to import metadata of the files already processed/produced in the current computing system. Overcoming these difficulties will enable all Belle and Belle II members to access and analyse the Belle data in the new computing system.

Another challenge is the duration of operation of the new system. The new computing system will be replaced in summer 2015. Therefore it will be necessary to iterate the resource requirements in the process of designing the next-next computing system and transfer the Belle data again, in order to reach the goal of preserving Belle I data until 2017.

## 4.6 The Data Preservation Programme at BESIII

### Motivation of data preservation

In early 1980s, IHEP decided to build an $e^+e^-$ collider running at the tau-charm energy region, called BEPC, which was completed in 1989. The detector at the machine is called Beijing spectrometer (BES). In the mid 1990s, there was a minor upgrade of the detector, which was then called BESII. The latest upgrade of BEPC was decided at the beginning of this decade, called BEPCII, which has a design luminosity of $10^{33}$ $cm^{-2}s^{-1}$, an increase of a factor of 100. The corresponding detector, BESIII, adopted the latest detector technology to minimise systematic errors in order to match the unprecedented statistics. The physics programme of the BESIII experiment includes light hadron spectroscopy, charmonium, electroweak physics from charmed mesons, QCD and hadron physics, tau physics and search for new physics. Due to its huge luminosity and small energy spread, the expected event rate per year is historical.

The BESIII experiment will last another 6 years. From the previous experience, it is clear that the physics potential of the data can be exploited for even longer time. It is expected the lifespan of the data can be more than 15 years. There is no clear decision whether there is new generation of similar experiment in the future.

### Data preservation model

The BESIII will take about 10 billion J/psi data and the data collected in other energy points such as Psi', Psi(3770), Psi(4040), etc. will be of equivalent size. The total amount of raw data is estimated to be about 3.6 PB. It is supposed the data reconstruction is repeated at least twice a year; the total size of the Rec. and DST will be about 1.8 PB. The size of Rec. and DST data from MC simulation will be at the same level as real data so that the total storage capacity should be 10 PB.

### BESIII software

The BESIII Offline Software System (BOSS) is developed using C++ language and object-oriented techniques on the operation system of Scientific Linux CERN. The CMT is used as a software configuration tool. The BESIII software uses lots of external HEP libraries including CERNLIB, CLHEP, ROOT among others, and also re-uses parts of code from Belle, BaBar, ATLAS and GLAST experiments. The whole data processing and physics analysis software consists of five functional parts: framework, simulation, calibration, reconstruction, and analysis tools.

The BOSS framework has been developed based on Gaudi, which provides standard interfaces for the common software components necessary for data processing and analysis. The framework employs Gaudi's event data service as the data manager. Reconstruction algorithms can access the raw event data from Transient Data Store (TDS) via the event data service. However, it is the raw data conversion service that is responsible for conversions between persistent raw data and transient raw objects. The detector's material and geometry information are stored in the GDML files. Algorithms can retrieve this information by visiting corresponding services. Through the DST conversion service, the reconstruction results can be written into ROOT files for subsequent physics analyses. Furthermore, the BOSS framework also provides abundant services and utilities to satisfy the requirements from different BESIII algorithms.

**Storage**

The storage resources consist of hierarchical storage management (HSM) system and parallel file system. The HSM of BESIII is based on Castor system developed by CERN that uses an IBM TotalStorage 3854 tape library system and LTO-4 tapes. Disk pool with about 200 TB capacities is used as file cache. The parallel file system is built on Lustre system to provide high throughput file access service to batch jobs. The capacity of the Lustre system is currently 1.6 PB and another 1 PB will be added in 2012.

**Batch system**

A PC farm is built with more that 5000 CPU cores as the computing resource for both reconstruction and MC simulation tasks. The PC farm and parallel file system are interconnected with 10G Ethernet so that each CPU core can support a job at the same time with high efficiency. The AFS and Kerberos credential system is deployed for user identification. Torque and Maui are chosen as the batch job system for BESIII jobs. Torque is an open source resources manager providing control over batch jobs and distributed compute nodes and Maui is the scheduler cooperate with Torque to control over when, where and how resources such as processors, memory and disk are allocated to jobs.

**GRID and cloud**

BESIII computing tasks are currently performed on the traditional batch system. But a grid environment is being developed using gLite middle. The job distribution and management on Grid has been successfully tested. But the data transfer and management services are being developed from scratch. A test-bed of Grid computing system has been established with a main site at IHEP and some satellite sites in China, Hong Kong and the USA. The use of virtualisation techniques is being investigated.

**Plan of data preservation**

In the period of the last year, and following discussions in the BESIII group, it is expected that the data preservation strategy will be established soon. BESIII is currently using a tape library as the backup system. From the experience in the last few years it was found that the operation of a tape system needs more human interventions. This may not be suitable solution when the experiment is concluded and support would be limited. New storage technology is being investigated. Besides the raw data, the way to preserve software and documentation is also being studied.

## 4.7 The Data Preservation Programme at CDF

The data preservation effort at CDF is in the early stages of R&D in which the primary goal is to define targets, understand requirements, and identify the major issues to address. Although many other experiments discussed within this document are at a more advanced stage, it is useful to describe the status of work at CDF and plans for further work to give some perspective on how experiments progress through the various phases of data preservation planning and implementation.

**Current computing model**

There are two major classes of processing within the CDF computing model: data and MC production, and data analysis. The scale of processing activity, level of automation, and the specificity relative to a particular physics analysis distinguishes these classes. Within the production class, there are two types of activities, raw data production and MC production that differ primarily by the source of input data to the reconstruction. Both activities are centrally managed and operated by a small group of collaborators. The output for both production activities is the same.

Raw data production itself involves several steps: extracting calibration constants from about 20% of the data stream, performing the full primary pattern recognition and reconstruction of all raw data, splitting the reconstruction output into primary production datasets, and performing a secondary reconstruction and initial data reduction to produce analysis-level ntuple files. An event is routed to one or more primary datasets based upon the triggers satisfied by that event. About 30% of all events are routed to more than one dataset. There are three separate ntuple flavours. Each primary dataset is ntupled by at least one flavour, while some are ntupled by more than one. The number of ntuple events on tape is about 50% larger than the number of primary production events due to this overlap.

MC production involves the generation of physics events and a full detector simulation to produce simulated raw data, followed by the standard raw data production processes. The final output of MC production is the fully reconstructed and n-tuple output files.

The second class of processing, the data analysis, is carried out by individual physicists working on specific analyses. A typical analysis will start with the production n-tuples to produce a set of smaller secondary and tertiary ntuples that contain the events and information of interest. The secondary and tertiary datasets are sometimes stored on tape, sometimes not, depending upon the discretion of the physics group. These data will be processed repeatedly in order to extract the physical measurements of interest. Most of this processing takes place on the computing resources at Fermilab. Most analyses will require that the creation of specific signal MC data samples in addition to the general background samples routinely created by the MC production group. Large signal MC samples that utilize the full detector simulation are typically submitted to the centrally managed MC production team. The resulting MC n-tuples will also be analysed repeatedly in order to understand systematic effects, obtain background estimates, etc.

Some additional CPU-intensive analysis processing is necessary in some cases. The most CPU intensive of these include calculations for matrix element analyses, complex likelihood fits, neural network calculations, confidence interval estimates, and MC pseudo-experiments. This processing is performed on a variety of resources, including those at Fermilab and off-site. At Fermilab, the various analysis activities consume between one half to two-thirds of the available computing cycles.

**Data sample**

All data are stored on tape within a dedicated Enstore library at Fermilab. Metadata describing the contents of each data file is stored in a data catalogue. The data handling system can be used to define datasets based upon metadata queries within the catalogue. The files within such a datasets can then be delivered upon demand from tape to worker nodes for processing via a 900 TB dCache-based disk cache. On a typical day, the system delivers between 50-150 TB to the processing farms at Fermilab. Table 4 shows the average event sizes for data taken at high luminosity, the total number of events on tape, and the total data volume within each category of data.

| Data type | Event size [kB] | Events logged ($10^9$) | Volume (PB) |
|---|---|---|---|
| Raw data | 150 | 14.7 | 2.2 |
| Production output | 150 | 26.0 | 3.8 |
| Production ntuples | 32 | 46.8 | 1.4 |
| MC data | 140 | 7.9 | 1.1 |
| MC ntuples | 67 | 8.9 | 0.6 |

*Table 4: CDF data stored on tape as of July 2011.*

**Reprocessing strategy**

CDF has attempted to minimize the need for large scale reprocessing of the raw data. With the notable exception of the tracking software, the event reconstruction software has been relatively stable since the early stages of the experiment. In the case of tracking, increases in the instantaneous luminosity of the beam and the resulting increase in the average number of collisions per beam crossing have threatened to reduce the track finding efficiency. The experiment has in most cases introduced improvements to the tracking software in advance of these increases so that no raw data reprocessing was needed.

Remaking ntuple datasets occurs as needed. Since the ntupling procedure includes a secondary reconstruction phase, many minor problems with the reconstruction can be addressed by remaking ntuples. Historically, the vast majority of these ntuple reprocessing episodes have been relatively small in scale. More recently, several problems and improvements have been discovered that will require large scale re-processing of certain datasets. In one case, for instance, a change introduced to address a high-luminosity tracking inefficiency introduced an unanticipated side effect in B-hadron tagging. As a result, ntuples for a large fraction of various signal and background datasets will need to be remade. The experiment currently expects that a final large-scale reprocessing of either the raw data or ntuples will be required after the end of data taking. The goal of this re-processing will be to incorporate a number of improvements and fixes now under discussion that will allow the experiment to better exploit the potential of the complete dataset.

The processing strategy for large-scale re-processing projects is either to increase the resources available to CDF temporarily via grid computing elements at Fermilab, or to tax CDF analysis users during the period of re-processing. Within either scenario, the re-processing tasks considered above will take several months to half a year to complete.

**Analysis resources**

The primary analysis resource available to CDF is a set of OSG-based computing elements at Fermilab purchased by the experiment. These grid farms provide approximately 10 MSi2k of processing power. CDF also has access to resource allocations at various OSG and LCG-based grid sites, and opportunistic access at others. The resources available from these sites varies depending upon local activities, ranging anywhere from 1 MSi2k to 5 MSi2k.

**Effort**

The experiment currently provides about 6 FTEs toward the coordination and operation of all phases of offline computing operations. The laboratory resources are more difficult to estimate, but are also in the neighbourhood of 6 FTEs. The fraction dedicated to data preservation is a small fraction of an FTE.

**The archival system**

Since CDF is still in the early R&D phase of the data preservation effort, it does not yet have a proposed long-term archival system. Present plans, however, call for DPHEP level 4 preservation (preserving all reconstruction, simulation, and analysis capabilities). All CDF data is stored in ROOT format. Since ROOT will remain supported throughout the LHC era, the current data format should be a suitable initial choice for archival storage. Similarly, the data handling system used by CDF is common to DØ and to a number of other existing and new experiments at Fermilab. All elements of this system should continue to be supported at least through the end of this decade. Continued access to the data through the end of the decade via the existing data handling system and experiment software is therefore subject only to available funding to maintain the systems at a scale appropriate to meet processing demand.

The choice of a possible archival analysis system is less clear. At present, the experiment is actively pursuing the goal of maintaining full analysis capability for at least five years after the end of data taking. While it is likely that many of the challenges faced during this time will be common to those faced for longer term analysis, successfully achieving the five year analysis goal will not ensure an ability to continue analysis into a longer-term archival period.

The experiment is mindful of this fact in planning for the five-year analysis period, and intends to align solutions to longer-term analysis problems toward those that will enhance sustainability further into the future. As an example, creating a deep knowledge base and a simplified analysis framework are essential elements of provisioning any long-term analysis capability. Some of the physics groups within CDF have recognized the utility of developing these elements as part of maintaining analyses that require regular incremental updates and improvements contributed by successive generations of graduate students and post-docs.

The objective of the CDF data preservation effort should be to leverage this work within the physics groups, and provide coordination and guidance across the collaboration so as to obtain a suite of common analysis framework tools, validation benchmarks, code archival procedures, and documentation. From this point, a model for longer-term analysis can be developed and the complete transition steps to an archival analysis system specified. Coordination by the data preservation project will be essential to ensure that each relevant analysis channel is completely specified using this suite of tools, benchmarks, procedures, and documentation.

**Resources for preservation**

At present a task force is being formed within CDF to specifically work on long-term data preservation, composed by members of CDF computing team, physics groups coordinators and experts from Fermilab Computing Sector. Investigation is on going to evaluate the possible contribution from offsite computing centres supporting CDF computing (CNAF and KISTI computing centres in Italy and South Korea). It is estimated that less than 1 FTE is currently devoted to data preservation; this number is expected to increase as soon as the task force will become fully operational.

## 4.8 The Data Preservation Programme at DØ

DØ is currently assessing its plans for data preservation, and discussing the format in which the data should be preserved, and how it will be accessible. To illuminate the extent to which data preservation methods developed for other experiments could be applicable to DØ, an overview of the computing system and data access is presented.

**DØ computing model**

Data from the detector are passed through a specialised reconstruction program that applies pattern recognition algorithms to form objects that are used for physics analysis, such as electrons, muons, jets, and missing energy. Based on triggers and the set of objects found, the events are written out to one of 13 "physical skims," the largest of which represents about 30% of the total number of events. Event flags are also added to each event, and these flags are used to rapidly filter the data when seeking a certain class of events. Another program converts the reconstructed output into a separate analysis data format in the form of a ROOT tree, which are typically used in physics analyses.

Simulated events are produced using one of several event generators, with the output passed through a GEANT3-based program that simulates the response of the DØ detector, and then through a program that converts the GEANT3 energy deposits into simulated raw data. During the final simulation stage, zero-bias events from data are overlaid on the simulated events to mimic the effect of multiple interactions per beam crossing. The output is then passed through the same reconstruction program as is used for data; however the simulated events are not separated into skims. The reconstruction output is converted into the same ROOT tree analysis format that is used for data.

Database access is required when reconstructing raw data, and therefore this is done exclusively at Fermilab. Simulated events are generated at external sites, which are either DØ-dedicated clusters or accessed opportunistically using grid technologies. Flat files that hold an approximation of the information available in the database are shipped with the executable to the worker node, and this information is used in the reconstruction of simulated events.

**DØ data sample**

Data taking at the Tevatron ended on September 30, 2011 and delivered 11.9 fb$^{-1}$ of integrated luminosity. DØ has collected a total of 10.2 billion raw data events in Run II, with 10.7 fb$^{-1}$ of integrated luminosity. All raw data, reconstructed data, skimmed data, and the analysis format data (ROOT tree format), along with all MC events are stored using a data handling system, SAM, into Fermilab Enstore tape storage. SAM is a set of servers that work together, communicating via CORBA, to store and retrieve files and associated metadata. Enstore is a robot-operated mass storage system developed and operated by the Fermilab Scientific Computing Division as the primary data storage for scientific data sets.

The total size of data stored in Enstore is 8.4 PB as of May 2012. It is expected that the total size will continue to grow with additional MC events needed by the physics analyses, which are still rather active, until 2015. Figure 17 shows the distribution of various types of DØ data in the Enstore tape system from October 2009 to September 2010, when the total volume was 1.8 PB.

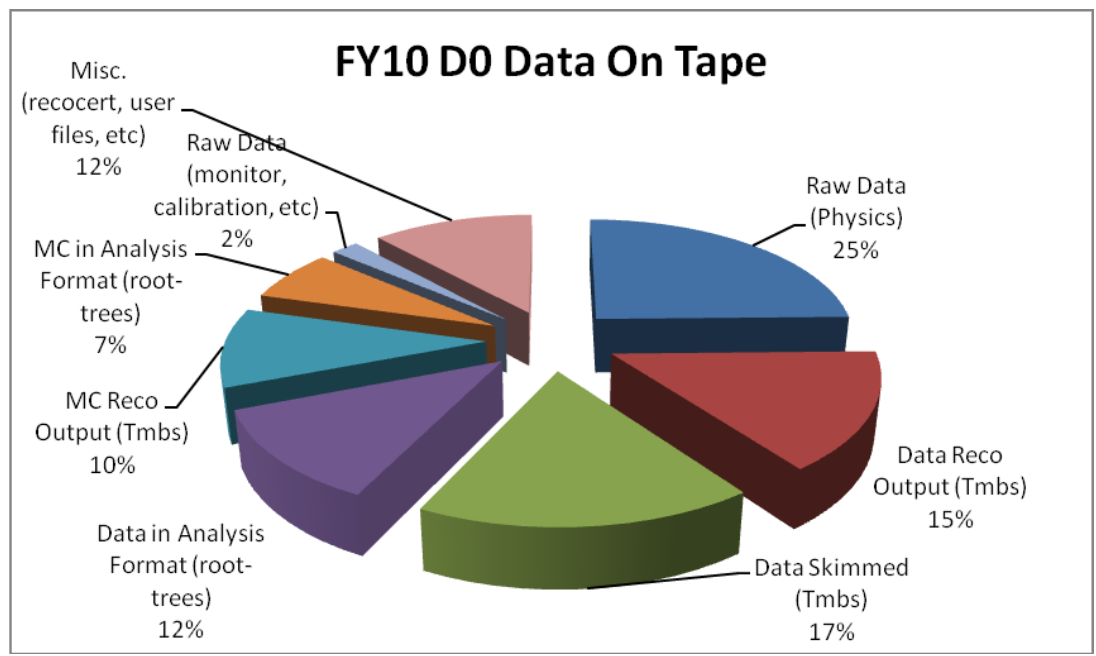

*Figure 17: Breakdown of DØ data stored on tapes between October 2009 and September 2010.*

**Reprocessing**

Data processing at DØ is CPU-intensive, owing primarily to the relatively small number of layers and high occupancy in the tracker. Sufficient CPU resources were available to process the data in "real time" as they were collected (with a few days delay to allow calibration constants to be calculated and stored in the databases), but not for reprocessing large samples of data in parallel. During periods when the Tevatron is running, the reprocessing was limited to a few months of data with updated calibration constants or to specific problems (for example, correcting swapped cables).

Figure 18 shows DØ reconstruction CPU time as function of the average luminosity in GHz-second/event. It increases rapidly with the increase in instantaneous luminosity.

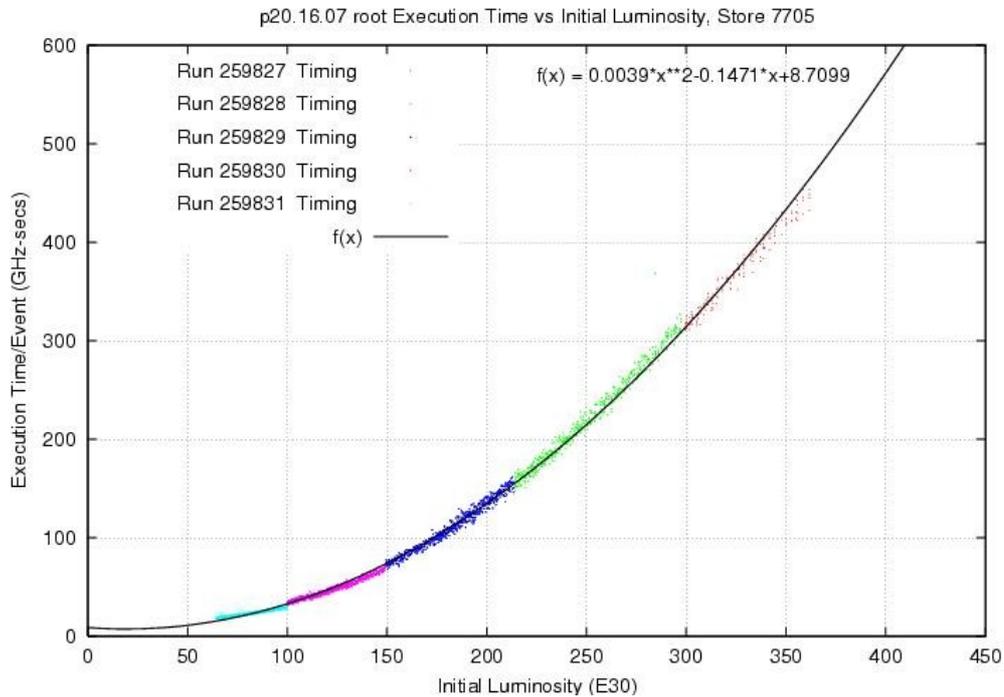

*Figure 18: DØ reconstruction CPU time versus initial luminosity of a run.*

After the end of Tevatron operations, the reprocessing process has used recently improved algorithms for a set of selected raw data, which mainly covers high transverse momentum related analyses. About 12% of the Run II data sample has been reprocessed to optimise the physics analyses.

**Effort**

DØ currently devotes about 10 FTEs to all aspects of offline computing, with support from the laboratory Computing Section. Recently some activities are directed towards long-term data archiving, but mainly at the discussion level, which amount to a fraction of an FTE.

**The archival system**

Storage and access to the DØ data in its present form will be supported for five years following the end of the Tevatron operations, which should allow the completion of all currently planned physics analyses. Long-term archival may prove useful if a signal is observed at the LHC, the interpretation of which is dependent on its effect in proton-antiproton collisions in a manner that was not illuminated in any of DØ's publications. Therefore the information archived should be sufficient to allow the re-analysis of the data for unexpected signatures; in addition the capability to generate new MC events should be retained. Proper analysis of DØ data requires that the MC be corrected to account for differences in object identification efficiencies. The corrections are applied in the form of event weights calculated based on the properties of the identified objects in the event (e.g. $p_T$ and $\eta$). Therefore these weighting factors and the infrastructure needed to apply them properly should also be retained. Fermilab Scientific Computing Division will take responsibility for long-term archiving of the Tevatron data, including migration of the data files to newer storage technologies.

**Resources for preservation**

Currently DØ commits a fraction of an FTE toward the task of preparing for data preservation, and the best way to achieve the goals listed above is still under discussion. It is very likely that the solution will be similar for DØ as it is for CDF and other high-energy physics experiments, so working in close coordination with other experiments within the DPHEP group will be of great benefit to DØ.

|  | BaBar | H1 | ZEUS | HERMES | Belle | BESIII | CDF | DØ |
|---|---|---|---|---|---|---|---|---|
| **End of data taking** | 07.04.08 | 30.06.07 | 30.06.07 | 30.06.07 | 30.06.10 | 2017 | 30.09.11 | 30.09.11 |
| **Type of data to be preserved** | RAW data<br>Sim/rec level<br>Data skims in ROOT | RAW data<br>Sim/rec level<br>Analysis level ROOT data | Flat ROOT based ntuples | RAW data<br>Sim/rec level<br>Analysis level ROOT data | RAW data<br>Sim/rec level | RAW data<br>Sim/rec level<br>ROOT data | RAW data<br>Rec. level<br>ROOT files (data+MC) | Raw data<br>Rec. level<br>ROOT files (data+MC) |
| **Data Volume** | 2 PB | 0.5 PB | 0.2 PB | 0.5 PB | 4 PB | 6 PB | 9 PB | 8.5 PB |
| **Desired longevity of long-term analysis** | Unlimited | At least 10 years | At least 20 years | 5-10 years | 5 years | 15 years | Unlimited | 10 years |
| **Current operating system** | SL/RHEL3<br>SL/RHEL 5 | SL5 | SL5 | SL3<br>SL5 | SL5/RHEL5 | SL5 | SL5<br>SL6 | SL5 |
| **Languages** | C++<br>Java<br>Python | C<br>C++<br>Fortran<br>Python | C++ | C<br>C++<br>Fortran<br>Python | C<br>C++<br>Fortran | C++ | C<br>C++<br>Python | C++ |
| **Simulation** | GEANT 4 | GEANT 3 | GEANT 3 | GEANT 3 | GEANT 3 | GEANT 4 | GEANT 3 | GEANT 3 |
| **External dependencies** | ACE<br>CERNLIB<br>CLHEP<br>CMLOG<br>Flex<br>GNU Bison<br>MySQL<br>Oracle<br>ROOT<br>TCL<br>XRootD | CERNLIB<br>FastJet<br>NeuroBayes<br>Oracle<br>ROOT | ROOT | ADAMO<br>CERNLIB<br>ROOT | Boost<br>CERNLIB<br>CLHEP<br>NeuroBayes<br>PostgreSQL<br>ROOT | CASTPR<br>CERNLIB<br>CLHEP<br>HepMC<br>ROOT | CERNLIB<br>NeuroBayes<br>Oracle<br>ROOT | Oracle<br>ROOT |

*Table 5: Summary of information from experiments*

## 4.9 Data Preservation and the LHC Experiments

The experiments at the LHC are foreseen to continue for at least 20 years, given the present schedule of the project. There is however a strong physics case to discuss data preservation now, in order to allow easy access to data collected in previous years, at different centre-of-mass energies, different pile-up conditions, or with lower trigger thresholds. Some use cases for these preservation activities can indeed become a reality in a year or two from now, requiring immediate attention. Examples of use of these data are precision measurements with new or improved theoretical calculations, cross checks for discoveries made at higher energy/higher luminosity, studies related to new models of physics beyond the SM. In addition, given the current and planned studies, the LHC data, being very rich, will have a large physics potential even after the active data taking.

Given the long lifetime of the LHC experiments and the large collected data volume, the issue of data preservation has to be addressed already during the active data taking. The LHC experiments can take advantage of the experience of the previous experiments' data preservation activities and apply timely the measures ensuring data preservation. Many of the challenges are directly addressed in the experiments' computing models which are designed to distribute and store the large data volumes in the computing centres connected via the worldwide grid. LHC experiments started with a fully distributed environment where the vast majority of the resources are located away from CERN. The LHC Computing Grid, which was approved by the CERN Council in 2001, has evolved into the Worldwide LCG (WLCG) with service support for all 4 LHC experiments.

The LHC experiments will need to address the risk of loss of data due to obsolescence of enabling technologies and due to physical damage. The risk due to physical damage is largely covered by the distributed storage in professional computing centres. The threat of obsolescence of hardware and software environment will require proactive measures ensuring that the data files will remain readable and usable in the long-term future.

A data preservation plan will be defined in order to prepare for the unavoidable migrations connected to software, external libraries, operating systems, storage media and the related hardware and in order to estimate the resources needed to take care of these migrations. A concrete stress test of a plan is to consider a use-case where an analysis done on reconstructed data from the first years of LHC running would need to be redone after the LHC long shut-down foreseen 2013-2014. Lessons learnt from such exercises will be incorporated in the long-term preservation of the data and associated software.

The details of the data preservation plan after the data taking will not be defined at this early stage, but the LHC experiments will follow with attention the procedures taken by the experiments which have recently completed data taking or are in the final analysis phase. This experience will be useful for a proactive planning of the long-term future.

While the preservation of the raw data is guaranteed by the experiments' distributed computing models, the physics results are preserved through publishing and storing them at external, persistent repositories. In addition to the written article, additional public data sets such as numerical values of the tables can be provided broadening the concept of the scientific publication. This is already being experimented and INSPIRE is planned to be the long-term platform for such additional information. Common efforts between experiments and theorists can be made more efficient if data are presented in a way that they can be combined and compared either with other experiments or with theoretical predictions. This will ensure a greater public re-use of the scientific data.

Between the raw data and the physics results, there is much valuable knowledge and know-how to consider. Preserving the relevant data and information during the many intermediate steps leading from the raw data to the final physics results will require attention. Most technical facts are recorded in experiments' internal notes but many well known and well defined details such as software versions and the set of updates, conditions, corrections, the identity of events with special properties and the location of the analysis-specific code may not be explicitly recorded. As all this is known when the analysis is on going it is matter of organisation and a limited amount of extra resources to preserve the full details. Part of the information is in collaborative media such as Twiki, posing an additional challenge to capture all relevant information. It is therefore important that the appropriate decisions are made to define the information to be preserved and the resources for the preservation activities are made available at this early stage of the experiment's lifetime. This will not guarantee that an earlier analysis can be redone in the future without technical modifications but it will guarantee that all technical knowledge connected to an analysis is preserved which is important for the internal efficiency of the experiment.

The LHC experiments will consider open access for their data with appropriate delays allowing each experiment to fully exploit the physics potential before publishing. The HEP data is complex and any public data will need to be accompanied with the software and adequate documentation. Simplified data is already provided by CMS making modest samples of selected interesting events available to educational programmes targeting high school students. The use of text-based file formats that are human-readable and largely self-explanatory, such as the ".ig" format which uses the JSON standard format, mean it is easier to read such files programmatically with for example C++ or Python, without the need for any experiment specific software. The use of common simplified formats for open access will be explored.

The LHC experiments are currently defining their data preservation and access policies and plans. The CMS experiment has recently approved a policy defining the data preservation and open access approach. Other experiments are discussing this issue and working towards policy statements based - similarly - on the levels of the data preservation model in the DPHEP context. The challenges are very similar across the experiments, and some nuances could exist at the access level, as a function of time. Each experiment will provide a plan how the policies will be implemented in the experiment-specific context.

# 5. DPHEP Common Projects

Among the projects and needs of the various experiments, certain areas have emerged as opportunities for common efforts to build preservation infrastructure within the community. Common infrastructures for data preservation can be jointly developed with the coordination and support of DPHEP, pooling the interests and resources of collaborations and other interested parties to eliminate duplication of effort.

In particular, technologies for data preservation, such as automated validation frameworks, are areas of potential collaboration. The development of a data storage solution for long-term preservation, given that day-to-day systems are unlikely to provide the level of data integrity security required may also be investigated. In addition, the RECAST project offers an alternative to maintaining the full data set.

Additionally, all collaborations have identified the need to preserve documentation and other high-level objects as a means of preserving critical know-how, and many are working with the existing knowledge management infrastructure, INSPIRE, to find appropriate permanent home for this information.

Finally, outreach and educational efforts using actual HEP data similarly benefit from pooling of resources, as many collaborations have made some efforts in this regard, and they can benefit from joint work. Within these areas some concrete projects have emerged, but the DPHEP group can also serve as a forum for the incubation of other relevant projects of common interest as they are identified.

## 5.1 A Generic Validation Framework for HEP Experiments

For data preservation to be useful, not only the data themselves must be preserved, but also the ability to perform some kind of meaningful operation on them. In the case of high-energy physics, this generally means preserving the software and environment employed to produce and analyse the data. While preservation of the analysis environment may be accomplished by freezing the experimental software and relying on the longevity of the frozen system, past experience suggests that this strategy would result sustain analysis capability for only a limited amount of time.

In order to preserve the analysis capability for longer period it would be beneficial to migrate to the latest software versions and technologies for as long as possible, substantially extending the lifetime of the software, and hence the data. It would therefore be beneficial to have a framework to automatically test and validate the software and data of an experiment against such changes and upgrades to the environment, as well as changes to the experiment software itself. A generic validation suite, which includes automated software build tools and data validation, would provide a solution to this problem that could be shared by different experiments. An illustration of such a validation system is given in figure 19.

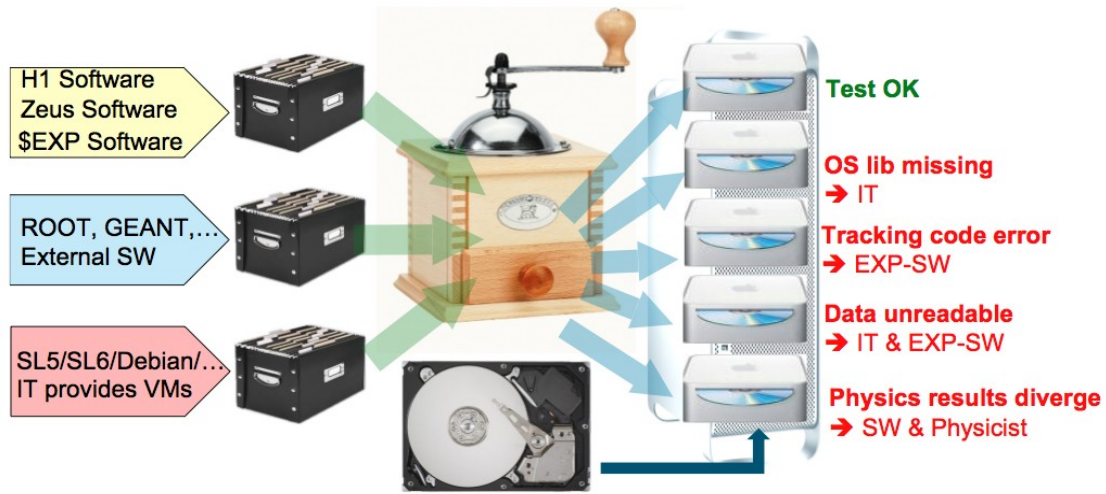

*Figure 19: An illustration of the basic idea behind the generic validation framework being developed at DESY.*

**Implementation**

Technically, this may be realised using a virtual environment capable of hosting an arbitrary number of virtual machine images, built with different configurations of operating systems (OS) and the relevant software, including any necessary external dependencies. Such a framework is by design expandable and able to host and validate the requirements of multiple experiments, and can be thought of as a tool to aid migration that will detect problems and incoherence, as well as identifying and reporting the reasons behind them. Such a framework would support a workflow such as the following:

1. In an initial, preparatory phase, the experimental software should be consolidated, the OS migrated to the most recent release, and any unnecessary external dependencies removed. Any remaining, well-defined necessary dependencies are then also incorporated. Analysis and data validation tests should then be defined and prepared, examining each part of the experimental software deemed necessary in the preservation model adopted.

2. A regular build of the experimental software is done automatically according to the current prescription of the working environment, and the validation tests are performed. At regular intervals, new OS and software versions will then be integrated into the system, under the supervision of experts from the host IT department and experiment.

3. If the validation is successful, no further action must be taken. If a test fails, any differences compared to the last successful test are examined and problems identified. Intervention is then required either by the host of the validation suite or the experiment themselves, depending on the nature of the reported problem.

4. The final phase occurs either when no person-power is available from the experiment or IT side or the current system is deemed satisfactory for the long-term need or stable enough. At this point the last working virtual image is conserved and constitutes the last version of the experimental software and environment. It should be noted however, that this now frozen system is unlikely to persist in a useful manner much beyond this point.

**Prototype**

A prototype version of the validation framework was successfully installed at DESY-IT during 2010. Contributions from the H1 (a precompiled analysis level executable), ZEUS (a precompiled MC production executable) and HERA-B (compilation) experiments, as well as a standard ROOT compilation test were incorporated into the framework and tested against three different OS (SL4, SL5, Fedora 13). The results of a test run of the framework are displayed graphically in figure 20, where successful (green) and failed (red) test results are shown.

|  |  | SL4 | SL5 | Fedora 13 |
|---|---|---|---|---|
| Compilation | ROOT V5.26 | -no F77 compiler gfortran found -libX11 MUST be installed | Estimated ROOTMARKS: 1534.29 | Estimated ROOTMARKS: 1512.76 |
| Run pre-compiled tgz using compat libs | H1 Data analysis | Processed 47243 events with J/Psi candidates Histogram written to jpsi_mods.root | Processed 47243 events with J/Psi candidates Histogram written to jpsi_mods.root | Processed 47243 events with J/Psi candidates Histogram written to jpsi_mods.root |
| Run pre-compiled tgz using compat libs | ZEUS MC prod | > ls -lh ZEUSMC.HFSZ627.E8954.GRAPE.Z01 4.2 MByte | > ls -lh ZEUSMC.HFSZ627.E8954.GRAPE.Z01 4.2 MByte | > ls -lh ZEUSMC.HFSZ627.E8954.GRAPE.Z01 4.2 MByte |
| Compilation | HERA-B | Compilation OK / DB connect fails | Compilation OK / DB connect fails | Compilation failed – needs code change |

*Figure 20: The results of the 2010 prototype generic validation framework at DESY. A series of tests were performed using software environments prepared with different OS configurations.*

**Towards the full scale implementation**

Following the implementation of the prototype described above, which demonstrated that experimental software could be run and validated in an isolated environment, the project has now entered the next phase. The design, development and implementation of the general version of the framework, including additional functionality such as an automated way of examining and reporting the results of the validation tests performed, is now underway. Although so far only the HERA experiments are involved, the framework should be generic enough that it may be further developed to include other experiments from other laboratories. The basic validation cycle employed by the system is illustrated in figure 21. Note the separation of the IT and experimental phases.

In addition to the common infrastructure provided by the IT division, the development and implementation of the tests by the participating experiments requires significant investment, even if basic validation structures already exist. As a first step, the number and nature of the experimental tests is surveyed, the level of which reflects the DPHEP preservation level aimed the participating collaboration.

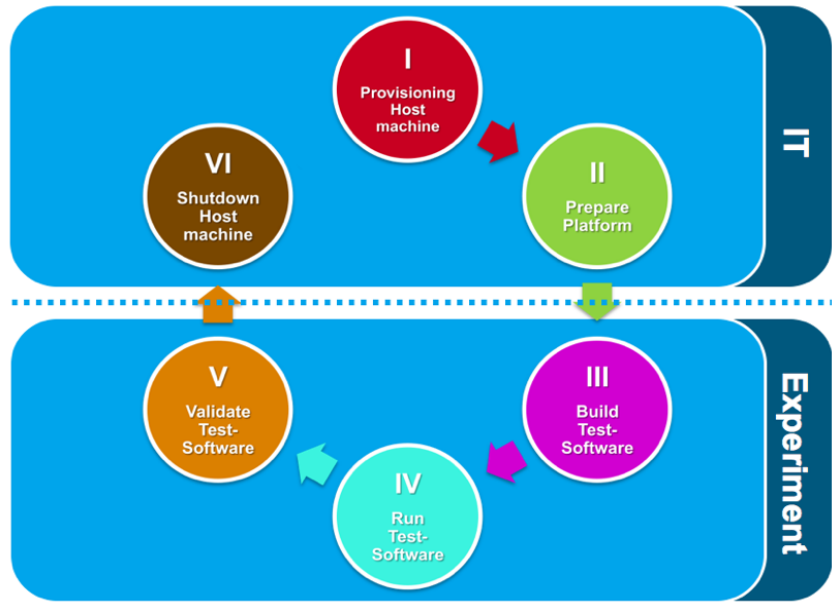

*Figure 21: A flow diagram showing the steps of the generic validation system under development at DESY. Note the separation of the experimental and IT parts.*

As an example, the preliminary structure of the tests to be installed by the H1 experiment is shown in figure 22. H1 plans to validate the full analysis chain, including compilation and execution of validation code for the full software environment, incorporating all aspects of production and analysis, as well as simulation and reconstruction software.

The left of figure 22 details the compilation of experimental and external software. This is considered as a series of tests, where the compilation of approximately 100 individual packages is carried out. The resulting binaries are stored as tar-balls on a central storage facility within the validation framework, where they are then accessible and used in the predefined tests, described on the right of figure 21. These tests are wide reaching, examining all areas of the H1 software including among others file production ("DST production" and "HAT/μODS"), comparison of analysis histograms ("Physics Analyses") and execution of experiment specific tools and macros ("h1oo/Fortran Executables"). H1 estimates a total of around 250 tests, including compilation, are required, to successfully validate the complete analysis chain, although it should be noted the implementation is still within the development phase.

The IT role in the validation project is key, and an initial 12 person months is required for the development and implementation, followed by about 6 person months per year for the maintenance and running of the framework. The person-power requirement for the implementation of validation schemes of the experimental software depends on which DPHEP preservation level is attempted. For a level 4 scheme, such as that envisaged by H1 as illustrated above, an initial 12 person months is required, again followed by 6 person months a year to provide the necessary support from the experimental side. It is important to note that such validation schemes require a long-term commitment from all involved parties, as previously illustrated in figure 12.

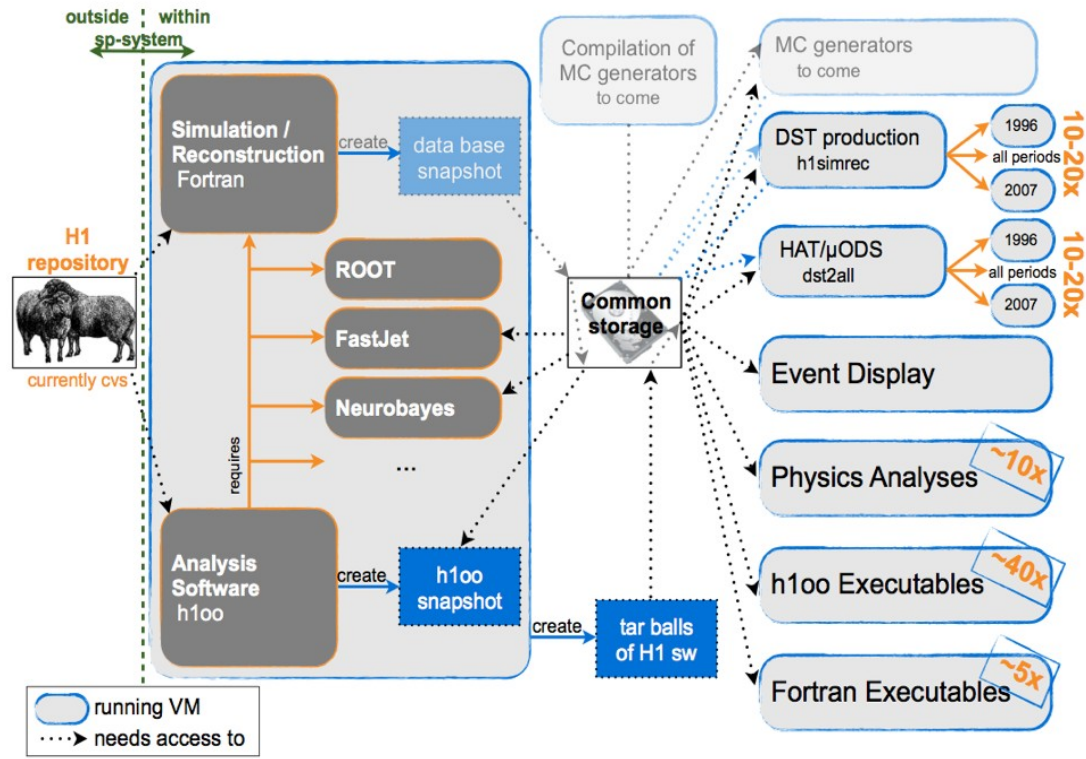

*Figure 22: A diagram showing the validation tests to be implemented by H1 within the generic validation system being developed at DESY.*

## 5.2 A Common Project on Archival Systems

The scope of the validation framework described in the previous section does not foresee an examination of the condition of complete data sets, but rather the use of smaller samples to test software changes. Mass storage solutions, such as dCache, are well established and employed by the majority of HEP collaborations. However, such systems were never intended to host data for a longer time period.

Taking the HERA data as an example, the DESY-IT division currently has no system able to fulfil all these requirements for the amount of data needed by the HERA experiments. Therefore, in conjunction with the larger DPHEP effort, the IT group are currently investigating the needs of the HERA experiments, establishing the necessary attributes required for such an archival system.

Commercial vendors have products for data archival which could fit the requirements. The current mass storage solutions are however not yet ready to easily integrate with them. One important point is that the data access is done via a HEP specific protocol: *dcap*. There are on going efforts by the dCache developers to offer other, standard data access protocols, where the most promising is certainly NFS 4.1. Other features of such a system would include: automatic migration to new media generation and technology, automatic data integrity checks, retrieval of the data themselves and metadata operations, built in redundancy for the most precious data and recovery from data corruption.

Initial studies are currently being carried out, and an estimated 12 person months is required within DESY-IT to develop an archival system for long-term, reliable storage of the HERA data. Dedicated person-power should also be foreseen for the maintenance phase.

## 5.3 The RECAST Framework

The ingredients needed for a typical analysis of HEP data include: the data itself and the software necessary to process it; selection criteria applied to the data (usually chosen to isolate some signal process); estimation of background processes satisfying the selection together with estimation of systematic uncertainties on those estimates; and an estimation of the amount of signal that will satisfy the event selection. Based on these ingredients a statistical statement may be made regarding the presence or absence of the hypothesised signal (perhaps in terms of the parameters of some underlying theory). Any meaningful result obtained from sharing HEP data will need to bring together these basic ingredients.

While there are many technical obstacles to preparing and using archived HEP data, one of the most challenging is the propagation of institutional wisdom needed to use the data properly. Even if the community is able to overcome the technical challenges in preserving HEP data, it will continue to be difficult to extract meaningful scientific results. The RECAST framework[56] provides a complementary approach.

The essence of the idea is to focus on analysis archival, as a given analysis encapsulates the event selection, observations in data, and estimates of backgrounds and systematic uncertainty. If these ingredients are archived and a pipeline is provided to process an alternative signal through the real detector simulation, reconstruction, and event selection, then any interested physicist would have all the ingredients necessary to arrive at new results from the data. The advantages of this approach are that it:

- Extends the impact of existing results from experimental collaborations
- Provides accurate interpretation of existing searches in the context of alternative models
- Does not require access to or reprocessing of the data (no need for new data access policies)
- Does not involve design of new event selection criteria
- Does not require additional estimates of background rates or systematic uncertainties

This method of recasting existing search results has been used to place constraints on alternative signals hypotheses, demonstrating that the RECAST framework provides a potential standard interface for processing alternative signals through archived analyses. Input has been sought from both the theoretical and experimental communities through a series of workshops and provided the original design of the framework. Furthermore, the use of RECAST was also discussed in the recommendations on the presentation of LHC results, released by the Les Houches working group on Searches for New Physics[57].

---

[56] K. Cranmer and I. Yavin, "RECAST: Extending the Impact of Existing Analyses, JHEP **1104** (2011) 038 [arXiv:1010.2506].

[57] S. Kraml et al, "Searches for new physics: Les Houches recommendations for the presentation of

The Perimeter Institute supported the development of a beta-version of the framework's database driven front-end and web-based API (application programming interface). The API is responsible for communicating new requests between the front-end (which collects requests for alternative signals to be processed through specific archived analyses) and the back-end (which is managed by an experimental collaboration and provides the library of archived analyses). The framework has been designed so that it:

- Standardises the format of such requests
- Maintains collaborations' control over the approval of new results with existing policies
- Allows new models to be considered even after a search is completed
- Complements data archival efforts

The heart of the RECAST framework is the library of archived analyses and the system for streamlining the processing of an alternative signal hypothesis through the analysis chain: the so-called "back-end" of the system. In the design, each participating collaboration would develop and maintain its own back-end, which includes ensuring that the collaboration approves any new results. Within ATLAS, CMS, and ALEPH efforts have begun to exercise this recasting technique.

It is worth noting that the technical requirements of RECAST do not directly fit into the DPHEP data preservation levels described in section 3. The back-end analysis archival system will require the reconstruction and simulation software (DPHEP level 4) as well as the analysis level software (DPHEP level 3); however, one need not preserve the full collision data as the recasting technique only relies on the events selected in the original publication. In some sense, RECAST is well described as additional information associated with a publication (DPHEP level 1) where the "information" is algorithmic in nature and requires level 4 functionality. The beta-version of the RECAST is now available[58].

## 5.4 Preservation of Documentation and High-level Objects

Not only the data and tools to analyse it, but also vast amounts of know-how must be kept to make preservation useful. Central infrastructures like INSPIRE, the successor to SPIRES, already exist and are working to preserve high-level data objects, and with additional effort can preserve internal notes, connections between internal documents, wikis, and other documentation. The primary challenges in this area are working to define and implement access policies for these materials and developing tools to ease the ingestion of material.

The HEP community, via the PARSE.Insight survey[59], and the DPHEP working group, recognises that preserving data in HEP involves not only the preservation of the raw data itself but also the preservation of vast amounts of know-how and existing data preservation activities in experiments have already identified documentation as crucial to these efforts. PARSE.Insight and DPHEP also highlighted the need to

---

LHC results", Eur. Phys. J. C **72** (2012) 1976 [arXiv:1203.2489].
[58] The beta version of RECAST can be found here: http://recast.perimeterinstitute.ca
[59] A. Holzner et al., "First results from the PARSE.Insight project", arXiv:0906.0485.

preserve and make accessible data objects with higher level of abstraction, in addition to the more conventional data and software artefacts. These are technically and logically much closer to publications than raw data, thus it makes sense to aggregate these high-level data sets with the corresponding literature in the scholarly record. INSPIRE is an existing third-party information system for HEP jointly developed and operated by CERN, DESY, Fermilab and SLAC. It is ideally situated to manage and preserve this additional know-how and documentation, as well as these higher-level objects for the entire field.

Since its public release in October 2011 at, INSPIRE is currently providing or planning general services:

- Complete search covering all preprints and published literature in HEP
- Google-like full text search of preprints (both arXiv and next pre-arXiv scans) and journal articles (via unique agreements with publishers)
- Coverage of theses and public experimental notes
- Storage of experimental notes from collaborations, hidden from public view when internal in nature
- Extraction and storage of figures from papers
- The capacity to deposit reasonably sized supplementary objects with papers (i.e. ROOT files, Mathematica, and similar)
- Cross linking with arXiv, NASA-ADS[60], HEPData[61], and other resources in the community
- Author identity management for similar names and for large collaboration papers
- Improved and seamless interlinking with HEPData and other services managing data
- Ability to issue industry-standard DOIs to documents and non-text files (figures, small data sets, additional tables) to allow the accurate tracking of their citations, together with the conventional papers
- Expansion of storage and access control for experimental notes and auxiliary material
- Increased coverage of theses and other material not in journals or arXiv, from past and present

These features, together with the links of INSPIRE with the four large labs and the established trust of the community give INSPIRE the opportunity to serve digital preservation in HEP in three areas:

1) Ingest and preserve more additional materials, such as documentation, internal notes and other supporting data from various stages of the evolution of a research result, integrate internal resources with other INSPIRE holdings, and link together with final papers and other objects that need to be shared within the research process.
2) Host high-level data files as part of the supporting information of papers. Make these directly citable and provide rich search and analysis functions

---

[60] The NASA-ADS service: http://adswww.harvard.edu
[61] The HEPData service: http://durpdg.dur.ac.uk/HEPDATA

       for non-text objects.
3) Aid in the creation of infrastructure for storing mid-level data such ROOT files, likelihood functions, together with the community. In particular, provide infrastructure for emerging technologies such as RECAST, enabling the connection of these projects to the scholarly record.

These areas are further described in the following sections, in addition to the common issue of access policies and availability, which is central to all of these areas.

**Access**

Internal materials ingested for preservation purposes may be sensitive or internal to varying degrees, depending on their nature, and their age, as well as the policies of the individual experiments to which they belong, which can also be function of time. INSPIRE, in conjunction with DPHEP and the collaborations, can implement policies, as defined by collaborations, to determine which materials would be publically accessible and which would be restricted, and how these would be restricted, and how those rights have to evolve with time, either from publication or connected to the lifetime of the collaboration. As an example, INSPIRE has already developed a system to manage author identity for HEP through web accounts linked to arXiv which are used to identify authors and their publication list. These accounts could be expanded, with additional effort which is currently not budgeted in the INSPIRE model, to accounts directly tied to laboratories and therefore collaboration. The simplest scenario would be the maintenance of an access list by the collaboration that applies to all materials, however this fails to account for policies that may be specific to certain materials, and crucially, breaks down as the collaboration ceases to persist as an entity. More complex scenarios could be implemented as well, such as "the authors of this paper should have access to this set of notes for as long as they have a computer centre account linked to this collaboration." Effectively, INSPIRE can implement polices tied to the duration of author's affiliation with a collaboration, with or without a "decay time" and detected automatically from computer centre accounts, paper authorship or feeds from collaboration-internal systems.

There could be significant development cost in determining and implementing access controls, which depends strongly on the nature of the controls desired, ranging somewhere between 6 and 12 person months according to the complexity of the scenarios and the sustainability of the effort, though very simple policies could be put in place right now. This has been already demonstrated for the H1 case for a single corporate account.

**Ingestion and preservation of notes, theses and publication histories**

In addition to secondary data files, INSPIRE is working with experiments to ingest and preserve documentation in a structured form such as internal notes so that they persist over time on a third party location, in access restricted form, as discussed above. At DESY the H1, HERMES and ZEUS collaborations, together with the DESY Library/INSPIRE staff have already worked on a project to upload internal restricted notes to INSPIRE, creating a subset of notes searchable within the same framework, and with the same feature set as public literature on INSPIRE. So far, the collections of notes are restricted to members of the collaboration, but now that they exist as INSPIRE records, making them available to the general public at some point in the future is a simple technological step.

The current access control mechanism is quite simplistic, but reasonably effective for short-term use, and is not substantially different from the procedures used by collaborations in private storage areas. This is a pilot project that could be extended with resources addressed above for access controls, along with similar effort for actual deposition and ingestion of shared among the experiment side and the INSPIRE/DESY Library side. Note that the DESY library resources could serve multiple experiments at DESY, and via INSPIRE, throughout the HEP community.

An effort is also on going at DESY to complete the list of Diploma and Doctoral theses containing results from the HERA experiments. This list goes back as far as the mid 1980s and often includes invaluable and unique information and research from the early days of the collaboration about the design of the experiments. In parallel to the digitisation of this information, an effort to ensure all such documents are made available on INSPIRE.

In addition to notes and theses, internal information such as presentations[62] might also be archived in this way. This would enable the presentation of the full history of a result, from the initial conference presentations and notes, through internal talks and notes, to a final submitted and refereed publication. Not only does this provide utility to the collaboration, but also provides the community as a whole the utility of linking conference presentations with the final published papers based upon them. There is already significant work and interest in this project at DESY with H1. This would be a direct continuation of the above project.

Some existing documentation, especially for older experiments, is still in paper form. INSPIRE and the laboratory libraries could offer a service of retrieving, cataloguing, scanning and passing this material through OCR (optical character recognition). Once ingested in INSPIRE this material would be preserved, made available in digital form in the long-term and, would be accessible via INSPIRE's Google-like search in the text of documents. While limited volumes of material fit readily within existing scanning/OCR projects at INSPIRE partner library, which are already in process, considerable volumes would imply additional resources. INSPIRE partner libraries can work with the collaborations to quantify the cost and obtain this service on their behalf, generating economy of scale and leveraging on their expertise in this field.

**Hosting of high-level data files**

INSPIRE can already preserve and index for the community high-level data (flat files containing additional numerical information, multi dimensional tables, ROOT files, simplified data formats for outreach or other purposes) that are submitted directly to HEPData, INSPIRE, arXiv or in some cases directly to the journals. These can be stored directly with the paper or papers to which they are associated, or even as stand-alone objects.

---

[62] Note that presentations in central conference servers (for example those held in Indico) might be made searchable by INSPIRE, while considerably more work is needed for identifying, ingesting and essentially rescuing from oblivion less visible conference materials. In all cases, additional effort would be needed to also enact long-term preservation, in common with all the other long-term preservation activities of INSPIRE.

As a part of its services to HEP, INSPIRE is already working closely with HEPData at the University of Durham to provide seamless interlinking between a paper-centric view of HEP, as represented in INSPIRE, with the high-level data which are either scraped by HEPData from the papers or, as it is becoming the norm more recently, directly submitted by the collaborations to HEPData. HEPdata records associated to a publication are now already presented as a separated tab with the publication record. They are stored as separate records facilitating an improved discoverability of these materials and also paving the way for value-added services (such as DOI assignments).

Another example of the expansion of INSPIRE services beyond bibliographic records as in the former SPIRES era is that INSPIRE currently extracts figures from HEP papers and indexes them. These are individually searchable through captions, or accessible from the papers. Further, if any of those figures have data in numerical form at HEPData, the two will be linked so that a single click could lead from one to the other.

Design and development of some of these services are in the INSPIRE core mission (such as the interoperability with HEPData and the citability and reuse tracking of high-level data sets). Further expansion and long-term sustainability would cost around 12 person months, in addition to what could be absorbed in the INSPIRE development roadmap. At the same time, a large-scale uptake of these services, and their long-term sustainability, would require additional development for their automation, between 9 and 12 person months, plus long-term curation resources to be possibly integrated with the data archivists effort or in close synergy with them.

**Infrastructures and services for data exchange and re-use**

One can easily imagine a particular data set, connected to a particular paper, is re-analysed in some form. INSPIRE can track this re-use, properly attribute citations, identify the individuals or the collaborations who have performed the analysis, and keep an incremental log of all these interconnections.

The INSPIRE team has decided to use industry standards to make the data sets on INSPIRE citable and traceable (as is the norm in several fields of the geo- and life-sciences). This service has now been implemented and is being operated as part of a global framework (DataCite). DOIs shall be used when reusing the data, data citation and reuse will be tracked on INSPIRE. The INSPIRE team is currently identifying best practices so that these can be implemented.

Since these solutions are newly emerging, the effort required for these projects is uncertain, however several months of effort to understand the requirements and several more to implement a framework around this type of data exchange and re-use would be a reasonable first estimate.

Emerging technologies such as RECAST offer opportunities to extend previous analyses to probe models beyond their original sensitivity (see section 5.3 for more details). Such enterprises bring to the fore some of the issues involved in the re-use of data such as the connection between existing literature and documentation of the existing data, as well as the literature produced by the new analyses. INSPIRE can offer RECAST and other similar solutions an infrastructure in which to connect old analyses, new analyses, documentation, and the data with the scholarly record.

Another area that has gained some attention is the publication of the combined LEP Higgs results in an extended format[63]. In their most general form, the published information is in the form of cross-section limits for physical Higgs masses for individual decay topologies. This information, can be thought of as DPHEP level 1, is used by tools such as Higgs-Bounds[64] to obtain limits on more general Higgs sectors. The limitation of this approach is that it is restricted to a single decay topology and cannot combine results from multiple decay topologies statistically. This statistical combination is done for the SM and restricted MSSM Higgs searches, where the relative contributions from the various signal topologies is specified. Using new technologies from RooFit/RooStats it is possible to publish digitally the combined likelihood function for the combined LEP Higgs search. An effort is on going from the LEP Higgs group to convert the results into this new format. Since this is the format used by the LHC, in principle, this information could even be used to combine results from the LHC.

**Long-term needs**

All the long-term preservation services outlined above have an upfront cost and a long-term curation and archival cost. The advantage of INSPIRE is that such upfront cost is strongly reduced compared to individual collaboration or laboratories, in that it would be incurred only once for developments benefitting the entire HEP community, and that it relies on an existing infrastructure and expertise. The long-term costs of maintaining the material are therefore also strongly reduced, and offer synergies with the role of the lab-based data archivists. Moreover, on going maintenance of the software and storage requirements are part of the lab-funded INSPIRE operations.

In addition to all upfront development costs, which drive the entire operation, there is an additional important cost, common to all the long-term preservation areas, which is to make INSPIRE compliant with all industry-standards for long-term preservation. The OAIS (Open Archival Information System) model defines the standards a system must meet to provide long-term preservation of digital information. This would then also allow receiving accreditation through international recognised standardisation bodies with which INSPIRE is in touch. In turn, this would allow credible long-term archival "data plans" to be presented to funding agencies which increasingly ask questions about data preservation. This effort would require around 24 person months, but part could be shared with the needs of providing an access infrastructure fitting the collaboration needs (such as answering the question "who can have access to which material and how long and where is this information stored", for each document preserved in INSPIRE).

---

[63] G. Abbiendi et al., [The ALEPH, DELPHI, L3, and OPAL Collaborations and the LEP Working Group for Higgs Boson Searches], "Search for the standard model Higgs boson at LEP", Phys. Lett. B **565** (2003) 61 [hep-ex/0306033].

[64] P. Bechtle et al., "HiggsBounds: Confronting arbitrary Higgs sectors with exclusion bounds from LEP and the Tevatron", Comput. Phys. Commun. **181** (2010) 138 [arXiv:0811.4169].

**Conclusion**

There is a unique opportunity to fully exploit the synergy between INSPIRE and DPHEP, and to optimise preservation resources by avoiding the repetition of many projects in many collaborations. Key areas identified are:

1) The ingestion and interlinking of additional data objects such as small data files, tables, figures, and other associated files in partnership with HEPData and arXiv.
2) The definition of the criteria for the long-term stewardship and access for these types of material, and the evaluation of the considerations needed to achieve the OAIS certification of INSPIRE as a preservation platform.
3) The expansion to other experiments of the existing pilot projects to ingest experimental notes, theses and additional documentation into INSPIRE. The experience gained from the demonstrator can also be used to enhance the access control and search mechanisms available for this information.
4) An analysis of the documentation existing in paper format today, for potential scanning and OCR projects, with ingestion in INSPIRE.
5) An investigation of the requirements for RECAST and other similar mid-level data exchange projects and an effort to understand possible extensions of existing infrastructures to integrate these projects with the scholarly record.

These projects promote the utility of INSPIRE, which is a resource that is centrally managed and relatively independent from experiments and collaborations, as well as from individual web pages, which tend to be unstable on a very short timescale. In partnership with long-term data archivists within the collaborations, these projects could be maintained and kept alive well beyond the timeframe of the native central infrastructures. In addition to providing independent long-term preservation, the powerful search capabilities of INSPIRE allow efficient discovery of information long after most of the contributors and experts are no longer actively involved.

## 5.5 Outreach

Scientists have a responsibility to teach others what we learn about our amazing universe. Having access to data from experiments all over the world can raise outreach efforts to the public to another level by letting non-experts interact with the scientific experience in a way not previously possible. The outreach tools developed for these efforts can also be used for undergraduate college courses and to train graduate students who will be the next generation of physicists at the frontier.

As our knowledge of the universe expands and new data are collected, we find it useful to return not only to our past conclusions, but also to the old data themselves and check whether or not it all survives in a consistent interpretation. This is one of the main thrusts of the DPHEP effort. But no one will understand the conclusions as deeply as those who interact with the experimental data. It would be a powerful tool in our arsenal, to have an infrastructure to store these data in a usable format so that others may walk through the same procedures and experiments that lead to our deeper understandings.

There are unique difficulties in presenting these vast and complex datasets to non-experts that are not faced by our colleagues in other scientific endeavours. But the onus is on us as scientists to rise to this challenge and learn how better to teach this material, thus new approaches need to be developed to educate and raise the overall awareness and appreciation of our work so that it becomes a more integral part of our culture.

A similar challenge faces graduate students who want to follow a career in particle physics and find themselves having to come up to speed very quickly on these complicated analyses. They may miss out on having the opportunity to learn the historical base that they need if they are expected to make the next big discoveries. By improving our educational tools for the general public, we will also develop better techniques for teaching new graduate students, who are our future collaborators, allowing them to more quickly contribute to the experiments. It may be that non-experts are able to provide an outside perspective, which benefits the HEP community in data visualisation, algorithm development or even the scientific analysis itself.

Well-defined, well-calibrated, and well-understood datasets can address the aforementioned outreach and education efforts. There are four main groups who can learn from and benefit from these data:

- The general interested public
- High school science students
- College students studying courses in particle physics or computing
- Graduate students in particle physics

Each group provides different challenges and each may interact with these data in different ways.

## Existing tools and projects

There is much work being done already by the Astronomy community in providing education web sites. The Sloan Digital Sky Survey (SDSS) provides education about their mission[65] as well as access to their data[66]. The Large Synoptic Survey Telescope (LSST) is still under construction, yet has an outreach programme in place[67]. The National Virtual Observatory[68] (NVO) provides access to a number of astronomical datasets along with tools and tutorials on how to interact with these data. The Particle Physics community has outreach web sites in the Contemporary Physics Education Project[69] (CPEP) and the Particle Adventure site[70], both run out of Lawrence Berkeley National Laboratory. While both provide a general education of our field and some basic exercises, they do not present data with which a non-specialist can work. The Quarknet project[71], run out of Fermi National Accelerator Laboratory, provides only limited access to data. Efforts for outreach have also been established at KEK[72]. Physics education research has shown that students develop a significantly better conceptual understanding of material through interactive learning. If interested parties can gain access to data, as well as tools and tutorials to walk them through projects of varying complexity, they can gain a far deeper understanding than if they were to simply read about these projects.

---

[65] The Sloan Digital Sky Survey: http://www.sdss.org/education
[66] Access to data from the SDSS: http://cas.sdss.org/dr5/en
[67] The Large Synoptic Survey Telescope: http://www.lsst.org/lsst/public/outreach
[68] The National Virtual Observatory: http://www.us-vo.org
[69] The Contemporary Physics Education: http://www.cpepweb.org
[70] The Particle Adventure: http://particleadventure.org
[71] The Quarknet project: http://quarknet.fnal.gov
[72] The B-lab project: http://belle.kek.jp/b-lab/b-lab-english

Compared to working with astronomical images, analysing HEP data presents different challenges and requires different tools. However, more and more web-based tools are being developed that encourage individuals to conduct their own analysis. The Gapminder project[73] makes available a wealth of data on international development and provides elegant plotting tools to demonstrate trends and correlations. This software was purchased by Google and has since been made available to the public for their own personal analysis[74]. In addition, the federal government in the USA has recognised that there is demand and have made much raw data available to the public[75], where people can download statistics on topics ranging from housing and construction to toxic emissions in different states. This trend is anticipated to continue and these web tools are available now to help physicists show how HEP data map onto scientific discoveries.

## Needs for HEP: New tools and common formats

Establishing how best to disseminate these data and developing tools that can be used to view the physics behind the numbers is another important aspect. While there are standard tools that most particle physicists use (e.g. ROOT), new tools need to be developed as a part of the outreach. One way to accomplish this is to provide templates that would allow high school students or undergraduate students to work with these data with more widely used tools (e.g. Matlab or Mathematica). Along with the tools, these data need to have a format with a low learning threshold for interested parties. The hope is that if these data are easy to access and the format is easy to understand, then others may develop even better visualisation tools. One major goal would be to provide a resource to which interested parties can upload their tools and projects for others to use.

More specifically, one possible implementation of a common data format for outreach would be to provide access to lists of particles that contain observable quantities like charge, momentum and particle type. With appropriate tutorials and analysis templates, it could be demonstrated how to manipulate momentum 4-vectors to produce histograms of invariant masses of the parent particles and show how new particles have been discovered in the past.

## Examples from BaBar

BaBar has put some initial effort into consideration of ways to make subsets of the BaBar dataset available for learning experiences. In the ten years that the BaBar experiment was running, this has become a very well understood dataset. There is a wealth of physics in these data due to the great variety of final states produced. Measuring the lifetime of long-lived states such as the $K_s$ or $\Lambda_0$ would introduce concepts from introductory physics courses (constant decay rates) and special relativity (time dilation), as well as more subtle experimental techniques (efficiencies and vertex finding). Searching for the D-meson by reconstructing it in both Cabibbo-favoured and Cabibbo-suppressed decay modes could supplement a more advanced course discussing the discovery of open charm and the recent Nobel Prize awarded to Makoto Kobayashi and Toshihide Maskawa. Individual modules could be developed to walk the user through a full analysis and allow them to check their results against published measurements. If the data are properly presented, we anticipate many others contributing analysis projects beyond those mentioned here.

---

[73] The Gapminder project: http://www.gapminder.org  
[74] Google tools: http://code.google.com/apis/virtualization  
[75] US government data site: http://www.data.gov

**Use of HEP data in the public arena**

The impact of having publicly accessible HEP data can be seen in the experience of one of this paper's authors in Science Hack Day[76] (SHD) events. In the words of one of the event's organisers, Ariel Waldman, "The mission of Science Hack Day is to get excited and make things with science! A Hack Day is a 48-hour-all-night event that brings together designers, developers, scientists and other geeks in the same physical space for a brief but intense period of collaboration, hacking, and building cool stuff".

In 2010, MC data from the BaBar experiment was brought to the SHD San Francisco event to provide the seed for a science-art crossover event. By the end of the weekend, the participants had used the data to produce a Particle Physics Wind Chime website, that allowed the user to map detector and particle properties (momentum, detector hits, particle-ID, etc.) onto sonic characteristics (volume, pitch, timbre, etc.). The excitement of the participants was palpable as they "heard" the sounds of particle physics! One could even hear the differences between different physics events. This experience was featured in a BBC science podcast and in Symmetry Magazine[77].

In 2011, the CMS experiment released a very small sample of real data for use in public outreach. These data are used in the I2U2 CMS e-lab website, but is mostly hidden behind the scenes. These data were discussed at the 5th DPHEP workshop at Fermilab and in November of that year, Matt Bellis took these data to the next SHD event. Again, people were excited about the idea of working with real live LHC data, and the participants hacked together a web page and online animation that described how di-muon events could be used to "discover new particles". The participants also learned how relativistic kinematics reveals new particles where Newtonian mechanics does not.

These data were re-used in an SHD event in Nairobi, Kenya. While Matt was unable to attend the event, he worked remotely with a group of computer programmers in Nairobi who had a keen interest in science. Again, the amateur enthusiasts were inspired about interacting with real data and across two continents the team hacked together another animation that displays these muons traveling over the world. This animation is available on YouTube[78].

These experiences show that there are non-scientists out there who want to go deeper into the science than simply what they see on TV. The threshold is high for them to be able to understand these datasets enough to actually do something with them, but with each experience, we learn more about how to explain how we do this challenging science. These experiences provide a real-life test bed of our data formats and our computational tools, which only improves both. If more data can be provided under the DPHEP aegis, we are certain that other unforeseen public science events will occur, to the benefit of us all.

---

[76] Science Hack Day: http://sciencehackday.com
[77] Symmetry Magazine, "An Ear for Science: The Particle Physics Wind Chime", June 23 2011.
[78] Muon Flight Paths: http://www.youtube.com/watch?v=ag7w0vgZj5g

**Conclusion**

Once there is a framework that generates interest in both the HEP and non-HEP community then the next step is to extend this work to other experiments. There are data from the world over which have led to a deeper understanding of the universe and it would be useful to have a central location to store these data so that others can walk through the same processes that led to these understandings. This effort would require input from all collaborations, but a central effort to coordinate formats and locations of approximately one person year would enable the smaller efforts from each collaboration to be leveraged into a common, useful outreach effort.

These uses for HEP datasets should be as much a part of the preservation effort as the re-analysis potential. The benefits to the community will resonate well beyond the experiment's running time and cannot be understated. The burden is on us, as scientists, to explain our evolving view of the universe. The usefulness of this archival, and maybe someday current, data for these educational efforts cannot be understated.

# 6. A Global Organisation for Data Preservation in HEP

**The landscape of data preservation in HEP**

Several stakeholders clearly emerge as concerned by data preservation in HEP:

1) The collaborations and the individuals that they comprise, who invested vast amounts of intellectual resources for the data taking and understanding.
2) The host laboratories, which have heavily invested in the accelerator infrastructures used for the production of the data.
3) The computing centres in the host laboratories and beyond, who are explicitly charged with the management of the data samples collected by the collaborations.
4) The national funding agencies that invested vast resources to support the construction and operation of the accelerators and the detector, as well as the data taking and data analyses process.
5) ICFA, as an overarching body overseeing global accelerator projects.
6) The public, government bodies, others interested in access to and use of data taken using public funding.

Since the beginning of 2009, with the launch of DPHEP, several initiatives have emerged both at the collaboration and at the host laboratory level, possibly under the aegis of funding agencies, to start addressing the issues of data preservation. These efforts, partly fostered by the activity of DPHEP, constitute an indispensable first step to address some of the most pressing problems.

At the same time, the field of HEP lacks a holistic vision for data preservation, ranging from technological solutions to the governance for the use and publication of results based on preserved data. Discontinuity in experimental programmes will lead, as it has done in the past, to a halt of preservation efforts. Even in the desirable cases in which individual groups were able to overcome their immediate difficulties, there is a strong risk of a lack of collective focus, ineffective transfer of knowledge and solutions, and no possibility to develop scalable discipline-wide solutions. This may lead to a duplication of efforts and lack of solid grounds for next-generation common projects pushing the borders of technology to solve present problems. This risk is made even worse by the extremely limited resources that can be allocated to preservation during or after the final stages of any single experiment.

The expected phasing out of the most active collaborations in data preservation in a few years may result in a rapid deterioration of the current situation. If no multi-experiment structure is created to take over the global supervision or coordination of the data preservation effort, any advantages gained so far may be lost due to a lack of sustainability. As such, a coordinated effort is extremely desirable. This section includes a proposal for a DPHEP organisation based on the current experience of a multi-experiment approach in high-energy physics.

**The DPHEP organisation**

In order to get the maximum impact out of the present projects, effectively multiplying the resources invested at the laboratory and collaboration level in data preservation, we believe in the necessity to have a multi-experiment and multi-laboratory approach for these activities, customised and consistent with the experiments and labs priorities and funding. This was also a strong recommendation of the intermediate document. Indeed, after the focused effort that is needed in each experiment to prepare the data sets for the long-term phase, the installation of an international coordinating body is recognised as the next priority. A possible form is to capitalise on the existing structure of the DPHEP ICFA study group and morph it from a fact-finding study into an operational model: the DPHEP organisation

**Activities of the proposed DPHEP organisation**

There are five main objectives for this DPHEP organisation:
1) Position itself as the natural forum for the entire discipline to foster discussion, achieve consensus, and transfer knowledge in two main areas
   a) Technological challenges in data preservation in HEP
   b) Diverse governance at the collaboration and community level for preserved data
2) Co-ordinate common R&D projects aiming to establish common, discipline-wide preservation tools
3) Harmonise preservation projects across all stakeholders and liaise with relevant initiatives from other fields
4) Design the long-term organisation of sustainable and economic preservation in HEP
5) Outreach within the community and advocacy towards the main stakeholders for the case of preservation in HEP

**Initial deliverables for the DPHEP organisation**

Within the first two years of becoming fully functional, the DPHEP organisation will achieve the concrete deliverables for each of its objectives described in table 6.

| Objective | Deliverable |
| --- | --- |
| 1.Positioning as forum | Catalogue of technical knowledge and practical solutions Description of possible alternatives for governance. |
| 2.Co-ordination of projects | Common R&D projects meet the expectations of the stakeholders. |
| 3.Harmonisation and liaison | Synchronisation of preservation projects in the field. Identification of areas where external knowledge needs to be transferred to HEP. |
| 4.Design sustainable future | Characterisation of discipline-wide toolkit for preservation Business plan for long-term preservation in HEP. |
| 5.Outreach and advocacy | Understanding of needs/opportunities for medium- and small-sized collaborations. Concrete discussions with funding bodies/laboratories. |

*Table 6: Deliverables of the DPHEP study group.*

**The operational model**

The DPHEP structure should be lightweight, yet able to harmonise existing activities, fostering new projects, informing all stakeholders. It is mostly built on the present one, with representatives from the laboratories, the experiments, the computing centres, officially appointed by their organisations, with oversight from the funding agencies. The newly created figures of laboratories or experiment data archivists would actively participate to activities, with one individual appointed by ICFA to chair the organisation. A crucial condition to ensure the success of the organisation and focus all resources, and thus multiplying their impact, is the creation of the position of a director of the initiative, with the responsibility to advance its agenda, harmonise all preservation activities in the discipline, and ultimately achieve the initiative deliverables. The organisation would continue to receive input from an advisory board, representing all stakeholders, and continue to report to ICFA.

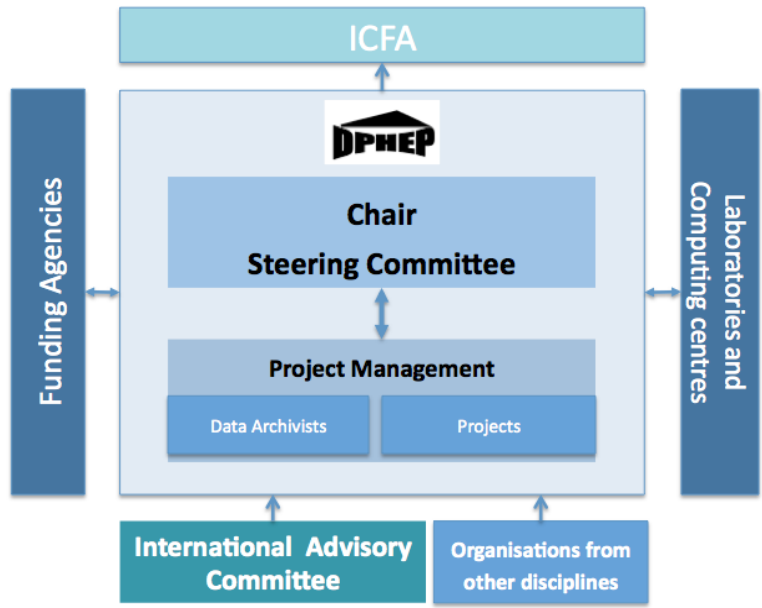

*Figure 23: DPHEP organisation and its associations.*

The following entities and their connections, as illustrated in figure 23, are to be defined in the operational model. A summary is also presented in table 7.

1) **DPHEP.** The organisation in charge with data preservation in high-energy physics. It is recognised as the unique body by the large laboratories represented in ICFA and gradually endorsed by national funding agencies. The organisation is chaired by a scientist, appointed by ICFA for a term of four years. The DPHEP Chair oversees the organisation, coordinates the Steering Committee, ensures the representation to the stakeholders and invites new members. It is a representative position ensured with a small fraction of an FTE. The operations are coordinated by a Project Manager (full time activity), who organises DPHEP events, coordinates funding proposals and ensures the information flow between the experiments and DPHEP. The organisation includes representatives from all member experiments (data archivists) as well as the personnel of the common projects. Contact persons ensure the link with similar organisations in other disciplines and favour common projects. The representatives are appointed by the participating experiments or other stakeholders to be part of the organisation.
2) **ICFA.** DPHEP is an organisation steered by the large HEP organisation represented by ICFA, and as such DPHEP reports to ICFA annually. The DPHEP chair is appointed by ICFA, after a proposal by the Steering Committee (SC). ICFA provides the oversight of the project and also ensures that the support committed to DPHEP is guaranteed.

3) **International Advisory Committee (IAC).** The IAC is formed by independent personalities in HEP as well as the wider data preservation community and provides DPHEP with specific advice on specific actions (workshops, alliances, projects, documents) or strategic plans. The IAC is nominated by the DPHEP Chair in agreement with ICFA.
4) **Steering Committee (SC).** The participating experiments, computing centres and other stakeholders directly involved in concrete actions for data preservation in HEP are represented in the SC. The SC is chaired by the DPHEP Chair and discusses and adopts medium-term strategy and supervises the production of the yearly progress reports. The SC encourages multi-laboratory projects and acts as a coordination board for the common projects.
5) **Funding Bodies.** The national funding agencies will participate to the direct funding of the DPHEP organisation. For instance, the post of Project Manager could be jointly funded by a combination of laboratories and agencies. Support for various projects could be attributed as well directly to DPHEP. The Chair is in charge with the communication with the funding agency. An annual meeting with funding agency representatives is envisaged, including representatives from large HEP laboratories.

| Organisational Body | Description | Input and positioning | DPHEP Output |
|---|---|---|---|
| **DPHEP** | Organisation for Data Preservation in High-Energy Physics | Projects in data preservation at experiment and laboratory level | Working groups on common projects, status report documents |
| **DPHEP Chair** | Overall coordination of DPHEP | Appointed by ICFA, represents DPHEP in relationship with other bodies | Yearly reports to ICFA, representation to other related scientific bodies |
| **DPHEP Project Manager** | Project management, administrative, technical, funding | Main operational coordinator, maintain contacts, organises meetings, lead proposals for funding | Reports to the steering committee |
| **Advisory committee** | Group of external personalities | Synergy with the wider HEP community, input from other fields and initiatives | Project proposals, documents for scrutiny |
| **Steering committee** | Internal executive body, chaired by the DPHEP Chair | Contributions from the participation members | Strategic and operational decisions |
| **Funding bodies** | Funding agencies are invited to take note on the progress reports and periodically analyse the relevance of the funding | Direct funding to the DPHEP organisation, under the supervision of the Project Manager | Quarterly progress reports |

*Table 7: Summary of the entities associated to the DPHEP study group.*

## Resources

The projects in data preservation develop at least three distinct levels, as discussed in the previous sections: experiment/collaboration level, multi-experiment initiatives and the global organisation DPHEP. General project may also encompass the theory-experiment interfaces or may be based on multi-domain collaborations. It is important to review the possible sources of funding for these activities in order to make sure that the proposed organisation is able to follow the existing funding schemes and is suited for an expected enlargement of the classical funding schemes used in HEP.

Resources for data preservation in the field come from different sources and are used at different levels. These can be described as follows:

1) Laboratory resources and/or funding agency resources through the experiments, address preservation problems specific to an experiment or a facility. These resources include the position of a data archivist. These resources are indispensable to prevent the on going catastrophic loss of data, and can their impact can be amplified by a coordinated approach.
2) Laboratories support in common the Project Manager position and commit the small fractions of personnel in the steering bodies.
3) Laboratories, individual funding agencies or a federation thereof support part of common R&D projects in preservation in the field, some initial aspects are described in the previous chapters. These projects, aligned to emerging industry standards in preservation will lay the foundation for a sustainable future in data preservation in the discipline, beyond ad-hoc and insufficient solutions.
4) Other institutions or grants that could be used to provide support for projects, operations, or access to data.

One could imagine that the common initiative and the common resources would be committed with the procedures customary in our field (expressions of interests, memoranda of understanding). In addition, it is expected that a strong cooperation is installed with other scientific domains, leading to common projects. These projects will most likely be managed in common structures. The proposed DPHEP organisation provides the necessary flexibility to initiate and adhere to new structures and funding schemes, in particular due to the executive structures managed by a dedicated full time position, the project manager.

The on going activities as well as the planned multi-experiment projects discussed in the previous sections lead to an estimation of the necessary resources, which is summarised in table 8.

| | Project | Goals and deliverables | Resources and timelines | Location, possible funding source, DPHEP allocation |
|---|---|---|---|---|
| **Experiment and laboratory** *Priority: 1* | Experimental Data Preservation Task Force | Install an experiment data preservation task force to define and implement data preservation goals. | 1 FTE installed as soon as possible, and included in upgrade projects | Located within each computing team. Experiment funding agencies or host laboratories. DPHEP contact ensured, not necessarily as a displayed FTE. |
| | Facility or Laboratory Data Preservation Projects | Data archivist for facility, part of the R&D team or in charge with the running preservation system and designed as contact person for DPHEP. | 1-2 FTE per laboratory, installed as a common resource. | Experiment common person-power, support by the host labs or by the funding agencies as a part of the on going experimental programme. A fraction 0.2 FTE allocated to DPHEP for technical support and overall organisation. |
| **Multi-experiment** *Priority: 3* | General validation framework | Provide a common framework for HEP software validation, leading to a common repository for experiments software. Deployment on grid and contingency with LHC computing also part of the goals. | 1 FTE | Installed in DESY, as present host of the corresponding initiative. Funding from common projects. Cooperation with upgrades at LHC can be envisaged. Part of DPHEP. |
| | Archival systems | Install secured data storage units able to maintain complex data in a functional form over long period of time without intensive usage. | 0.5 FTE | Multi-lab project, cooperation with industry possible. Included in DPHEP person-power. |
| | Virtual dedicated analysis farms | Provide a design for exporting regular analysis on farms to closed virtual farm able to ingest frozen analysis systems for a 5-10 years lifetime. | 1 FTE | The host of this working group should be SLAC. Funding could come from central projects and can be considered as part of DPHEP. |
| | RECAST contact | Ensure contact with projects aiming at defining interfaces between high-level data and theory. | 0.5 FTE | Installed with proximity to the LHC, the main consumer of this initiative, with strong connections to the data preservation initiatives that may adopt the paradigms. |
| | High level objects and INSPIRE | Extend INSPIRE service to documentation and high-level data object. | 0.5-1.5 FTE | Installed at one of the INSPIRE partner laboratories. |
| | Outreach | Install a multi-experiment project on outreach using preserved data, define common formats for outreach and connect to the existing events. | 1 FTE central + 0.2 FTE per experiment | A coordinating role can be played by DPHEP in connection with a large outreach project existing at CERN, DESY or FNAL. The outreach contributions from experiments and laboratories can be partially allocated to the common HEP data outreach project and steered by DPHEP. |
| **Global** *Priority: 2* | DPHEP Organisation | DPHEP Project Manager | 1 FTE | A position jointly funded by a combination of laboratories and agencies. |

*Table 8: Resources required by projects of the DPHEP study group.*

# 7. Outlook

This document is the result of the first large scale effort to coordinate at an international level the activities of data preservation and long-term analysis for a selection of large HEP experiments. The initial focus of the DPHEP group concerns data and experiments from colliders, a decision which was driven by an objective situation: in a time window of a few years, several flagship HEP collider programmes stopped their data taking: HERA at DESY in 2007, PEP-II at SLAC in 2008, KEK in 2010, and the Tevatron in 2011. Within the same time window, the successful start of the LHC experimental programme has shed new light on the historical paradigms and problems of data management in HEP, but also left room for new hopes. Such large, coherent and powerful communities as the ones gathered around the LHC experiments may already now discuss, enrich and adopt new programmes for long-term data preservation, following the first ideas described in this document.

Innovative approaches, resulting from technological developments and organisational progress, will certainly change the landscape of HEP data collection and analysis and will hopefully facilitate the preservation of the data. Without trying to look into a crystal ball, a few avenues can be imagined already now. The activity of the DPHEP study group over the last three years has lead to an overall awareness of the data preservation issue in HEP, but also made evident to all its members and for the community at large that there is a need for more action to be taken, in particular:

1) **Coordination**: There is a clear need, expressed since the very beginning for international coordination. In fact, all local efforts profit from an inter-laboratory dialog, from exchange in information at all levels: technological, organisational, sociological and financial. The organisation proposed in section 6 is not an artificial construction, it is just the reality achieved by a group of enthusiasts that should be brought to a long-term perspective by solid, commensurate and courageous decisions of the funding and coordination bodies responsible for the wealth of HEP experimental data produced so far.

2) **Standards**: One of the first systematic investigations of the DPHEP study group was to compare the computing models. With no great surprise, these were found to be quite similar. However, the technical solutions adopted in each experiment are diverse. This diversity originates from various constraints, most of them related to the local configuration, available expertise, resources, funding models and so on. These reasons are objective and part of the construction process. However, there is a strong need for more standard approaches, for instance in what concerns data formats, simulation, massive calculation and analysis techniques. An increased standardisation will increase the overall efficiency of HEP computing systems and it will also be beneficial in securing long-term data preservation.

3) **Technology:** The computing technologies used during the lifetime of an experiment are not necessarily adapted to a long-term access to a preserved data set. Furthermore, there is a common feeling that the preparation of long-term data analysis improves the stability and the reliability of the computing projects. It is striking to observe that the usage of some of the cutting edge paradigms like virtualisation methods and cloud computing have been probed systematically in the context of data preservation projects. These new techniques seem to fit well within the context of large scale and long-term data preservation and access.

4) **Experiments**: The reflections and the projects presented in this document do not cover the whole spectrum of problems and opportunities in HEP. Indeed, the DPHEP study group was initiated around the obvious complexity of the collider experiments, but has also systematically consulted other types of experiments. If the fixed target hadronic experiments presented issues very similar to those of collider experiments, reports from neutrino and astro-particle experiments revealed a rather different picture of the critical issues on data management. However, the main issues revealed by the DPHEP study group are easily extendable to other experiments. Conversely, the recent experience shows that new aspects revealed by different computing philosophies in general do improve the overall coherence and completeness of the data preservation models. Therefore the expansion of the DPHEP organisation to include more experiments is one of the goals of the next period.

5) **Cooperation:** High-energy physics has been at the frontier of data analysis techniques and has initiated many new IT paradigms (web, farms, grid). However, it is likely that other disciplines will equal or even overtake HEP in the coming years and decades in what concerns the quantity and the complexity of the necessary computing. In addition, in several cases (astrophysics, life sciences), the data is more shared: the open access and long-term preservation paradigms are more evolved than in HEP. In the context of an explosion of scientific data and of the recent or imminent funding initiatives that stimulate concepts as "big data", the large HEP laboratories will need to collaborate and propose common projects with units from other fields. Cooperation in data management: access, mining, analysis and preservation; appears to be unavoidable and will also dramatically change the management of HEP data in the future.

The new results from LHC and the decisions to be taken in the next few years concerning LHC upgrades and other future projects will have a significant impact on the HEP landscape. The initial efforts of the DPHEP study group will hopefully be beneficial for improving the new or upgraded computing environments as well as the overall organisation of HEP collaborations, such that data preservation becomes one of the necessary specifications for the next generation of experiments.

# Appendix A: Committees of the DPHEP Study Group

## International Steering Committee
DESY-IT: Volker Gülzow (DESY)
H1: Cristinel Diaconu (CPPM/DESY, chair)
ZEUS: Aharon Levy (Univ. Tel Aviv)
HERMES: Gunar Schnell (UPV/EHU)
FNAL-IT: Victoria White (FNAL)
DØ: Dmitri Denisov (FNAL), Gregorio Bernardi (LPNHE)
CDF: Giovanni Punzi (FNAL), Robert Roser (FNAL)
IHEP-IT: Gang Chen (IHEP)
BES III: Yifang Wang (IHEP)
KEK-IT: Takashi Sasaki (KEK)
Belle: Hisaki Hayashii (NWU), Leo Piilonen (VPI), Yoshihide Sakai (KEK)
SLAC-IT: Richard Mount (SLAC)
BaBar: Michael Roney (SLAC/Victoria)
SLAC: Amber Boehnlein (SLAC)
CERN-IT: Frederic Hemmer (CERN)
ATLAS: Fabiola Gianotti (CERN)
CMS: Guido Tonelli (CERN)
LHCb: Pierluigi Campana (CERN)
ALICE: Paolo Giubellino (INFN/Torino)
CERN/Scientific Information Service: Salvatore Mele (CERN)
CLEO: David Asner (Carleton)
JLAB: Graham Heyes (JLAB)
BNL: Michael Ernst (BNL/IT)
STFC: John Gordon (RAL)

## International Advisory Committee
Jonathan Dorfan (SLAC, co-chair)
Siegfried Bethke (MPI Munich, co-chair)
Young-Kee Kim (FNAL)
Hiroaki Aihara (U.Tokio)
Dominique Boutigny (IN2P3)
Michael Peskin (SLAC)
Gigi Rolandi (CERN)
Alex Szalay (JHU)

## Contact persons from other communities and projects
Fabio Pasian (Trieste) International Observatory for Astrophysics
Sayeed Choudhury (John Hopkins Univ. USA) Data Conservancy/Blue Ribbon
Adil Hassan (QMU London) DRESNET
Robert Hanisch (STSCI USA) IVOA

# Appendix B: Workshops of the DPHEP Study Group

**DESY January 2009**

http://indico.cern.ch/conferenceDisplay.py?confId=42722

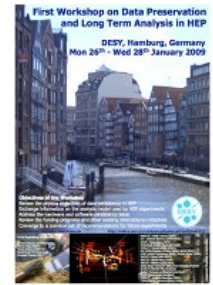

**SLAC May 2009**

http://indico.cern.ch/conferenceDisplay.py?confId=55584

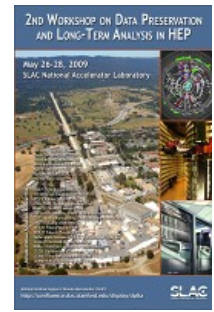

**CERN December 2009**

http://indico.cern.ch/conferenceDisplay.py?confId=70422

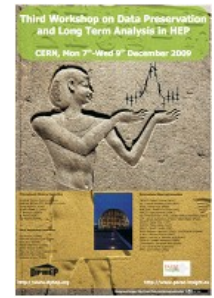

**KEK July 2010**

http://indico.cern.ch/conferenceDisplay.py?confId=95512

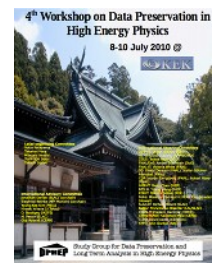

**Fermilab May 2011**

http://indico.fnal.gov/conferenceDisplay.py?confId=3977

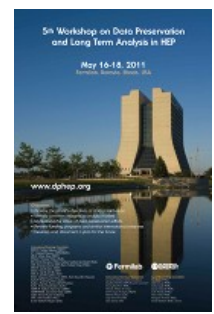